\documentclass[preprint,authoryear,12pt]{elsarticle}

\usepackage[T1]{fontenc}
\usepackage{lmodern}
\usepackage[utf8]{inputenc}
\usepackage{amsmath,amssymb,amsthm}
\usepackage{graphicx}
\usepackage{bm}
\usepackage{longtable} 
\usepackage[colorlinks=true,allcolors=blue]{hyperref}
\usepackage{microtype}  
\graphicspath{{./}}

\newcommand{\Zt}{\mathbb{Z}_2}
\newcommand{\R}{\mathbb{R}}
\newcommand{\Hone}{H^{1}}
\newcommand{\dz}{\delta^{0}}
\newcommand{\Minc}{M_{\mathrm{inc}}}
\newcommand{\Mcap}{M_{\mathrm{cap}}}
\newcommand{\CF}{\mathrm{CF}}
\newcommand{\CNT}{\mathrm{CNT}_2}
\newcommand{\Vs}{V^{\star}}
\newcommand{\code}[1]{\texttt{\def\_{\textunderscore\allowbreak}#1}}
\journal{Journal of Mathematical Psychology}

\begin{document}

\begin{frontmatter}

\title{Binding-Motivated Contextuality:
A Cross-Domain Cyclic Test in Perception and Judgment}

\author[cuny]{Adam Y. Shavit}
\ead{as1127@hunter.cuny.edu}
\affiliation[cuny]{organization={Hunter College and the Graduate Center, CUNY},
  country={USA}}

\begin{abstract}
Psychophysics and decision research study perceptual binding and judgment contextuality separately. We argue a binding-motivated perceptual task and judgment contextuality are scored against the same cyclic
noncontextuality inequalities and share one \emph{convex} global-consistency geometry (the correlations that mixtures
of globally consistent assignments can produce), not one cohomology class. Only
\emph{contextuality} is tested, since binding's own residual vanishes here. Building on sheaf
formulations of predictive coding (Seely 2025) and contextuality (Abramsky \&
Brandenburger 2011), a cyclic set of pairwise judgments without signalling admits a noncontextual
explanation exactly when the cyclic inequalities hold (judgments
do signal, so the test uses the disturbance-corrected form), and
its severity is measured by the \emph{complete} contextual fraction CF, the share of
responses no noncontextual model reproduces (Abramsky,
Barbosa \& Mansfield 2017) rather than by the \v{C}ech invariant, which can miss it
(Car\`u 2017). The perceptual arena uses
\textbf{two binary judgments per} cyclic-dominance pairing; an appendix rejects the one-bit alternative on
construct-validity grounds. The central test is \textbf{cross-domain}: one cohort performs
both arenas, and a shared latent tolerance predicts a positive correlation between their
signed cyclic margins $V^{\star}$, the pre-clamp quantities behind CF. Both are
\emph{inconsistency} scores, and general response consistency confounds a bare
correlation, so each arena's
margin is first regressed on its own variance- and reliability-matched control, and the
residuals are correlated. A positive result would support a shared
residual association, not a common causal mechanism, which we prove this design cannot
identify at any sample size. All three tests are designed but unrun, and the perceptual
arena awaits pilot gates. This is a \textbf{methodological proposal},
not a confirmatory report.
\end{abstract}


\end{frontmatter}

\emph{Rigor tags:} \textbf{[R]} proved/verified, \textbf{[R-refuted]} proved
\emph{false} --- a claim this paper made and has now disproved, kept visible rather than
quietly dropped, \textbf{[C]} conjecture, \textbf{[A]} analogy.

\section{Two problems, one shape}\label{sec:two}
\begin{figure}[htbp]\centering
\includegraphics[width=0.92\linewidth]{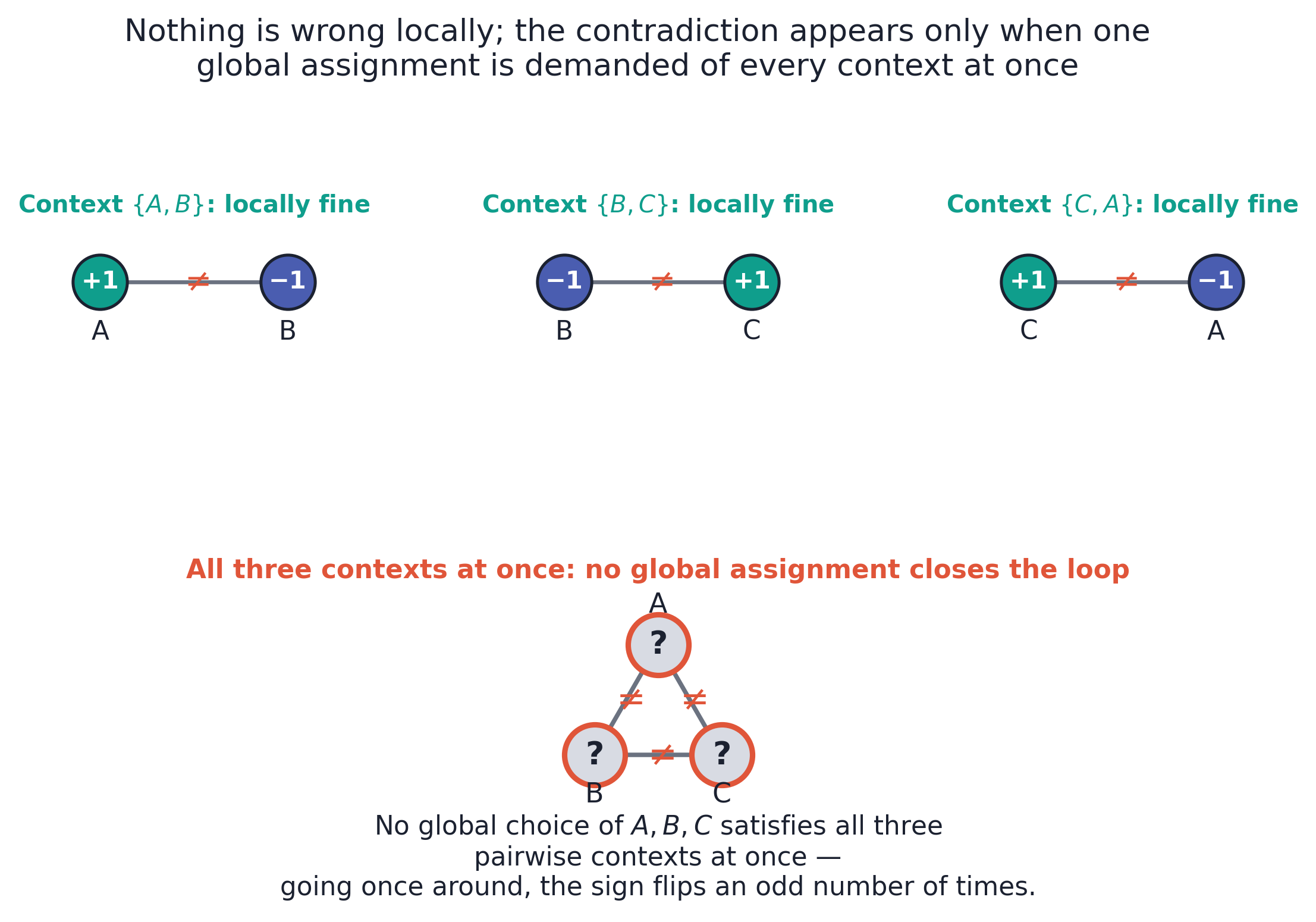}
\caption{The core thesis without notation. Three contexts of a frustrated cyclic
judgment, each perfectly interpretable on its own (\emph{top row}) --- but no single
global assignment to all three contents satisfies every pairwise context at once
(\emph{bottom}), because the required sign flips an odd number of times going once
around the loop. That odd number of flips is the same $n{=}3$ parity contradiction proved
in \S\ref{sec:ncycle} and drawn again, with the monodromy computed explicitly, in
Figure~\ref{fig:monodromy}. The panels above stage it as three separate local views
rather than one graph, before $\Hone$, $\CF$, or Contextuality-by-Default are
introduced.}
\label{fig:local-global}
\end{figure}

The binding problem asks how asynchronous features become one experienced event.
The contextuality of judgment asks why locally-coherent responses ---
question-order effects, the conjunction fallacy --- admit no single joint
probability distribution. Both are failures of \emph{global consistency over
consistent local parts}. Sheaf-theoretic methods answer such failures with a family of
tools rather than with one theorem covering them all. In a specified extension or
torsor problem a cohomology class can obstruct global compatibility; for empirical
probability models, global extendability is instead tested by the noncontextual
polytope, and the standard cohomological witnesses need not be complete. Abramsky \&
Brandenburger (2011) showed contextuality corresponds \emph{exactly} to the failure of
a global section; the finer \v{C}ech class of Abramsky et al.\ (2015) only
\emph{witnesses} it (\S\ref{sec:ncycle}). Seely (2025) built the operational sheaf for
predictive-coding inference. Our contribution places the perceptual and cognitive tasks under
the same \emph{convex} cyclic global-consistency obstruction, with one computable severity scale
and two matched experiments --- argued informally here, proved in \S\ref{sec:ncycle}.
Figure~\ref{fig:local-global} illustrates the thesis without notation: three locally
consistent views and a loop that cannot be closed.

\noindent One scoping condition belongs here rather than only in the appendix that
establishes it. The binding sheaf's \emph{geometric} residual does \textbf{not} carry this
obstruction: for the frustrated odd cycle its coboundary is invertible, so that residual is
identically zero while the contextual fraction $\CF$ sweeps $(0,1]$ (\ref{app:prop15a},
part (i)). ``One obstruction'' is supported for the \emph{convex} obstruction defined by
violation of the noncontextual correlation polytope (\S\ref{sec:ncycle}). The binding sheaf
enters as motivation for the severity split, not as the object the unification runs through.

\noindent\textbf{The paper's genre, stated once so that every later claim can be
read against it.} Two genres were available here and they license different sentences. One
is a \emph{confirmatory test}: a study that has been run, whose named statistic has known
operating characteristics under the nulls that actually threaten it, and which reports
whether a hypothesis survived. The other is a \emph{theoretical and methodological
proposal}: a claim about what two literatures share, made precise enough that a specific
study could refute it, together with the design, the statistic, the simulated calibration
and the pilot gates that study would need. \textbf{This paper is the second.} No data are
reported; the perceptual arena is a proposed instantiation whose stimulus parameters a
pilot must fix; the named cross-domain statistic is calibrated only against the
pure-confound null it was simulated under, and the two alternatives to it are not
calibrated at all; and the design provably cannot identify a common causal mechanism at any
sample size (\S\ref{sec:tests}). We therefore say \emph{primary pre-specified test} rather
than ``primary confirmatory test'' for the analysis \S\ref{sec:tests} names: the phrase
means the one analysis fixed in advance, ahead of any data, whose verdict the paper agrees
to be bound by --- not an analysis that has already confirmed anything. ``Confirmatory
use'' still appears where it is the correct statistical term: a \emph{future} tier whose
requirements a statistic has not yet met. The limitations this framing makes explicit are stated in
\S\ref{sec:tests} and \S\ref{sec:prior} rather than softened: choosing the proposal genre
is what lets them be stated plainly instead of being carried as tension against a
confirmatory claim the evidence does not support.

\section{The framework in brief}\label{sec:framework}
\emph{Plain-language on-ramp (for readers fluent in Contextuality-by-Default,
CbD, but not sheaves). A
\textbf{sheaf} attaches local data to each part of a system together with rules
for how overlapping parts must agree (Bredon 1997); a \textbf{global section} is one assignment
consistent everywhere at once --- here, a single coherent percept, or a single
joint distribution reproducing every context. The \textbf{first cohomology}
$\Hone$ is where the obstruction lives, and the \textbf{obstruction class}
$[a]\in\Hone$ of the data at hand is the obstruction itself: in a specified extension
problem, $[a]=0$ means the local pieces glue into a global whole and $[a]\neq0$ means they
are locally consistent yet globally impossible. For the probability models tested here a zero
class does not by itself rule contextuality out (\S\ref{sec:ncycle}). The distinction is load-bearing here, because
the group does not move with the data --- $\Hone(C_{n};\Zt)\cong\Zt$ for
\emph{every} cycle, odd or even --- so what separates a frustrated triangle from
an unfrustrated square is the class, not the group. The contextual fraction $\CF$
is a third object again: a real-valued convex measure of how much of the data
resists any global model, not a cohomology class. \textbf{$\Hone$ is reserved
throughout for this literal cohomology group, and $[a]$ for a class in it}; where later sections use the phrase
``shared obstruction'' for the binding and contextuality readings, it names the
convex correlation-polytope obstruction ($\CF$ and its relatives) and never
$\Hone$ itself. The two do not coincide anywhere in this paper: \ref{app:prop15a}
proves they cannot. Table~\ref{tab:notation} at the end of this section
collects every symbol with the distinctions that are load-bearing. In the operators
below, the coboundary $\dz$ records
how much adjacent local pieces disagree, and its Laplacian $L_{P}$ diffuses those
disagreements.}

Model perception as a cellular sheaf $\mathcal{F}$ over a bounded causal base $W$ --- a
finite poset of moments, of small and fixed size. Stalks are momentary sensory states;
restriction maps are the lossy reductions that compress a high-dimensional input to a
low-dimensional percept. Earlier drafts attached numerical ranges to both the capacity
and the compression ratio. They are removed here rather than sourced: no cyclic result
below depends on them, and quoting empirical endpoints without a citation does the
modeling choice no good. The construction needs only a bounded base and lossy restrictions. The
bound present moment is a \textbf{global section}; a percept's coherence is
governed by the coboundary $\dz$ and two sheaf Laplacians that must not share a
symbol: the \textbf{Euclidean} quadratic-form Laplacian
$L_{E}:=(\dz)^{\mathsf{T}}\dz$, which is what the binding energy penalises, and the
\textbf{precision-metric} generator $L_{P}:=P^{-1}(\dz)^{\mathsf{T}}\dz$ --- $P$ is the
positive-definite precision (inverse-variance) matrix for the cue coordinates --- which is
what diffuses the disagreements in the metric perception actually uses. They have the
same kernel, $\ker L_{P}=\ker L_{E}=\ker\dz$, but, unless $P$ commutes with $L_{E}$ (as at
$P=I$, where the two coincide), only $L_{P}$ is self-adjoint for
$\langle x,z\rangle_{P}=x^{\mathsf{T}}Pz$, so only $L_{P}$ gives the $P$-orthogonal
rather than the Euclidean harmonic projection. Soft binding minimizes
$J(s)=\tfrac12\|s-c\|_P^2+\tfrac12\lambda\|\dz s-y\|^2$, which trades fidelity to
the cues $c$ against a penalty on failing the relational demand $y$ the overlaps
impose; $\lambda\ge0$ sets the strength of that penalty, with $\lambda=0$ leaving the
cues unbound and larger $\lambda$ enforcing the overlap constraints more strongly.
Minimizing it splits the response
into a \textbf{capture shift} $\Mcap$ (diffusive pull between cues) and an
\textbf{incoherence residual} $\Minc=\|\Pi_{\operatorname{coker}\dz}\,d\|$, where
$\Pi_{\operatorname{coker}\dz}$ projects the mismatch onto the directions no global
assignment of vertex states can reach --- $\Minc$ is the
$\lambda\to\infty$ hard-binding limit of the $\lambda$-dependent form derived in the
full appendix, where
$d:=y-\dz c$ is the edge-wise mismatch between what the overlaps demand and what the
raw cues supply, and $\Minc$ is the part of that mismatch no single percept absorbs.
\textbf{The demand $y$ is what makes incoherence possible.} Were the conflict written as $\dz c$ it
would be exact by construction, its cokernel component identically zero, and the
whole incoherence channel empty; only relations the overlaps impose, and that no
global assignment of vertex states generates, can leave a residue.

\noindent\textbf{The hard-binding limit, and the condition under which it is cue
fusion.} Minimizing $J$ at finite $\lambda$ gives
$s_{\lambda}=(P+\lambda(\dz)^{\mathsf{T}}\dz)^{-1}(Pc+\lambda(\dz)^{\mathsf{T}}y)$.
As $\lambda\to\infty$ the minimization first makes $\lVert\dz s-y\rVert$ as small as it
can be, then picks the $P$-closest such $s$ to the cues:
\begin{equation}
  s_{\infty}=c+P^{-1}(\dz)^{\mathsf{T}}
             \bigl(\dz P^{-1}(\dz)^{\mathsf{T}}\bigr)^{\dagger}(y-\dz c).
  \label{eq:hardbind}
\end{equation}
In words: the percept starts at the raw cues and is moved only as far as the
unmet relational demand $y-\dz c$ forces, in the direction the precisions make
cheapest. That move absorbs only the part of the demand lying in the image of $\dz$. The
irreducible edge residual it
leaves is $r_{\infty}=\bigl[I-\dz{\dz}^{\dagger}\bigr](y-\dz c)$, whose norm is the
$\Minc$ defined above. \textbf{If $y=0$}, and more generally exactly when $(\dz)^{\mathsf{T}}y=0$,
Equation~\eqref{eq:hardbind} collapses to a projection of the cues themselves,
\begin{equation}
  s_{\infty}\big|_{y=0}=\Pi^{P}_{\ker\dz}\,c,
  \qquad
  \Pi^{P}_{\ker\dz}=I-P^{-1}(\dz)^{\mathsf{T}}
                     \bigl(\dz P^{-1}(\dz)^{\mathsf{T}}\bigr)^{\dagger}\dz ,
  \label{eq:vertexproj}
\end{equation}
the $P$-orthogonal harmonic projection onto the globally consistent sections. That
restricted case, and not the general relational demand, is what equals the
precision-weighted cue-fusion estimate --- the maximum-likelihood estimate under
independent Gaussian likelihoods with known precisions $P$. The identity coincides with
the Bayesian posterior mean only under a flat prior on the latent; with a proper prior
the Bayesian estimate carries an extra prior-precision term and the identity fails
\textbf{[R, given the $P$-metric and given $y=0$]}.

\noindent\textbf{Two projectors on different spaces.}
The maximum-likelihood identity above follows from the chosen precision-weighted energy,
not from topology alone, and the two operators it uses must not be run together. The
coboundary carries vertex states to edge demands, $\dz:C^{0}\to C^{1}$, so
$\Pi^{P}_{\ker\dz}$ of Equation~\eqref{eq:vertexproj} acts on \emph{vertex} states and
returns a percept, whereas $I-\dz{\dz}^{\dagger}$ acts on \emph{edge} demands and returns
the part of the conflict no percept absorbs. At $P=I$ the vertex projector is the
unweighted Moore--Penrose form $I-{\dz}^{\dagger}\dz$, and it is that one --- never the
edge projector $I-\dz{\dz}^{\dagger}$ --- that equals the maximum-likelihood fused
percept. The two are different operators on different spaces, and on a general sheaf
base they need not even have the same dimensions. The edge projector, unlike the vertex
one, does not depend on $P$ at all: the binding energy weights only the vertex space, so
$I-\dz P^{-1}(\dz)^{\mathsf{T}}(\dz P^{-1}(\dz)^{\mathsf{T}})^{\dagger}=I-\dz{\dz}^{\dagger}$
for every positive-definite $P$, which is why $\Minc$ can be written without a precision
weight (verified, with negative controls on both directions:
\code{scripts/p32\_projector\_direction\_check.wl}). Modulo the precision reweighting on
the vertex side, the severity split is Seely's Hodge decomposition. Novelty therefore
cannot come from fusion; it must come from a configuration in which no global section
exists at all.

{\footnotesize
\setlength{\tabcolsep}{5pt}
\setlength{\LTcapwidth}{\linewidth}
\begin{longtable}{lp{0.71\linewidth}}
\caption{Notation, with the distinctions that are load-bearing. Three pairs are easy to
run together and are kept apart throughout: $\Hone$ against $\CF$ (a group against a
real-valued convex measure), $\Minc$ against $\Minc^{\star}$ (a seminorm against a facet
slack), and $V$ against $\Vs$ ($V$ maximises over facets; $\Vs$ scores one fixed in
advance --- and in \ref{app:prop15a}, where the model is non-signalling, $V$ additionally
carries no $\Delta$ because none is needed there). The
word ``obstruction'' is not used alone in any paragraph containing more than one of
these.}
\label{tab:notation}\\
\hline
\textbf{Symbol or phrase} & \textbf{Reserved for, and for nothing else}\\
\hline
\endfirsthead
\multicolumn{2}{l}{\textit{Table~\ref{tab:notation}, continued.}}\\
\hline
\textbf{Symbol or phrase} & \textbf{Reserved for, and for nothing else}\\
\hline
\endhead
\hline
\multicolumn{2}{r}{\textit{Continued on the next page.}}\\
\endfoot
\hline
\endlastfoot
$\Hone$ & A named cohomology group of a specified sheaf or coefficient system --- never
a synonym for ``has no global section,'' and never a convex statistic. No dataset moves it.\\
$[a]$ & A specified obstruction or cohomology class in such a group. $[a]=0$ means the
local pieces glue.\\
$\dz$ & The coboundary $C^{0}\to C^{1}$: vertex states to edge demands.\\
$P$ & The positive-definite precision (inverse-variance) matrix for the cue coordinates.
A \emph{metric on the vertex space}, never a projector and never an operator on edges. As
a matrix it appears only as a weight --- in $\langle x,z\rangle_{P}=x^{\mathsf{T}}Pz$, in
$L_{P}$, and as the superscript of $\Pi^{P}_{\ker\dz}$.\\
$\Vs_P$, $F_P$, $\Delta_P$ & A \emph{second} use of
the letter, and the only other one: as a plain subscript on a scalar it names the
\textbf{arena}, $P$ for perceptual against $J$ for judgment --- so these run against
$\Vs_J$, $F_J$, $\Delta_J$ (\S\ref{sec:tests} onward).
Nothing is weighted by the precision matrix there. The two uses never meet --- the matrix
is a matrix and never a subscript on a scalar, the arena label is a subscript and never a
matrix --- but a reader who met only the row above would be entitled to misread
$\Delta_P$, so it is defined here rather than left to context.\\
$\Pi^{P}_{\ker\dz}$, $\Pi_{\operatorname{coker}\dz}$ & The two projectors, on two
different spaces, kept apart throughout. The first is $P$-weighted and acts on
\emph{vertex} states, returning the fused percept in the restricted case $y=0$. The second
acts on \emph{edge} demands, returning the part of the conflict no percept absorbs; it
equals $I-\dz{\dz}^{\dagger}$ for every positive-definite $P$, which is why it carries no
superscript. Both are written $\Pi$, so that no projector is ever written $P$.\\
$L_{E},L_{P}$ & The Euclidean quadratic-form Laplacian $(\dz)^{\mathsf{T}}\dz$ and the
precision-metric generator $P^{-1}(\dz)^{\mathsf{T}}\dz$. Same kernel, different adjoints.\\
$\Mcap$ & Capture shift: the diffusive pull between cues in the binding energy.\\
$\Minc$ & The \emph{fixed linear} edge residual from the binding energy, $\lVert(I-\dz{\dz}^{\dagger})(y-\dz c)\rVert$.
A seminorm. Identically zero on the frustrated odd cycle (\ref{app:prop15a}(i)).\\
$\Minc^{\star}$ & The \emph{positive convex} cycle-facet slack $\max\{0,s_{\max}-(n-2)\}$.
A different functional from $\Minc$, and the one that co-vanishes with $\CF$.\\
$\gamma$ & The adjacent-correlation vector of a cyclic empirical model, in $[-1,1]^{n}$.\\
$s_{\max}$ & The maximum of $\sum_i\varepsilon_i\gamma_i$ over odd sign patterns
$\varepsilon$.\\
$\Delta$ & Measured direct influence (signalling): the summed absolute shift of a
content's marginal between its two contexts.\\
$V$ & In \ref{app:prop15a}, where the model is non-signalling, the raw facet violation
$s_{\max}-(n-2)$, whose positive part is $\Minc^{\star}$. In \S\ref{sec:tests}, the signalling-corrected
$s_{\max}-(n-2)-\Delta$. The two coincide exactly when $\Delta=0$.\\
$\Vs$ & The signed cycle-facet margin after the stated disturbance correction, on a facet
\emph{fixed in advance}: $\sum_i\varepsilon^{\star}_i\gamma_i-(n-2)-\Delta$. Negative below
threshold. The primary pre-specified statistic.\\
$\CF$ & Contextual fraction of an empirical model, or of its consistification when the
system signals. Clamped at zero: $\CF=\tfrac12[\Vs]_{+}$ when $\varepsilon^{\star}$ is the active facet.\\
$\CNT$ & The Contextuality-by-Default measure, related by $\CF=2\,\CNT$ for cyclic
systems (Cervantes 2023).\\
$c^{\ast}$ & The uniform-anticorrelation threshold $-(n-2)/n$.\\
``shared geometry'' & The common global-assignment and cycle-polytope structure the two
arenas are scored against. Claimed here.\\
``shared mechanism'' & A causal claim requiring an identification step this design does
not carry (\S\ref{sec:tests}). \emph{Not} claimed here.\\
\end{longtable}
}

\section{The $n$-cycle obstruction [R]}\label{sec:ncycle}
A ring of pairwise judgments has one consistent explanation exactly when the
cyclic inequalities hold; the contextual fraction measures how badly they fail.
Consider $n$ dichotomous judgments $M_0,\dots,M_{n-1}$ (outcomes
$\pm1$), each evaluable only against its two neighbours, so that the
compatibility graph is the $n$-cycle and the contexts are the adjacent pairs. An
empirical model $e$ gives each context a distribution over $\{\pm1\}^2$.

\medskip\noindent\textbf{The object, in signed-graph terms, before any probability
enters.} A cyclic scenario is a \emph{signed cycle}: $n$ edges, each carrying an expected
product $\gamma_i$ of its two endpoint variables. A deterministic global assignment is a
$\pm1$ labelling of the $n$ contents, and the edge signs it induces multiply to $+1$
around the loop --- an even-parity constraint. So the deterministic solutions are exactly
the \emph{even-parity cut vectors} of the cycle, a sign pattern is \emph{frustrated}
exactly when no labelling realizes it (Harary 1953), and the frustration index of signed-graph theory
counts the edges a best labelling must break. Probabilistic noncontextual
behaviours are then mixtures of those deterministic solutions --- the convex hull of the
even-parity cut vectors --- and a violation is support-function excess over one of its
facets. The whole of \S\ref{sec:ncycle} is that one picture, with the transition made
precise in three steps: \textbf{parity defines the vertices, convexification defines the
empirical region, facet slack measures violation.}

\medskip\noindent\textbf{Theorem (the deterministic-to-convex bridge).}
\emph{For a binary $n$-cycle, the deterministic global assignments generate the
even-parity cut vectors, and their convex hull is the noncontextual correlation
polytope $\mathrm{CUT}(C_n)$. Inside the empirical correlation cube $[-1,1]^n$, at most
one odd cycle facet can be violated. If its signed excess is $V>0$, then the distance
from $\gamma$ to the noncontextual polytope is $V n^{1/p-1}$ in every $\ell_p$ norm,
$1\le p\le\infty$; and in the consistently connected cyclic system the contextual
fraction is $\CF=V/2$; for a signalling system, with $\CF$ the Contextuality-by-Default fraction of
the consistification defined below, $\CF=V/2$ still holds with $V$ replaced by its
disturbance-corrected form (\S\ref{sec:ncycle}, via Cervantes 2023).}

\medskip\noindent Every part is proved in \ref{app:prop15a}: the facet characterisation
from Barahona \& Mahjoub (1986) and Araújo et al.\ (2013), the uniqueness of the violated
facet in part (iii), the $\ell_p$ scaling of the distance from Dzhafarov, Kujala \&
Cervantes (2020, Eq.~(56)), and the contextual-fraction normalization in
part (v) via Cervantes (2023). Stating it this way places the paper's best formal content
in a recognizable neighbourhood rather than in a claim about cohomology: signed-graph
balance (Harary 1953) and the frustration index; Ising and spin-glass frustration on
cycles, where finding a minimum-frustration labelling is the same combinatorial problem
(Barahona 1982); cut and correlation polytopes, whose facet structure is the subject of a
standing literature (Deza \& Laurent 1997); marginal polytopes in graphical models, of
which $\mathrm{CUT}(C_n)$ is the binary pairwise instance; and linear-programming dual
witnesses, which is what a violated facet is. None of these threatens the paper's
novelty. They say what the cyclic result is an instance of.

\medskip\noindent\textbf{Proposition (global section $\Leftrightarrow$
noncontextuality; the frustrated triangle).}
(a) A single joint distribution over all $n$ judgments reproducing every context
distribution exists iff $e$ satisfies the cyclic inequalities (the $n=3$ triangle,
for $\pm1$ variables with zero means: Suppes \& Zanotti 1981; the $n=4$ case:
Fine 1982; general $n\ge4$: Araújo et al.\ 2013). The obstruction is the contextual fraction
$\CF(e)=1-\mathrm{NCF}$,
$\mathrm{NCF}=\max\{\sum_g b_g : \sum_{g|_C=s} b_g \le e_C(s),\ b_g\ge0\}$. The
linear program returns $\mathrm{NCF}$, the largest fraction of the data a mixture
of global assignments reproduces; $\CF$ is the fraction that resists every global
model, and $\CF=0$ iff a global section exists.

\noindent This linear program presupposes \textbf{consistent connectedness}: that a
measurement's marginal distribution is the same in both of its contexts. If instead a judgment's marginal shifts with what it is paired against, $\CF$ denotes throughout
the Contextuality-by-Default contextual fraction of the \emph{consistification} (Cervantes
2023), the generalization for which the relations below are proved. The plain fraction is
formulated for consistently connected systems only, and its generalization is not unique
(Kujala \& Dzhafarov 2019, \S7.1); we adopt Cervantes's throughout. Consistent
connectedness is the precise point at which the sheaf-theoretic and
Contextuality-by-Default accounts have to be related rather than merged, and Dzhafarov
(2023) sets out that relation directly; the positioning in \S\ref{sec:prior} follows it.

\noindent This $\CF$ is a \textbf{complete} obstruction measure: the faithful
contextual fraction of Abramsky, Barbosa \& Mansfield (2017) vanishes
\emph{exactly} when a global section exists. It is \textbf{not} the \v{C}ech
cohomological class of Abramsky et al.\ (2015), which is only \emph{sufficient}
for contextuality: a nonzero class certifies contextuality, but a zero class does
not certify its absence, so a zero class does not rule contextuality out (the
Hardy model is the standard witness; Car\`u 2017). $\Hone$ is not a synonym for
non-globality, and it is not used as one here: a convex statistic is never called an
$\Hone$ class anywhere below.

\noindent\textbf{$\CF$ and the perceptual $\Minc$ are scored against the same convex
cyclic geometry, and the co-vanishing must be read carefully} (resolved in
\ref{app:prop15a}, Prop 15a). The \emph{literal} biconditional
$\Minc=0\Leftrightarrow\CF=0$ for the fixed-coboundary
\textbf{cokernel-projection norm} is \textbf{false [R-refuted]}: the frustrated
odd-cycle coboundary is invertible, so $\Minc\equiv0$ while $\CF$ sweeps $(0,1]$.
Nor does another choice of stalks repair it: a fixed linear residual is a
\emph{seminorm} and vanishes on a subspace, whereas $\CF$ vanishes on the
full-dimensional cut polytope --- the correlations that mixtures of global $\pm1$
labellings can produce --- so no linear $\Minc$ can equal $\CF$. The co-vanishing
\textbf{does} hold \textbf{[R]} under the non-signalling hypotheses of
\ref{app:prop15a}, for the \emph{convex} obstruction $\Minc^{\star}:=\max\{0,\,s_{\max}-(n-2)\}$ --- the amount by which the
correlation vector $\gamma$ overshoots the cycle inequality that defines
$\mathrm{CUT}(C_n)$.

\noindent The positive excess $\Minc^{\star}$ is the unnormalised \emph{facet slack}.
Under the hypotheses of \ref{app:prop15a} (the empirical correlation cube
$\gamma\in[-1,1]^n$) it is also exactly the $\ell_1$ distance to $\mathrm{CUT}(C_n)$,
with $d_p=\Minc^{\star}n^{1/p-1}$ for $1\le p\le\infty$ --- note this is
$\Minc^{\star}$, not the signalling-corrected $V$ defined below. \textbf{The $\ell_p$
scaling is prior art and is not claimed here.} Dzhafarov, Kujala \& Cervantes (2020) read
cyclic contextuality measures as an $\ell_1$ distance to the noncontextual polytope, their
Theorem~15 identifies that distance with the excess of the cyclic inequality's left side
over its right --- which is $\Minc^{\star}$ in the coordinates used here --- and their
Eq.~(56) states the all-$p$ form outright as
$\mathrm{CNT}_2^{(p)}=n^{(1-p)/p}\,\mathrm{CNT}_2$, whose exponent $(1-p)/p$ is the
$1/p-1$ written above. This paper adds the route and the witness: the
cut-polytope derivation, and an explicit nearest point that attains the distance in every
$\ell_p$ norm at once (\ref{app:prop15a}(iii)). $\mathrm{CUT}(C_n)$'s nontrivial facets are
exactly the cycle inequalities (Barahona--Mahjoub 1986, who characterise the facets for
any graph with no $K_5$ minor); its remaining facets are the trivial box constraints
$\lvert\gamma_i\rvert\le1$, which every empirical correlation vector satisfies
automatically. Those cycle inequalities are the ones the contextual fraction $\CF$ tests
for this family (the Abramsky--Barbosa--Mansfield (ABM) tests, 2017).

\begin{figure}[htbp]\centering
\includegraphics[width=\linewidth]{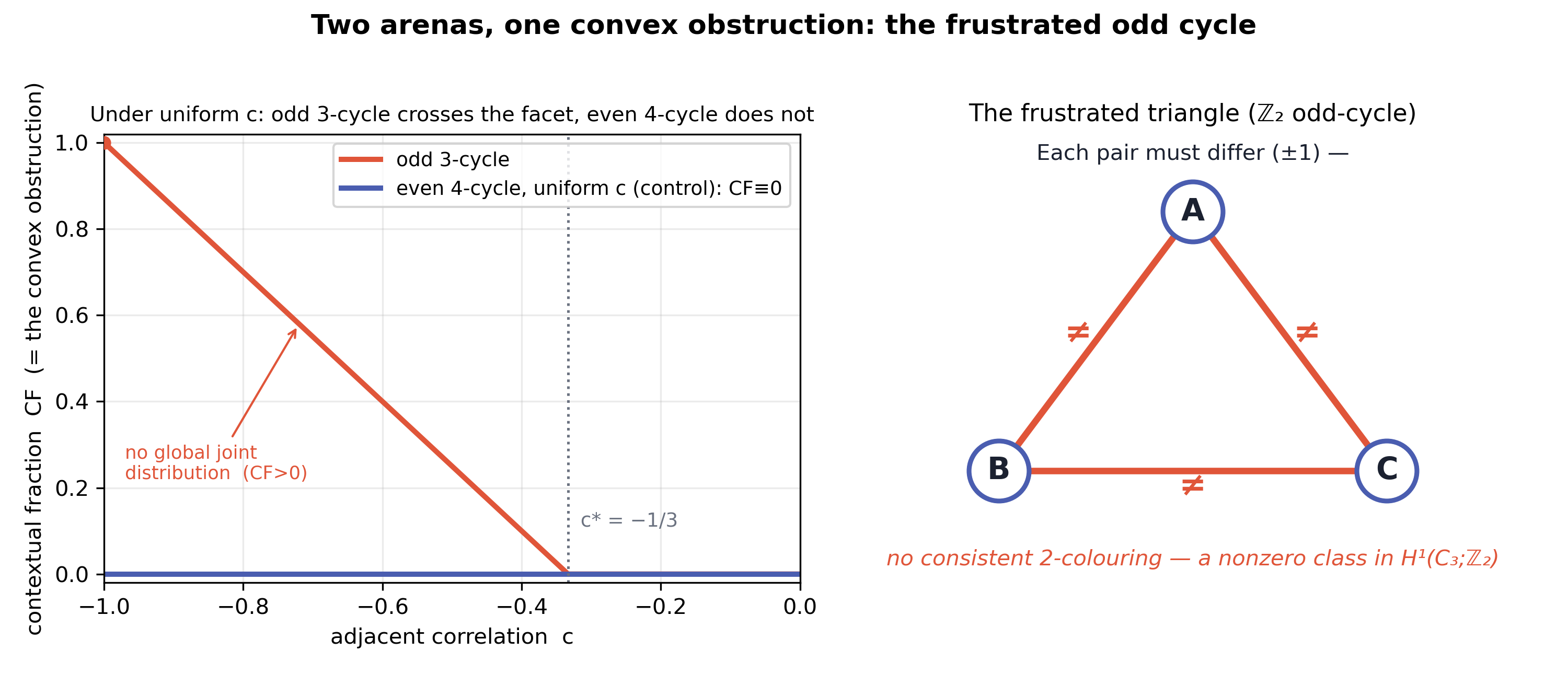}
\caption{\emph{Left:} the contextual fraction $\CF(c)$ rises below the threshold
$c^\ast=-1/3$ for the odd 3-cycle but is identically zero for the even 4-cycle
\emph{under uniform $c$} (verified against the linear program;
\code{check\_contextuality.py}; the even
4-cycle is contextual off this family --- CHSH). \emph{Right:} the frustrated
triangle --- every adjacent pair must differ, with no consistent 2-colouring ---
is, in the deterministic limit, the two-valued ($\Zt$) obstruction shared by a binary cyclic-ordering percept and the
judgment cycle (the \emph{continuous} Penrose tribar is the $\R$-holonomy version,
\S\ref{sec:arenas}).}
\label{fig:unification}
\end{figure}

\noindent\textbf{The two arenas' shared object, stated exactly.} They are scored
against the same family of cyclic noncontextuality inequalities and share one convex
global-assignment geometry --- the cut polytope of the $n$-cycle and the facet slack on
it. Two cohomological statements accompany that geometry. They point opposite ways, and the paper keeps them apart. \emph{They do share} a $\Zt$ obstruction in the
deterministic limit: the frustrated odd cycle carries a nonzero class in
$\Hone(C_{n};\Zt)$ in either substrate, which is the shared object named in
\S\ref{sec:arenas} and drawn in Figure~\ref{fig:unification}. \emph{They do not share} a
real-coefficient class, and \ref{app:prop15a} proves they cannot: the frustrated odd
cycle's real coboundary is invertible, so its literal linear $\Hone_{\R}$ vanishes while
$\CF$ is positive. The $\Zt$ class they do share is nonetheless the wrong instrument for
the empirical question, and that is why the tests below are not built on it: it is the
same nonzero class for every frustrated cycle, so it carries none of the between-subject
variation the cross-domain test needs. The severity that moves with the data is the
convex facet slack, which is not a cohomology class at all. Shared geometry is the claim this paper makes
and can support; shared mechanism is a causal claim needing an identification step this
design does not carry (\S\ref{sec:tests}).

\noindent For this cyclic family the ABM contextual fraction $\CF$ and the
Contextuality-by-Default measure $\CNT$ --- which quantifies excess cyclic
inconsistency after correcting for direct influences (signalling) --- are related by a constant
factor: $\CF=2\,\CNT$, the proven relation for cyclic systems (Cervantes 2023). So
the survey's test statistic and its magnitude carry the same information up to
scale. The script \code{check\_contextuality.py} numerically verifies the linear-programming
half at $c=-0.4,\dots,-1$: $\CF=\max(0,V)/2$. The
normalization $\CNT=V/4$ is taken from the cited theorems rather than recomputed
from its own quasi-coupling definition. The design quantity plotted below is the
signalling-corrected cyclic-inequality violation $V=s_{\max}-(n-2)-\Delta$, where $\Delta$ is the
measured signalling: how far a judgment's marginal shifts with its context. The
system is contextual iff $V>0$. On that side $\CF=\max(0,V)/2$, so $V=2\,\CF$;
on the noncontextual side $V$ runs negative while $\CF$ is pinned at zero.

(b) With uniform adjacent anticorrelation $c$ and unbiased marginals, the odd-$n$
closed form is $\CF(c)=\max(0,-((n-2)+nc)/2)$ (threshold $c^\ast=-(n-2)/n$, slope
$-n/2$; verified against the linear program for $n=3,5,7$). For $n=3$ this is
$\max(0,-(1+3c)/2)$: zero
for $c\ge-1/3$, rising to $1$ at $c=-1$. The even $4$-cycle with uniform $c$ never
obstructs --- it 2-colours. \emph{(Verified: \code{check\_contextuality.py}.)}

(c) The $n=3$ frustrated triangle --- ``every adjacent pair must differ, yet no
global $\pm1$ labelling does'' --- is an odd cycle that cannot be 2-coloured; that
is the obstruction, and the even cycle \emph{under uniform $c$} is the matched
control (Fig.~\ref{fig:unification}).

\medskip\noindent\textbf{One caution the framing must respect: parity is not the
criterion in general.} The even $4$-cycle is the Clauser--Horne--Shimony--Holt (CHSH) scenario and \emph{is} contextual for
suitable (non-uniform) correlations; the even-cycle vanishing in (b) is specific
to the uniform-$c$, unbiased-marginal family we use as a control. The real
discriminator is \textbf{frustration past the threshold $c^\ast$}, not oddness ---
the even control works because uniform anticorrelation on an even cycle stays
sub-threshold, not because even cycles cannot obstruct. With that scoping,
\S\ref{sec:arenas}
exhibits the \emph{same} thresholded cyclic frustration in two different
substrates.

\section{Two arenas --- perception and judgment}\label{sec:arenas}
The $\Zt$ thresholded-frustration cycle of \S\ref{sec:ncycle} can be instantiated in two
substrates. \emph{(A note on the impossible triangle: Penrose (1992) classifies the
tribar by $\Hone$ of an annulus with coefficients in the multiplicative group
$\R^{+}$ of depths. The same paper treats a second figure --- Schr\"oder staircases
ringing a heptagonal annulus --- whose ambiguity group is that of the Necker cube,
and states its obstruction as $\Hone$ with $\Zt$ coefficients; Ghrist \& Ghrist
(2026) develop that $\Zt$ setting into a hierarchy for bistable systems. Penrose
therefore used two coefficient groups for two figures in one paper: the
coefficient group is a modeling choice, not an intrinsic constraint. We build the
perceptual arena from the $\Zt$ (binary cyclic-ordering) version; the continuous
holonomy plays no role in the tests below and is kept here only as the visual
reason impossible figures} look \emph{like the cognitive effect, not as a claim
the design below tests.)}

\paragraph{Perception --- a proposed binary cyclic-ordering instantiation. {[}C, unpiloted{]}}
\textbf{One arena: relative depth.} Earlier drafts offered relative depth, brightness or size
as interchangeable relational features. They are not interchangeable, and offering three
is a way of specifying none, so this paper fixes \textbf{relative depth} and treats the
brightness and size variants as untried alternatives rather than as part of the design.
Present three stimuli whose depth dominance is arranged to be cyclic ($A\succ B$,
$B\succ C$, $C\succ A$) --- an \textbf{intransitive} percept --- and elicit, \emph{in
each pairing}, \textbf{two} binary judgments, one about each stimulus in that pair. Each
of the three stimuli is therefore judged in exactly two of the three pairings, which is
what makes the design an instance of \S\ref{sec:ncycle}: a genuine $\pm1$
frustration over $\Zt$ on the thresholded 3-cycle, scored against \emph{the same
cyclic noncontextuality inequalities} as the survey and read out by the same
per-subject $\CF$.

\noindent\textbf{Stimulus parameters this specification leaves to a pilot.} Static
scalar stimulus values are transitive: three fixed depths admit a total order, so a cycle
has to be \emph{induced}, by context-dependent appearance or by
multidimensional tradeoffs that make the pairwise comparison depend on which pair is
shown. A criterion that tracks the co-presented stimulus can produce the same observable
(in B.8's draws a cut that merely shifted with the pairing did not); the
external-criterion format below discourages that route but cannot exclude it. The cue conflict this paper assumes will do that work --- pitting a pictorial
depth cue against a binocular one so the winning cue changes with the pairing --- is
stated here as a design intention, not as a calibrated stimulus set. The physical
parameter values, the disparity and occlusion magnitudes, and the viewing geometry are
\textbf{not fixed in this paper and must come from a pilot}. Until one is run, the right
description of this arena is a \emph{proposed perceptual instantiation}, and the paper
uses that phrase rather than claiming the arena realizes the cyclic obstruction.

\noindent\textbf{Pilot gates, stated as pass/fail before the main study.} A pilot
is required and must clear all five of the following, each of which the paper's own
simulations already give a number for. \textbf{(1) Frustration in range:} the population
of within-context correlations straddles the threshold $c^{\ast}=-1/3$, with sample mean
$\bar c$ in $[-0.70,-0.40]$ --- far enough past threshold to be contextual, not so far
that every subject saturates (Design note below). \textbf{(2) Disturbance small enough:}
the measured $\Delta$ leaves $V>0$ attainable, i.e.\ $\hat\Delta<3\lvert\bar c\rvert-1$ at
the observed $\bar c$ (Fig.~\ref{fig:survey}, left). \textbf{(3) Facet stability:} the
all-negative sign pattern $\varepsilon^{\star}$ is the maximising facet for at least
$95\%$ of piloted subjects, the precondition \S\ref{sec:tests} fixes it on.
\textbf{(4) Between-subject variance:} per-subject $\Vs$ has standard deviation at least
$0.30$ on the margin scale, against the ${\approx}0.54$ the design model predicts on that
same scale, so the cross-domain correlation is not range-restricted to nothing. The two
scales must not be mixed here: the design model's ${\approx}0.23$ (Design note below) is the
spread of $\CF$, and because $\CF=\tfrac12[\Vs]_{+}$ clamps every sub-threshold subject to
zero, the margin's spread is the larger of the two by construction. \textbf{(5) Reliability:}
split-half reliability at least $0.80$ on the frustrated load at the planned trial count,
and at least $0.86$ on the control. \textbf{Stop rule:} if reaching gates (1) and (4)
requires a stimulus manipulation whose measured $\Delta$ then fails gate (2) --- that is,
if the only way to make the percept cyclic enough is to make it signal --- the perceptual
arena is not viable as specified and the cross-domain test should not be run on it.

\noindent\textbf{Two binary judgments per pairing, not one --- and the difference is not
cosmetic.} A single forced choice per pairing (``is A nearer than B?'') does not
instantiate the \S\ref{sec:ncycle} object. Read as one variable per context, the three
responses always admit a joint assignment, so the obstruction is to a \emph{total order}
--- the linear-ordering polytope --- and not to a global section at all. Read instead as one
choice expanded into complementary outcomes (winner $+1$, loser $-1$), it \emph{is} a valid
cyclic system, but a degenerate one: every within-context correlation is then $-1$ by
construction, the frustration term is pinned at $F\equiv2$, and the margin collapses to
$\Vs=2-\Delta$ --- so all individual variation would come from signalling rather than from
frustration. That degeneracy inverts the psychological reading this design depends on: with
uniform choice probability $p$, contextuality requires $\tfrac13<p<\tfrac23$, so fair random
responding would be maximally contextual while strong deterministic cyclic dominance --- the
very effect the task is built to elicit --- would be \emph{non}contextual after correction
for direct influence. The paper's own reproduction of published data meets exactly this
degeneracy in seven of the eight published forced-choice systems it recomputes.
\textbf{Collecting two separately calibrated binary judgments per pairing avoids it}: the
within-context correlation then becomes a free design parameter rather than a constant,
random responding gives $c\approx0$ and is safely noncontextual, $\Delta$ remains separately
measurable, and the frustration the power analysis below is indexed by ($c$ straddling
$c^\ast$, mean ${\approx}-0.55$) is the quantity actually being estimated. This is the
encoding all simulations in \S\ref{sec:tests} assume.

\noindent\textbf{The two judgments, concretely, and why $c$ is free rather than fixed.}
In each pairing the two stimuli are presented together, and each is judged
\emph{independently against a fixed external criterion shared across all three pairings} ---
not against the other stimulus in the pair. The criterion is fixed: ``does this object
appear nearer than a reference depth held constant for the session,'' asked once per
stimulus. The two judgments are collected \emph{simultaneously} within a pairing --- both
responses to one display, order of the two response prompts counterbalanced within
subject --- so that neither judgment is designed to be the context for the other; any
influence of the first response on the second is a prompt-order effect, which the
counterbalancing makes measurable. Pairing
order is counterbalanced across subjects and pairings are interleaved rather than
blocked, which is also the adaptation control: no pairing is repeated consecutively, and
the reference depth is re-shown between blocks so criterion drift is measurable rather
than absorbed. This follows the
Contextuality-by-Default double-detection paradigm (Cervantes \& Dzhafarov, 2017a), in which two
stimuli are judged together in each context, restricted here to a cycle: each
stimulus is a content measured in exactly two pairing-contexts (in their experiment each
content appeared in three, a non-cyclic system), and the pair of
same-context judgments about the two contents sharing that context is exactly the
$(R^{c_i}_{q_i},R^{c_i}_{q_{i\oplus1}})$ object \S\ref{sec:ncycle} requires. Because each
judgment is anchored to the external criterion rather than to the co-presented stimulus, the
response format does not mechanically force the within-context correlation to any value: it
is free to be near zero, if the two absolute judgments are made independently, or strongly
negative, if a coherent (here intransitive) percept pushes the pair's joint criterion
placement apart. Cervantes and Dzhafarov's own double-detection data show correlated,
non-degenerate joint responses under exactly this paired-judgment format, which is the
precedent for treating $c$ as estimable rather than assumed. Whether $c$ in this design in
fact straddles $c^\ast$ is a piloting question this specification makes answerable, not one
it answers in advance.
\textbf{The one-bit alternative is worked out in full in
\ref{app:encodingb}}, rather than dismissed here: it has an exact margin for
arbitrary, unequal response probabilities, $\Vs=2-(|q_1+q_2|+|q_2+q_3|+|q_3+q_1|)$ in the
signed dominance strengths $q_i=2p_i-1$, and a within-arena test with honest size and good
power. Direction, not intractability, decides against it. Four results there decide
against it, three of them sufficient on their own: the frustration term is pinned at $2$, so between-subject
variation in the quantity this paper is about does not exist; the transitive control scores
$\CF=1-s$ against the frustrated arm's $\max(0,1-3s)$ at every dominance strength $s$, so the
within-arena contrast runs backwards; the cross-domain test loses false-positive control
in a way that \emph{worsens} with sample size, reaching $0.91$ at $N=400$ in a world with no
shared mechanism at all, a cost of control precision rather than a defect of the encoding
(\ref{app:encodingb}, B.6); and the one-bit statistic turns out to measure the wrong quantity
outright. That last is the sharpest, and it is proved rather than asserted
(\ref{app:encodingb}, B.3(iv)) \textbf{[R]}: Encoding B's contextual region is
\emph{exactly} the interior of the linear-ordering polytope, so a respondent is scored
contextual precisely when their pairwise choices \emph{can} be written as a mixture of
total orders. The margin measures success at order-representability, not cyclic
frustration, and its maximum sits at indifference on all three comparisons. A respondent
whose choices admit no such representation --- stochastically intransitive in the
linear-ordering sense, which is the obstruction the one-variable-per-context reading above
is about --- is scored \emph{non}contextual. The two readings of the same forced choices
therefore never fire together; the caveat, stated in B.3(iv), is that this is a statement
about the polytope and not about \emph{weak} stochastic transitivity, which a respondent
can violate --- running a full majority cycle --- while Encoding B still scores them
contextual.
Perceptual contextuality has been measured before (double-detection, Cervantes \& Dzhafarov
2017a, 2017b), and Zhan et al.\ (2024) report a related result, a
violation of a contextual reformulation of the Leggett--Garg inequality for bistable objects. But no one has, to our knowledge, run the
purpose-built \emph{cyclic-relational} perceptual design the theory points to; the
theory predicts $\CF>0$ there. This is a \textbf{principled prediction [C]}, not a
documented effect. The premise's main risk is that apparent intransitivity is only
variability among transitive preferences (Regenwetter et al. 2011, for \emph{preference}), and under the
\emph{adopted} two-judgment encoding that risk is now closed by proof rather than by
assertion \textbf{[R]}: any mixture of transitive orders, read out by a criterion fixed
across contexts and corrupted by independent per-judgment lapses, admits a global joint
distribution and is therefore scored $\Vs\le0$ --- it cannot produce a positive margin at
any response probabilities (\ref{app:encodingb}, B.8). The proof turns on each judgment
being anchored to an external criterion rather than to the co-presented stimulus, which is
a design feature of this arena and not an assumption of convenience: the comparative
alternative fails the bound immediately, at $\Vs=2$. A positive result here is therefore
structural, not variability. The proposition does \emph{not} cover a criterion that
drifts within a session. So far as the drift shifts a content's marginal between contexts it
enters as measured direct influence, handled by the $\Delta$ correction and priced in
\S\ref{sec:tests}; a criterion that tracks the co-presented stimulus can instead change the
joint response with no marginal shift, which the $\Delta$ correction does not catch. Such a
criterion can produce the same observable as the pairing-dependent appearance this arena is
built to elicit (B.8's comparative readout is the extreme case: under a uniform mixture of
orders it gives $\Delta=0$ and $\Vs=2$), and no score in
the design separates the two: the external criterion discourages comparative responding but
cannot exclude it. A positive margin therefore rules out mixtures of transitive orders, not
a criterion that tracks the co-presented stimulus.

\paragraph{Judgment --- cyclic contextuality. {[}C, on established math{]}}
Three attributes in pairwise contrast on a cyclic feature, each adjacent pair
anticorrelating \textbf{past $c^\ast$}. By \S\ref{sec:ncycle} the responses then admit no single
joint distribution; $\CF>0$ --- the cognitive face of order effects and the conjunction
fallacy.

Both arenas instantiate the same frustrated-cycle constraint structure: its deterministic
limit carries a nonzero $\Zt$ class in $\Hone$, and $\CF$ (with its perceptual dual
$\Minc^{\star}$) measures, for the noisy case, departure from the corresponding noncontextual
polytope --- the convex extension of that same obstruction, not the same cohomology class.
Whether the two arenas' margins covary across people, after prespecified adjustment, is
the empirical question \S\ref{sec:tests} puts at risk. Whether \emph{one} mechanism drives
both is a further claim that no design built from these six scores can settle
(\S\ref{sec:tests}).

\section{Three tests --- only the cross-domain one tests the
\emph{unification}}\label{sec:tests}
Each single-arena test, once run, would show only that one domain carries a cyclic
contextuality violation; \textbf{it would not test that the two are one mechanism.}

\begin{itemize}
\item \textbf{Within-arena, perceptual.} Binary cyclic-ordering (\S\ref{sec:arenas}): measure
$\CF$ from the \textbf{two binary judgments elicited in each pairing}; the
frustrated cycle is contextual and the sub-threshold and transitive controls are
not (under uniform $c$).
(\code{penrose-experiment.md}, discretised variant.)
\item \textbf{Within-arena, cognitive (the least costly to run).} The between-subjects
cyclic-judgment survey; frustrated cycle $\CF>0$, uniform even-4-cycle and
absolute-rating arms as null controls. The even-4-cycle control is null
\textbf{only under uniform, sub-threshold anticorrelation} (\S\ref{sec:ncycle}): an even cycle
whose empirical correlations drift off-uniform can itself become contextual (Snow
Queen) \emph{even when the mean correlation looks safely sub-threshold} --- e.g.\
$(0.6,0.6,0.6,-0.6)$ has mean $0.3$ yet $\CF=0.2$ (verified,
\code{check\_vanishing\_equivalence.py}). The design must therefore \textbf{verify
the control arm's measured correlations stay uniform and below $c^\ast$}, with the
explicit anisotropy tolerance $\lVert c-\bar c\,\mathbf{1}\rVert_\infty\le(1-\lvert\bar
c\rvert)/2$ as a preregistered rejection rule: every edge correlation must stay within
half the mean correlation's remaining distance from the $\pm1$ boundary, or the
uniform even-cycle control is rejected as too anisotropic. (\code{contextuality-experiment.md}.)
\item \textbf{$\bigstar$ Cross-domain (the test of the thesis).} One cohort
performs \textbf{both} arenas. We define each individual's perceptual and
judgment contextuality as \emph{the same} quantity on the binary ($\Zt$) cycle --- the contextual
fraction $\CF$ from the \S\ref{sec:ncycle} linear program applied per subject to the
\textbf{two binary dominance judgments elicited in each pairing} of a frustrated cyclic
triple (\S\ref{sec:arenas}). (This is
deliberately the discrete $\CF$, \textbf{not} a continuous holonomy threshold;
``loop-tolerance'' is operationalized as this per-subject $\CF$, so both arenas
contribute the same-class object.) The hypothesis under test is that a single
latent obstruction-tolerance contributes to both, predicting a \textbf{positive
individual-differences correlation} between the two arenas' margins. A positive result
establishes a \emph{shared residual association consistent with} that latent tolerance,
not that one mechanism generates both: the two are separated by an identification step
this design does not carry, and the paragraphs below on observational equivalence and on
rival traits show the separation is impossible from these scores at any sample size.
\end{itemize}

\noindent\textbf{A bare correlation is not diagnostic.} Both scores are \emph{inconsistency} measures. Because the control
alone needs about eight times the frustrated arm's trials (below), the two arenas are
collected in \textbf{separate sessions}, one arena per session, order counterbalanced across
subjects (total burden and the split: see the Design note below). A general factor of
response consistency and attention ($g$) --- state-level within a session, and, if present, a
stable trait-level component across the two --- would produce a positive correlation with
\textbf{no} shared sheaf mechanism either way.
\textbf{Residualization} gives the design its discriminance. It needs a control chosen with care, and the control's own
test-retest correlation across the two sessions is reported to check whether it tracks a
stable trait or only same-day state.
\textbf{(i) A $g$-capturing control, variance- and reliability-matched.} The
control is a cyclic triple whose within-context correlation the stimuli fix at one
noncontextual value for every subject, so that its between-subject variance comes from $g$
alone. It is scored on the same signed margin $\Vs$ as the arms (item~2 of the
specification below), on the same fixed facet $\varepsilon^{\star}$. For this control
$\varepsilon^{\star}$ is not the maximising facet, so its score tracks the control's
mean correlation, and through it $g$, rather than serving as a contextuality margin; pilot gate~(3) applies to the arms only. On that scale such a triple is \emph{not} degenerate: careless
responding pulls its correlations toward zero and moves its margin, so it varies with $g$.
Scored on the clamped $\CF$ the same triple reads zero for every subject, has no
between-subject variance, and leaves a null correlation forced by range restriction, so
$\CF\equiv0$ holds only for the clamped score. Every simulation of the primary statistic in
this paper fixes the control's correlation at $c=+0.50$. No odd facet can carry three
positive correlations of that size past the noncontextual bound, so the simulated control is
noncontextual at every lapse rate; careless responding raises its margin from $-2.50$ toward
$-1$, the value under fully random responding, while it lowers the arms' margins, and
a linear adjustment does not need the two signs to agree. Whether a stimulus triple can fix
that correlation for every subject is a piloting question. That the control carries no
obstruction tolerance is an assumption of the simulations, not something the design checks
(see the null reading below). The superseded $\CF$-scored comparisons reported below need a
control with variance under the clamp, so they use a past-threshold triple at $c=-0.50$, which
is itself contextual. Tuning it so its per-subject inconsistency score has the \emph{same
variance and reliability} as the frustrated-load $\CF$ is the design's safeguard against
range restriction and attenuation. It is neither necessary nor sufficient for unbiased
residualization. An exactly measured control is \emph{sufficient} to remove a linear confound at
any scale when that confound is \emph{one combined factor}: the task and the control weight $g$'s
stable trait and its same-day state in the same proportion. It is not necessary. The exact
zero-bias condition, $\gamma(\kappa^2+\psi)=\lambda\sigma\kappa$ in the notation defined below,
also holds for an error-free control that carries no same-day state at all. Where the condition
fails, an error-free control still leaves a bias, upward when the two arenas share their loadings --- see \emph{An error-free control still leaves a bias if the control splits $g$
differently from the task}, below. A matched but noisy control leaves the reliability-dependent bias reported below. Matching
also does not by itself make the control a pure estimate of $g$, since a trait that loads on the
obstruction itself can satisfy the same matching --- see \emph{One violation of that
second assumption forges the result the study is looking for}, below, where that trait is
constructed and shown to be unidentifiable from the design's own scores. Matching rules
out one failure mode, not all of them. \textbf{(ii) Partial out the control.} Regress each
arena's frustrated-load $\CF$ on its control-load score and correlate the
residuals: this targets the shared $g$, and removes it only to the extent that the
control load measures it, in the same trait-to-state proportion as the frustrated load. Under a shared mechanism the residual
correlation is positive; under two independent sheaves (or a pure $g$ confound) it
is $\approx0$.

\noindent\textbf{(iii) The single-latent claim, as a structural test.} The
strongest form of the prediction is not merely that the two arenas share
\emph{variance} (leg ii) but that \emph{one} obstruction-tolerance drives both ---
a shared factor accounting for the entire cross-arena link, with negligible
arena-specific residual. This is a \textbf{latent-variable (structural equation
model, SEM)} hypothesis
(a general-consistency factor and a shared-obstruction factor as \emph{separate}
latents, multiple indicators per arena); it is \textbf{not} recoverable by a
reliability-disattenuation of the correlation. A naive
disattenuation $r_{\mathrm{true}}=r_{\mathrm{obs}}/\sqrt{\rho_p\rho_j}$, where $\rho_p$
and $\rho_j$ are the split-half reliabilities of the perceptual and judgment scores, is
\emph{not} confound-immune --- this correction addresses attenuation only, and does not
remove a reliable shared confound, because a general-consistency factor is reliable, so
split-half reliability counts it as true variance and a reliable $g$ \textbf{forges}
a disattenuated correlation of $1$ with no shared obstruction at all
(\code{crossdomain\_linking.py}); residualizing on a \emph{noisy} control before
disattenuating does not repair it. The control-adjusted test we rely on is therefore
leg (ii)'s \textbf{plain} residualized correlation --- computed on the \emph{signed
obstruction margin} $\Vs$ rather than on $\CF$, for the reason set out below; the
single-latent claim is its fuller multi-indicator extension, estimable under the loading
and control assumptions stated next.

\noindent With \textbf{two frustrated-load indicators plus a control per arena}, a
method-of-moments latent-variable estimator recovers the shared-obstruction
correlation $(S_{\mathrm{cross}}-L_g^2)/(S_{\mathrm{within}}-L_g^2)$, where
$S_{\mathrm{cross}}$ is the cross-arena covariance of the frustrated indicators,
$S_{\mathrm{within}}$ the corresponding within-arena covariance, and $L_g^2$ the
covariance component attributed to the general-consistency factor $g$ --- in words, the
estimator subtracts the $g$-attributed covariance from both the cross-arena and
within-arena moments before forming the shared-obstruction ratio ---
\textbf{confound-safe in simulation only when $g$ has no same-day component}
(\code{crossdomain\_sem.py}, whose generator draws one $g$ shared across sessions). In that
world it returns
$\approx0$ under independent obstruction with a reliable $g$, where naive
disattenuation forges $\approx1$; it returns $1$ under one shared latent and $0.5$
when half the obstruction is arena-specific. Its bootstrap confidence interval
separates these cases at $N\approx85$--$120$ on the power side. On the null side that interval
under-covers: the nominal $90\%$ confidence interval contains $0$ in $83$--$88\%$ of
independent-world runs, so the separation is better read as a detection claim than as a coverage
guarantee. The method-of-moments form assumes
equal $g$-loadings across
arenas and control noise capturing \emph{only} $g$; where those fail it can bias
$\rho_{\mathrm{obs}}$ outside $[0,1]$. \textbf{Those two assumptions are not enough, and the screen
needs a third: that $g$ has no same-day state at all.} The design itself allows such a state
(above), and with it the screen fails however $g$ is read, including as one combined factor
that task and control load in the same proportion. Within a session the frustrated--control
covariance carries state as well as trait, while the cross-session control--control covariance
carries trait only, so $L_g^2$ is mis-estimated. With both listed assumptions holding, one shared
latent, and task and control each loading trait $1$ and state $\tfrac12$, the population value
is $7/11=0.64$ rather than $1$ (Monte Carlo median $0.633$ over $2000$ cohorts at $N=400$);
under independent obstruction with every trait and state loading $\tfrac12$ it is $-3/2$
rather than $0$. Remove the state and it returns $1$ exactly (Monte Carlo median $0.9997$)
(\code{scripts/p32\_bverify\_plim.wl}, \code{scripts/p32\_bverify\_a\_f3\_mc.py}). The screen
therefore stays secondary and is reported only as conditional on that third assumption. The
control's test-retest correlation tests it for the controls only, not for the frustrated
arms.

\noindent\textbf{One violation of that second assumption forges the result the study is looking
for, and the design invites it.} If a trait loads on the two \emph{control} tasks and on nothing
else, their covariance is inflated, the $g$ contribution $L_g^2$ is under-subtracted, and a shared
obstruction appears where there is none. The size is exact rather than simulated. With no shared obstruction at all, a trait loading
$0.6$ on each control (about a fifth of control variance) drives the population value of
$\rho_{\mathrm{obs}}$ to $0.209$, and at $N=400$ about half of cohorts return a value above
$0.2$. \code{p32\_control\_trait\_forgery.py} derives that population value in closed form
and confirms it over $400$ cohorts; with the trait removed, the estimator returns to
$\approx0$. The exposure is structural rather than hypothetical: the two
controls are deliberately matched in form across arenas, which is precisely what would give them
a shared task-specific trait the frustrated arms lack. \textbf{No check on these scores' covariances can rule
that trait out.} The trait's variance enters the observed second moments (the variances and
covariances) only through the
control--control covariance, where it is exchangeable with the shared-obstruction loading: raise
one and lower the other, and every variance and covariance stays where it was. The world just described
--- no shared obstruction, a trait loading $0.6$ --- and a world with no trait at all and a
genuine shared-obstruction correlation of $9/43=0.209$ have the \emph{identical} population
covariance matrix on all six scores, and give $\rho_{\mathrm{obs}}$ the identical population value
(\code{p32\_control\_trait\_identifiability.py}, exact in rational arithmetic and re-derived in an
independent computer-algebra engine). A covariance-based test that the controls' residual
covariance after $g$ is null is therefore unavailable rather than underpowered: \textbf{no statistic
computed from these six scores' covariance matrix has discriminating power above its own size.}
At two parameter points with the same covariance matrix, a level-$\alpha$ procedure built on that
matrix rejects with the \emph{same} probability in both worlds, for every
$N$ --- which is the precise statement, and a stronger one than ``zero power'': a test can always
reject at rate $\alpha$, it simply cannot do so more often in the world the design cares about.
Gaussian latent variables and errors are sufficient for the same to hold of every statistic of the
six scores, because the covariance matrix then fixes their whole distribution. That condition is
not necessary for the covariance result, and the design does not guarantee it. When a latent
variable is skewed, third moments can separate the two worlds. With centred-exponential latents at
the two parameter points above, the controls' third cross-moment, cut at a size of $0.046$ in the
trait world, rejects in $0.190$ of cohorts in the trait-free world at $N=400$ ($3000$ cohorts per
world). With Gaussian latents the same statistic rejects in $0.052$, no more than its size
(\code{scripts/p32\_bverify\_a\_f3\_mc.py}). That route is weak and rests on a skewness the design
neither assumes nor measures, so nothing below relies on it. Without it, the
design can report only the contamination a positive result would require. The share of the
\emph{cross-arena} covariance of the frustrated indicators that would have to come from this
trait rather than from $g$, for the entire observed correlation to be forged, is
$1-L_g^2/S_{\mathrm{cross}}$ --- $26.5\%$ in the case above. ($S_{\mathrm{cross}}$ is the
cross-arena covariance defined above, not the control--control covariance through which the
trait enters the moments; the trait's route in and the quantity it contaminates are two
different covariances, and the formula is the second.) That bounds the
positive reading rather than testing it, and it is a second reason, alongside the rival-trait
argument below, that a positive result corroborates rather than confirms.

\noindent\textbf{Exact nonidentification calls for partial identification, not a
better point estimator.} A design that cannot separate
two parameters should report the set they could jointly occupy rather than a single
number with a confidence interval around it, so the analysis reports four things and
none of them is a point estimate of a shared mechanism. \textbf{(1) A sensitivity
curve}: the implied shared-obstruction correlation as a function of the fraction of the
control--control covariance assigned to method variance, swept over its whole range
rather than evaluated at one point. \textbf{(2) Bounds}: the interval the shared target
component can occupy under a maximum method loading fixed in the preregistration, which
is what keeps the set narrower than the whole admissible range. \textbf{(3) The $26.5\%$ figure as a sensitivity
threshold, not a test result} --- it is the contamination share at which the observed
correlation is entirely forged, so it answers ``how much method variance would it
take,'' never ``is there method variance.'' \textbf{(4) What would contract the set}: a
validated pure anchor, or randomized crossed forms of the control, each of which
identifies only under explicit exclusion and invariance assumptions that the
preregistration would have to state in advance. Adding a third and fourth control moves
the design up the hierarchy set out in the Limitations, and is the structural fix; the
four items above are what can be reported without it.

\noindent A full six-indicator SEM is the natural full-model extension: fit by robust
maximum likelihood or generalised least squares (ML/GLS) with free per-arena loadings. It is
\textbf{planned rather than established} --- the implementation in
\code{crossdomain\_sem\_ml.py} is not yet calibrated for confirmatory use. Its \emph{point estimates} are sound --- across forty
independent cohorts at $N=400$ with half the obstruction shared it recovers $\hat\rho=0.502$
(SD $0.045$), so the single low cohort reported in an earlier draft, $\hat\rho=0.379$, was a
sampling draw and not estimator bias. \textbf{Two hypotheses have been run together here and must be
separated.} One-latent ($\rho=1$) against two-latent ($\rho<1$) is also a boundary comparison,
$\rho=1$ being the edge of the correlation's range; no
shared obstruction ($L_\theta=0$) against some shared obstruction ($L_\theta$ free) is a
\emph{boundary} one, and it is the second that \code{crossdomain\_sem\_ml.py} actually implements.
Calibration for either is not yet established: the operating characteristics rest on twenty
cohorts per cell. The boundary geometry of $L_\theta=0$ motivates a
$\tfrac12\chi^2_0+\tfrac12\chi^2_1$ reference distribution --- a 50--50 mixture of a
point mass at zero and a one-degree-of-freedom chi-square distribution --- rather than
$\chi^2_1$, but \textbf{that
mixture is asserted here and neither derived for this model nor implemented}: the code compares
its likelihood-ratio statistic to a plain $\chi^2_1$. Until the null geometry is derived and
numerically confirmed, the statistic should be referred to a \textbf{parametric bootstrap} null
rather than to any assumed closed form. \textbf{The primary pre-specified test is therefore leg~(ii)'s residualized correlation,
computed on $\Vs$ below and on nothing else} (specified item by item at the end of this
section). The method-of-moments screen (\code{crossdomain\_sem.py}) and the full SEM
(\code{crossdomain\_sem\_ml.py}) are both \emph{secondary}: reported alongside, unable to
change the primary verdict, and calibration pending.

\noindent\textbf{Simulations back leg (ii)} (\code{crossdomain\_estimability.py},
\code{crossdomain\_linking.py}). The \textbf{naive} cross-domain correlation is
strongly positive in \emph{both} a shared-mechanism and an independent world
($r\approx0.98$ vs $0.63$ --- the $g$ confound), so it is non-diagnostic; the
\textbf{residualized} correlation, in these simulations, is positive only under a shared mechanism
($r\approx0.91$ vs ${\approx}0$). Its residual bias under a pure-$g$ confound
\emph{depends on how well the control measures $g$}: in \code{crossdomain\_linking.py}
the residualized correlation is $0.161$ at control reliability $0.70$, $0.089$ at
$0.80$, and $0.033$ at $0.90$. At $N\approx200$ a bias of that size is not negligible,
which is what forces the control-precision requirement below.

\noindent\textbf{An error-free control still leaves a bias if the control splits $g$
differently from the task.} Those simulations draw one $g$ shared across sessions, with no same-day
part. The design allows both parts (above), and the arenas run in separate sessions, so only
$g$'s stable trait crosses from one arena to the other; its same-day state does not. The
regression on the control removes trait and state together, in the proportion the control
carries them. If the frustrated margin carries them in a different proportion, the regression
removes the wrong amount of trait, and the leftover trait reaches both arenas' residuals and
correlates them with no shared obstruction at all. Write one arena's frustrated margin as
$\gamma T+\sigma S+e$ and its control as $\lambda T+\kappa S+u$, with $T$ the trait, $S$ that
session's state, $e$ and $u$ independent errors, and $\psi$ the variance of the control's error
$u$. After residualization the trait loading left on the frustrated margin is
\[
  \gamma-\beta\lambda=\frac{\gamma(\kappa^2+\psi)-\lambda\sigma\kappa}{\lambda^2+\kappa^2+\psi},
  \qquad \beta=\frac{\gamma\lambda+\sigma\kappa}{\lambda^2+\kappa^2+\psi},
\]
where $\beta$ is the population regression slope. In words: the slope is set by trait and
state together, but only the trait part of what it removes matters across sessions. The
cross-arena residual covariance is the product of the two arenas' leftover loadings. So, in a
design whose two arenas carry the same loadings, the test is unbiased under no shared
obstruction if and only if $\gamma(\kappa^2+\psi)=\lambda\sigma\kappa$. The case that matters is
the one the earlier sentences assumed: task and control load \emph{one combined} $g$, the same
trait-to-state proportion, and the control is error-free; then the condition holds and the
bias is zero. That case is sufficient, not necessary: the condition also holds for an
error-free control with no same-day state ($\kappa=\psi=0$), whatever the task's proportion, and
at isolated points where control error and a mismatched proportion cancel. Error-free
measurement alone is not sufficient: with $\psi=0$ and $\kappa\neq0$ the condition requires the
shared proportion. When the arenas share their loadings the leftover enters both residuals with the same
sign, so the bias is upward, toward the hypothesis; when they differ it can take either sign
(an exact case with one arena over-adjusted gives $-2/25$). Its size is not small. Let $g$
carry half of each score's variance, place the task's $g$ at $45^\circ$ in the trait--state
plane and the control's at $60^\circ$, $15^\circ$ apart, and measure the control without error.
The population residual correlation is then $0.047$, and at $N=200$ the test rejects in
$15.9\%$ of null cohorts at the one-sided $1.645$ cutoff, three times the nominal $5\%$. The
calibrated cutoff of item~7 below, at its upper value $1.76$, lowers that only to $13.3\%$, because its simulated
null has no session state and so cannot register this bias. The rate grows with sample size, as a
fixed bias does: $12.0\%$ at $N=100$, $23.7\%$ at $N=400$, $37.6\%$ at $N=800$. A control at
reliability $0.90$ at the same angles gives $22.3\%$.
 Both controls of this check hold the nominal
rate: $4.9\%$ with the control's direction parallel to the task's, and $5.1\%$ at a
non-proportional point with control error that satisfies the condition
(\code{scripts/p32\_bverify\_f1\_mc.py}, $20{,}000$ cohorts per row, $5{,}000$ for the
sample-size rows; the formula in closed form,
\code{scripts/p32\_bverify\_plim.wl}). The control's test-retest correlation (above) estimates
the control's trait share but not the task's, so it does not settle the condition on its own.
\textbf{The primary test is unchanged} (items below). A positive result is now read as conditional
on the one-combined-$g$ assumption, and alongside the secondary check that follows.

\noindent\textbf{A secondary check that does not depend on how $g$ splits: the cross-session
tetrad difference.} With $S(\cdot,\cdot)$ the sample covariance across subjects and $C_P$, $C_J$
the two control-load margins, the check computes
\[
  \tau=S(\Vs_P,\Vs_J)\,S(C_P,C_J)-S(\Vs_P,C_J)\,S(C_P,\Vs_J),
\]
a product of covariances taken only \emph{between} the two sessions, so the same-day state and
the control's error never enter it. In the population, with $L_{\theta,P}$ and $L_{\theta,J}$
the two margins' loadings on a shared obstruction, $\tau=\lambda_P\lambda_J
L_{\theta,P}L_{\theta,J}$ for any state loadings, any control error, and unequal loadings across
arenas. In words: $\tau$ is zero with no shared obstruction (and no control-only trait, below), and
nonzero under one whenever both controls carry trait $g$. It is tested one-sided with an influence-function standard error,
in this division-free form, which needs no ratio of covariances and so no division by a
control--control covariance that may be near zero. In the setting above at $N=200$ its null rejection rate is $3.3\%$ with an error-free
control, $3.2\%$ at control reliability $0.90$ and $4.5\%$ with parallel directions: at or below
nominal, not inflated. It detects a shared-obstruction loading of $0.3$ in $38.7\%$ of cohorts
(\code{scripts/p32\_bverify\_tau\_mc.py}, $20{,}000$ cohorts per row). \textbf{A trait that
loads on the two controls and on nothing else forges $\tau$ completely}: such a trait adds a
term proportional to the square of its loading, and at a loading of $0.6$ on each control, with
no shared obstruction, $\tau$ rejects in $95.2\%$ of cohorts. The method-of-moments screen (above) is
forged by the same control-only trait, so $\tau$ removes the session-state bias and
does nothing about the forgery. Its power also vanishes as the controls' trait loading goes to
zero, and its power against the primary test at equal trial burden has not been computed. It is
\textbf{secondary}: reported alongside the primary, unable to change the primary verdict.

\noindent\textbf{A positive result corroborates; it does not confirm.} Residualizing on
the control can remove general response consistency and nothing else, and two things limit how
much of it goes. One is control error: with targets and controls matched in variance and
reliability ($1/2$ each) and no shared obstruction at all, control error alone leaves a residual
correlation of $1/6$. The other is the split of $g$ between trait and state, for which an
error-free control is not enough (two paragraphs above). Any \emph{other}
trait loading on both arenas --- an extreme-response bias, task-switching fatigue, a
general tolerance for ambiguity --- would survive that residualization and produce a
positive correlation with no shared obstruction. The positive direction therefore
corroborates rather than confirms, and the latent-variable form of leg~(iii) does
\emph{not} narrow it: that model defines the shared obstruction as whatever cross-arena
covariance the controls fail to absorb, so a rival trait the control misses is
indistinguishable from it. Narrowing the positive reading needs discriminant
indicators --- direct measures of the candidate rival traits, with the obstruction
factor shown to survive partialling them out --- which this design does not carry.
The \emph{null} is the sharper of the two readings, and only under two conditions: it is
clean only if the control loads on nothing \emph{but} $g$ (a control that accidentally
loads on the obstruction over-adjusts and manufactures a null, and when the two arenas'
controls split $g$ and the obstruction in the same proportion, as this design matches them,
no check on these scores' covariances can rule that out; only the discriminant indicators
named above, which this design does not carry, could), and even then a null without a preregistered minimum
correlation $r_{\min}$ only bounds the shared-mechanism account from above rather than
refuting it outright --- see \emph{Falsifier --- and it needs a number, not a direction},
below, which is where $r_{\min}$ is specified and left unfixed.

\noindent\textbf{The simulations assume a confound that is linear in the lapse rate.} Every
simulation in this section is generated from the same measurement model the estimators
assume: the confound moves the arms and the control linearly on the score scale. With the
noncontextual control the control's margin rises with the lapse rate while the arms' margins
fall. A linear adjustment does not depend on that sign, and the earlier past-threshold
control, whose margin fell with the arms', gives exactly the same distribution of every
residualized $\Vs$ statistic. The mechanism the design is actually guarding against ---
careless responding --- can also reach a control through a score that is not linear in the
lapse rate: a transitive control scored by its observed intransitivity rate, which rises as
$f(1-f)$ with $f=\lambda/2$. None of the runs above simulate that. ``Confound-safe''
therefore means consistent under its own assumptions.

\noindent\textbf{That misspecification has now been run on the null side, and the choice of
statistic survives it.} The transitive control is a triple scored by its observed
intransitivity rate, which \emph{rises}, nonlinearly, with the lapse rate where the frustrated
arms' scores fall. Both statistics were carried through it end-to-end on the same cohorts
(\code{scripts/p32\_vstar\_robustness\_checks.py}, $4500$ cohorts per cell). Two things
change and one does not. The clamped statistic fares \emph{better} under this confound than
under the linear confound, where it is adjusted on the past-threshold control at $c=-0.50$
($0.075$ against $0.094$ at matched control reliability), so part of
the case against it was specific to the confound model. But $\Vs$ still runs lower at a control
reliability of $0.87$, just above the $0.86$ the design asks for ($0.064$ against $0.075$),
and the ordering reverses
only where the control is measured badly ($0.143$ against $0.136$ at reliability $0.62$) --- a
regime the control-precision requirement above already excludes. Neither statistic reaches the
nominal rate under this confound: both sit near $0.063$--$0.065$ even at control
reliability $0.98$, so the calibrated cutoff is required under either confound model and its
size depends on which one holds. \textbf{The power side survives the misspecification too}:
with each statistic at its own calibrated cutoff, so that both run at a true $5\%$, $\Vs$
detects the target effect with power $0.964$ against $0.928$ at control reliability $0.87$, and
$0.970$ against $0.923$ at $0.92$ --- advantages of $3.6$ and $4.7$ percentage points. Those runs
leave the disturbance $\Delta$ out of both statistics; under the linear confound on the same
cohorts with $\Delta$ left out the advantage is $3.9$ ($0.967$ against $0.929$, control reliability
$0.88$), and once $\Delta$ is scored in both, as item~2 of \S\ref{sec:tests} prescribes, it is
$18$ ($0.927$ against $0.745$ on the same cohorts; \code{scripts/p32\_delta\_included\_rerun.py}, $24{,}000$ cohorts
per cell). The choice of statistic therefore holds on both axes under both confound models.

\noindent\textbf{Curvature settles the choice of statistic and raises a sharper worry
about the design.} If the confound attenuates the
frustrated arms through a \emph{power} of the lapse rate, $(1-\lambda)^p$, while the control
tracks $\lambda$ linearly, no linear adjustment can follow the bend. $\Vs$'s advantage does
not shrink there --- it widens, from $2.4$ percentage points at $p=1$ to $24$ at $p=2$
($0.110$ against $0.353$) and $44$ at $p=3$ ($0.283$ against $0.721$). Curvature does not
penalise the unclamped statistic; it compounds with the clamp, so the statistic carrying both
nonlinearities is hurt roughly twice over. \textbf{But at $p=3$ the margin itself runs at
$0.283$}, so curvature threatens the \emph{design} rather than the choice between statistics,
and it is the largest such exposure measured here. Whether the attenuation is anywhere near
linear is an empirical question about the task, and
this work has not answered it. Inspect the relation between the design's own lapse estimate
and the frustrated-load score for curvature before trusting the primary analysis: a
strongly curved relation invalidates the linear residualization for either statistic.

\noindent\textbf{Falsifier --- and it needs a number, not a direction.} A
variance/reliability-matched, $g$-residualized cross-domain correlation near zero ---
on the signed margin $\Vs$ --- counts against the \emph{unification} (each arena may
still be sheaf-shaped alone). That correlation is the constraint we are not aware of
either field having measured (\S\ref{sec:prior}). \textbf{But ``$\hat\rho\approx0$'' cannot by itself falsify
``$\rho>0$'',} because the qualitative prediction contains arbitrarily small positive
effects and no finite sample excludes them. The falsifiable version is quantitative:
preregister a smallest effect of interest $r_{\min}>0$, and reject the prediction only
when a calibrated one-sided upper confidence bound $U_{1-\alpha}$ falls below
$r_{\min}$. Absent a defensible $r_{\min}$, the honest reading of a null result is that
it \emph{places an upper bound} on the shared cross-domain component rather than
refuting the unification. \textbf{This paper does not yet fix $r_{\min}$:} the shared component of $30\%$
of the target trait's variance used in the power simulations below is an
alternative the design can detect, which is a different thing from an effect the theory
commits to, and promoting one to the other is an author decision the preregistration
must record before data collection.

\begin{figure}[htbp]\centering
\includegraphics[width=\linewidth]{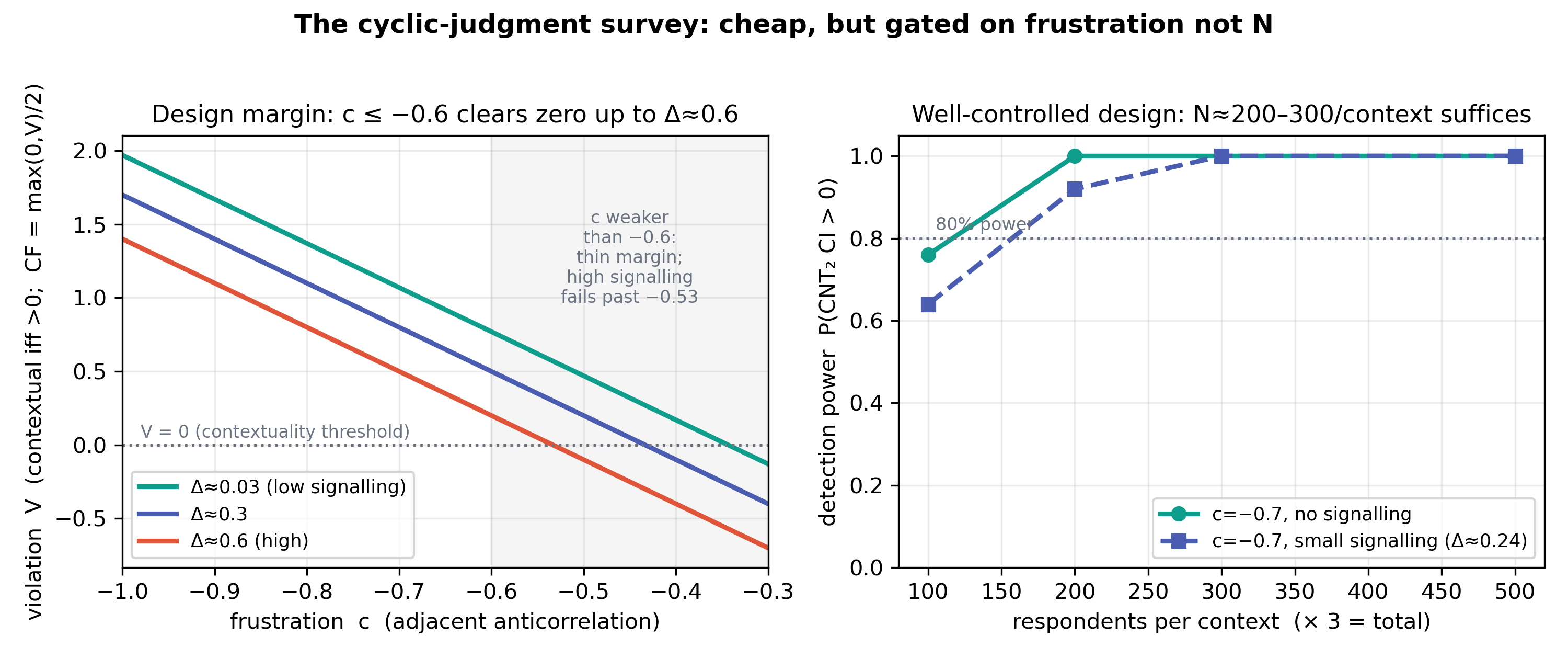}
\caption{The survey is gated on frustration, not sample size. \emph{Left:}
the signalling-corrected cyclic-inequality violation $V=3|c|-1-\Delta$ must exceed zero for the
system to be contextual. The contextual fraction is recovered from it as
$\CF=\max(0,V)/2$, so $V=2\,\CF$ on the contextual side only; the curves run
negative where $\CF$ is pinned at zero. Clearing zero requires $|c|>(1+\Delta)/3$, so
the three lines cross zero at $c\approx-0.34$, $-0.43$ and $-0.53$ for $\Delta\approx0.03$,
$0.3$ and $0.6$. The design's $c\le-0.6$ is therefore a sufficient margin, not a necessary
condition: it keeps $V\ge0.2$ even at the highest signalling plotted, $\Delta\approx0.6$,
where a system at $c=-0.5$ is not CbD-contextual ($V=-0.1$). \emph{Right:} for a well-controlled design
($c=-0.7$, low signalling) detection power reaches ${\sim}1.0$ by $N\approx
200$--$300$ participants contributing to each context (verified simulation,
\code{survey\_pipeline.py}).}
\label{fig:survey}
\end{figure}

\medskip\noindent\textbf{Design note (range restriction).} For the per-subject
$\CF$ to correlate it must \emph{vary} across subjects, which conflicts with
pushing frustration so far past $c^\ast$ that everyone saturates; the same applies
to the $g$-capturing control. Simulation confirms this is
workable: at a population straddling $c^\ast$ (mean $c\approx-0.55$), per-subject
$\CF$ has usable spread (SD${\approx}0.23$) --- and on the margin scale the same
population gives SD$(\Vs){\approx}0.54$, the figure pilot gate (4) is written against,
larger for the clamping reason given there
(\code{scripts/p32\_v4\_pilot\_gate\_scale\_check.py}) --- with split-half reliability
in the range $0.81$--$0.85$ by $40$ trials/context and $0.90$--$0.92$ by $80$. The interval is
across three independent implementations of the same two-judgment encoding rather than across
seeds alone: $0.851$ and $0.920$ from \code{crossdomain\_estimability.py} (mean over $20$ seeds
at $N=400$), and $0.807$ and $0.896$ from the separate generator in
\code{p32\_vstar\_control\_noise\_sweep.py}. \textbf{Those figures are
for the frustrated load and do not carry over to the control}, which needs about eight times
as many trials for the same reliability --- see the requirement below, and budget the
two arms separately rather than by a single trial count. The
cross-domain frustration $c$ is chosen for this spread and is \emph{not} the
survey's $c=-0.7$ (Fig.~\ref{fig:survey}). An end-to-end \textbf{simulation pilot}
(\code{pilot\_simulation.py}) separates the two tests sharply: the \emph{naive} test
is forged by a reliable $g$ (false-positive ${\approx}1.0$), while the
$g$-residualized test is not.

\smallskip\noindent\textbf{Scored on $\CF$, the residualized test does not run at its
nominal rate, and extra control precision will not make it.} Residualizing on a
control estimated less precisely than the frustrated arms under-adjusts, so the
rate falls as the control is measured better --- but it falls to a floor above
the true $2.5\%$ nominal size (\S\ref{sec:tests}) rather than down to it. Averaging over
independent seeds at $N=200$ (\code{scripts/p32\_verify\_residualization.py --only precision
--seeds 8 --cohorts 500 --ctrl-trials 80,160,320,640,1280,2560,5120}; eight seeds, $500$
cohorts each). A confound-off negative control run alongside every cell confirms the excess is
genuinely attributable to the confound rather than to noise: the off-arm sits at $0.023$ at
every control length, and each on/off gap clears its own pooled standard error by $z\ge4.6$
(worst case, the $5120$ cell; the $80$ cell clears at $z\approx17$):

\begin{center}\small
\setlength{\tabcolsep}{4pt}
\begin{tabular}{lccccccc}
\hline
control trials/context & $80$ & $160$ & $320$ & $640$ & $1280$ & $2560$ & $5120$\\
multiple of a frustrated arm & $1\times$ & $2\times$ & $4\times$ & $\mathbf{8\times}$ & $16\times$ & $32\times$ & $64\times$\\
false-positive rate & $0.121$ & $0.065$ & $0.049$ & $0.042$ & $0.043$ & $0.042$ & $0.043$\\
standard error over the $8$ seeds & $0.005$ & $0.003$ & $0.003$ & $0.003$ & $0.003$ & $0.003$ & $0.004$\\
\hline
\end{tabular}
\end{center}

\noindent The last three columns agree within one standard error, so this run's rate
asymptotes near $0.043$ --- but the third decimal is not settled, and we do not report it as
though it were. An independent re-derivation of the same asymptote through a different
random-number path (\code{scripts/p32\_mcverify\_asymptote\_0058.py}) returns $0.049$; the
disagreement is checked and open, with the worst single cell separating the two runs by
$z\approx2.6$ and a line-by-line comparison of the two implementations finding no divergence in
the formulas. We therefore quote the interval the two runs bracket rather than either endpoint:
\textbf{the residualized test runs at about $0.04$--$0.05$, between roughly $1.7\times$ and
$2\times$ the true $2.5\%$ nominal size}. The choice between the two implementations changes the multiple;
it does not bring any point of that interval near nominal. The excess is structural, not a measurement
problem. $\CF$ is clamped at zero, and regressing one clamped variable on another
leaves a residual dependence that no precision removes, because the clamp belongs
to the estimand and not to the measurement. The point is sharper than a limit on
precision: residualizing $\CF$ on the \emph{true} confounding variable itself ---
not a well-measured control, the latent it estimates --- still leaves a rate of
$0.089$ at $N=200$ (\code{scripts/p32\_vstar\_control\_noise\_sweep.py}). No
improvement to the control can reach a defect that lives in the outcome.

\begin{figure}[htbp]\centering
\includegraphics[width=\linewidth]{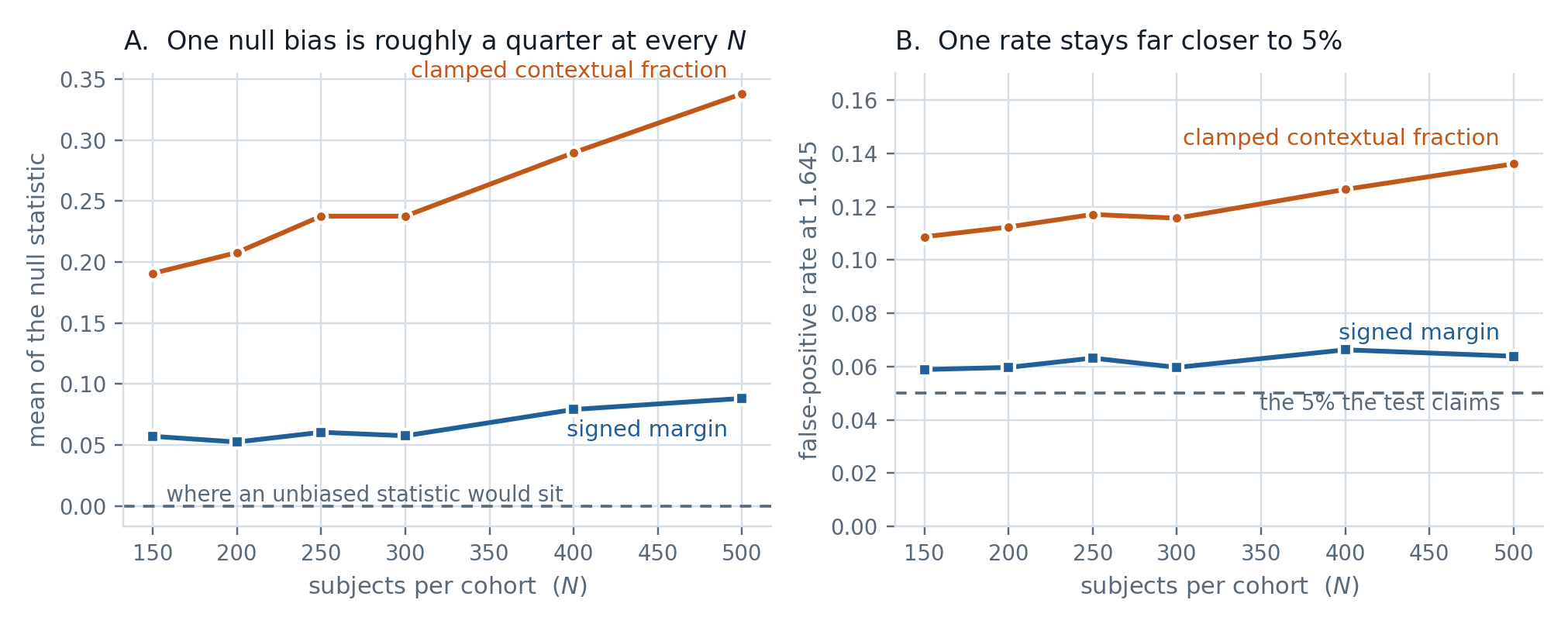}
\caption{The change of primary statistic, and its limits.
Both statistics under the pure-confound null, at the settled one-sided $5\%$ comparator
$1.645$, with the control at $640$ trials/context --- the eight-times floor the
control-precision requirement sets --- and scored the same way as the arms it
adjusts: the noncontextual control at $c=+0.50$ for the margin, the past-threshold control at
$c=-0.50$ for the clamped statistic, whose clamp would score a noncontextual control zero for
every subject. Every arm and control is scored with its own plug-in disturbance $\Delta$, as
item~2 of \S\ref{sec:tests} prescribes. \textbf{A.}~Neither null statistic sits at zero, and
both drift upward with $N$, because a fixed bias enters the Fisher-$z$ scaled by $\sqrt{N-3}$.
Size separates them: the margin's bias runs about a quarter of the clamped statistic's
at every sample size ($0.05$ against $0.21$ at $N{=}200$; $0.09$ against $0.34$ at $N{=}500$;
ratio $0.24$--$0.30$).
\textbf{B.}~The consequence over the range this design plans for. The margin runs between
$0.059$ and $0.066$ against the $5\%$ it claims, while the clamped statistic runs from
$0.109$ to $0.136$, roughly double. \textbf{The margin is not unbiased}: the control's own
plug-in $\Delta$ adds noise unrelated to the lapse rate, so the control under-adjusts, and the
primary test compares the margin against the $95$th percentile of its own simulated null at
the realised design point ($1.74$ at $N{=}200$, $1.77$ at $N{=}500$ here) rather than against
$1.645$. At $320$ trials/context, the four-times floor this design previously specified, the
same simulation puts the margin's rate at $0.074$--$0.087$ and its cutoff at $1.84$--$1.94$,
rising with $N$, so the property the margin was adopted for is bought by the control arm's
trials. Six sample sizes, each averaged over three independent seeds at $8000$
cohorts; values in \code{manuscript/fig\_calibration\_data.tsv}, generated by
\code{scripts/p32\_fig\_calibration.py}.}
\label{fig:calibration}
\end{figure}

\smallskip\noindent\textbf{The excess is a biased statistic, and it therefore grows
with the sample.} Under the null the residualized correlation is not centred on
zero but carries an upward bias of about $0.03$. Since the Fisher-$z$ statistic
scales that bias by $\sqrt{N-3}$, the error rate \emph{rises} with sample size
rather than holding --- the opposite of the reassurance a larger study usually
gives (\code{scripts/p32\_calibrate\_residualized\_test.py}, with the control at the $160$
trials/context this design previously specified). With the control at $320$, and $\Delta$ left out as in that
script, the same statistic's null mean is about half as large ($0.19$ against $0.37$ at
$N=200$); the difference is control precision alone, since rerunning either simulation at the
other's control trial count reproduces the other's mean. (Figure~\ref{fig:calibration} scores
$\Delta$ as well and draws the control at $640$.) The $160$-trial values:

\begin{center}\small
\begin{tabular}{lccc}
\hline
$N$ & $200$ & $300$ & $400$\\
mean of the null statistic (should be $0$) & $0.37$ & $0.48$ & $0.60$\\
error rate at the Fisher-$z$ value $1.96$ & $0.067$ & $0.094$ & $0.117$\\
critical value giving a true $5\%$ & $2.11$ & $2.29$ & $2.39$\\
power at $1.96$ / at the calibrated value & $0.94/0.93$ & $0.99/0.97$ & $1.00/1.00$\\
\hline
\end{tabular}
\end{center}

\noindent\textbf{That table's statistic and comparator are both superseded.} Every number in it is a correct computation of the \emph{superseded}
$\CF$-scored statistic scored against $1.96$ --- the two-tailed $5\%$ value, which for this
directional test is the one-tailed $2.5\%$ value. The comparator adopted throughout the rest
of the paper, and in Figure~\ref{fig:calibration}, is the one-tailed $5\%$ value $1.645$
(\S\ref{sec:tests}). The miscalibration is worse than it looks: a realised $0.067$ against a
true nominal $0.025$ is a $2.7\times$ inflation, not the $1.34\times$ that reading $0.067$
against ``the $5\%$ it claims'' would suggest. The table is retained as the record of why
the statistic changed, not as a specification of the primary test.

\begin{figure}[htbp]\centering
\includegraphics[width=0.62\linewidth]{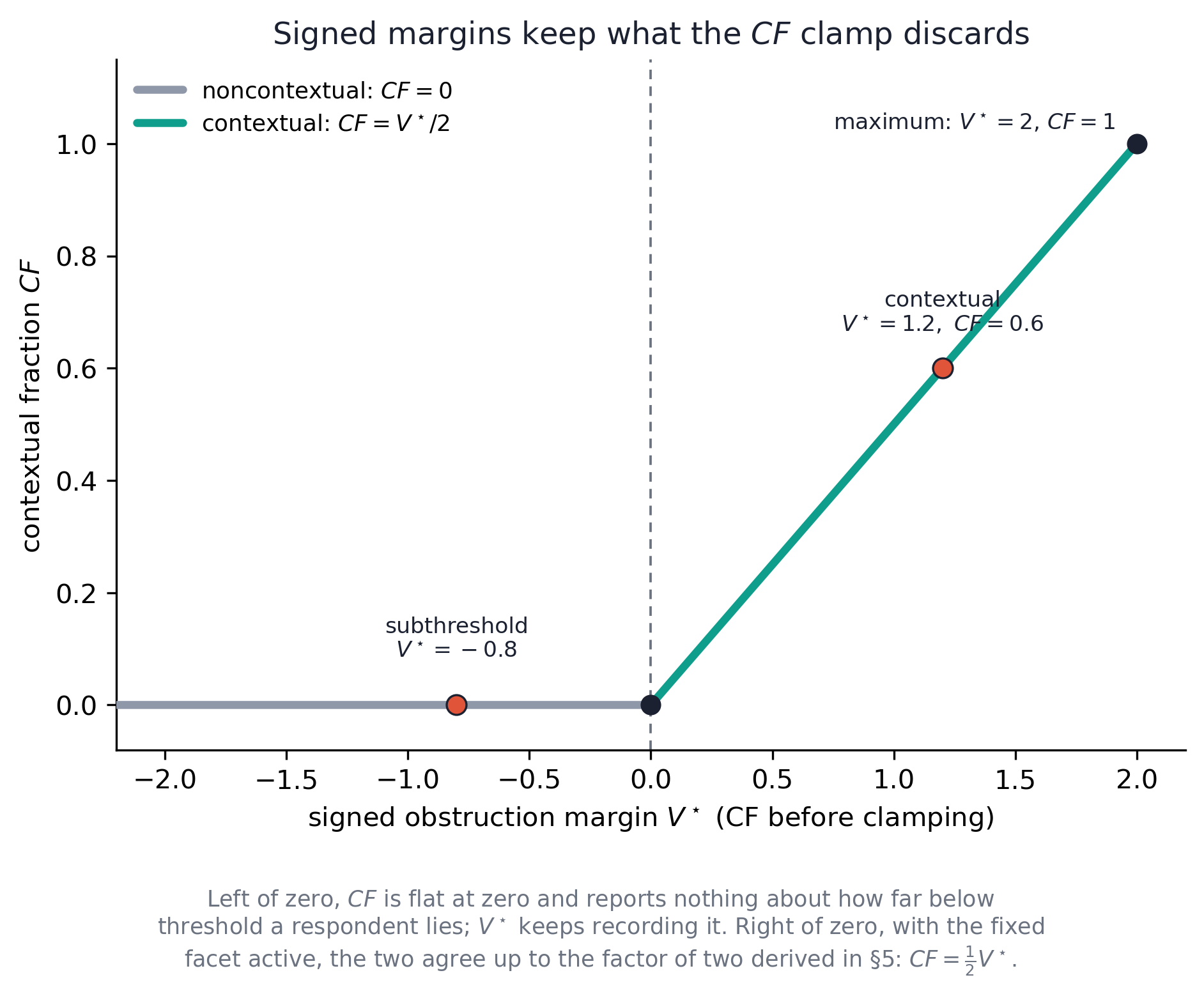}
\caption{The hinge this section's repair turns on, drawn directly from the identity just
stated. Left of zero $\CF$ is flat and reports nothing about how far below threshold a
respondent lies; $\Vs$ keeps recording it. Right of zero, with the fixed facet active, the
two agree up to the factor of two, and the line stops at $\Vs=2$, $\CF=1$, the largest
margin the cyclic bound allows. The hinge identity $\CF=\tfrac12[\Vs]_{+}$ explains why the cross-domain analysis in
\S\ref{sec:tests} is scored on $\Vs$ (``$\CF$ before clamping'') rather than on $\CF$
itself: clamping both arenas at zero manufactures exactly the correlation-attenuating
floor effect this section's simulations diagnose.}
\label{fig:vstar-hinge}
\end{figure}

\noindent \textbf{The repair is the statistic, not the threshold.} A clamped score is
what creates the bias, so the primary test is computed on the quantity $\CF$ is clamped
\emph{from}. For a cyclic system with correlation vector $\gamma$, disturbance $\Delta$ and
a facet fixed in advance by its odd sign pattern $\varepsilon^{\star}$, the \textbf{signed
obstruction margin} is
\[
  \Vs \;=\; \textstyle\sum_i \varepsilon^{\star}_i\gamma_i \;-\; (n-2) \;-\; \Delta ,
  \qquad\text{and}\qquad \CF \;=\; \tfrac12\,[\Vs]_{+}
\]
when $\varepsilon^{\star}$ is the active facet.
$\Vs$ is negative below threshold, zero on the boundary and positive above it: while
$\varepsilon^{\star}$ is active it carries all of $\CF$'s information plus the sub-threshold
variation $\CF$ discards. Off that facet the complete $\CF$ maximises over other odd sign
patterns, and $\Vs$ is then a well-defined pre-specified statistic but not $\CF$ before
clamping: $\gamma=(-0.9,0.9,0.9)$ with $\Delta=0$ has $\CF=0.85$ while $[\Vs]_+=0$. In the odd
anticorrelation design every population correlation is negative, and the all-negative facet is
then the unique maximiser, so fixing $\varepsilon^{\star}$ in advance costs nothing and
removes the selection bias of maximising over facets in the same sample. Strict negativity
of all population correlations is what guarantees the all-negative facet is the unique
maximiser, and that precondition is
not merely assumed: in this design's own population model it holds for $99.88\%$ of subjects,
failing only where the latent draw pushes the frustration up against its ceiling of zero. The
selection bias it removes is in any case small \emph{here} --- an alternative odd facet beats
the fixed one when some adjacent pair of measured correlations sums above zero (e.g.\ for
$\varepsilon^{\star}=(-1,-1,-1)$, the alternative $(-1,+1,+1)$ wins iff
$\gamma_2+\gamma_3>0$), which occurred for $3.15\%$ of $400{,}000$ simulated subjects at the
design point, with zero of those disagreeing with the fixed facet's own contextuality verdict
(\code{scripts/p32\_vstar\_robustness\_checks.py}). \textbf{The clamp, not the maximisation,
is what the change is for}, and Figure~\ref{fig:vstar-hinge} is that clamp in one picture:
flat on the left, where $\CF$ stops recording anything, and a straight line on the right,
where the two statistics differ only by a factor of two.

\noindent\textbf{The margin is a difference of two measured things, so it carries a
confound the $g$-control does not close.} Write $\Vs=F-\Delta$, where
$F=\sum_i\varepsilon^{\star}_i\gamma_i-(n-2)$ is the raw frustration carried by the facet and
$\Delta$ is measured direct influence. Bilinearity then gives
\[
  \operatorname{Cov}(\Vs_P,\Vs_J)=\operatorname{Cov}(F_P,F_J)+\operatorname{Cov}(\Delta_P,\Delta_J)
  -\operatorname{Cov}(F_P,\Delta_J)-\operatorname{Cov}(\Delta_P,F_J).
\]
In words: a positive cross-domain association in the margin can come from shared
\emph{frustration} --- the shared-mechanism claim above --- or from shared \emph{signalling},
which is not. The second route is not hypothetical --- a subject trait acting only on the
disturbance terms, with $\Delta_P=\Delta_J=T$ and genuinely independent $F_P,F_J$, forges
$\operatorname{Cov}(\Vs_P,\Vs_J)=\operatorname{Var}(T)>0$ on its own (verified symbolically and
by simulation, \code{p32\_confound\_and\_sensitivity\_check.py}). \textbf{Unlike the control-load trait of
\emph{One violation of that second assumption} above, this one is repairable, because
$\Delta_P$ and $\Delta_J$ are already observed in the trial data.} The preregistered analysis should therefore report all four
components $F_P,\Delta_P,F_J,\Delta_J$ and require the cross-domain association to survive
adjustment for the two measured disturbances alongside the nuisance-control scores. \textbf{One
caveat keeps this a strengthening rather than a fix:} the adjustment is exact only for a
$\Delta$ measured without error, and this design's $\Delta$ is an estimate with a known upward
bias --- a sum of absolute values has one --- so the adjustment attenuates the forged component
without guaranteeing its removal. For the one shared trait the design currently models, a
per-subject lapse rate, the measured coupling into the margin is $r=-0.006$ (SE $0.005$,
$40{,}000$ draws), so the exposure is empirically small \emph{for that trait} and untested for
a general disturbance-loading one.

\noindent\textbf{The case for $\Vs$ over a residualized-$F$ correlation.} An alternative test would
discard $\Delta$ from the statistic entirely and instead regress $F_P$ on $\Delta_P$ and $F_J$
on $\Delta_J$ separately, then correlate the two residuals. Nothing in the paper adopts that
route: it would need its own threshold, effect-size and power calibration, distinct from the
one built around $\Vs$ in Figure~\ref{fig:calibration} and \S\ref{sec:tests}. $\Vs=F-\Delta$
instead subtracts $\Delta$ directly, with the coefficient fixed by $\Vs$'s own definition
(\S\ref{sec:ncycle}) rather than estimated, so it reuses that existing calibration.
\textbf{That defence does not extend to the three-quantity adjustment just proposed}: requiring
the cross-domain association to survive adjustment for $\Delta_P$ and $\Delta_J$ together with
the nuisance-control scores is a different, uncalibrated statistic. Every size and power figure
quoted in this section is for $\Vs=F-\Delta$ alone, not for the adjusted version.

\medskip\noindent\textbf{One primary analysis; all alternatives are secondary.} Two candidate procedures have been carried side by side above, only one of
them calibrated. Leaving both live is not a design; the paper therefore fixes the
primary pre-specified test as follows, and every other cross-domain statistic in this
paper --- the three-quantity disturbance-adjusted version, the cross-session tetrad
difference $\tau$, the method-of-moments screen, the full six-indicator SEM, and the
descriptive $\CF$--$\CF$ correlation --- is
\textbf{secondary}, reported alongside but not able to change the primary verdict.

\begin{enumerate}
\item \textbf{Estimand.} The population Pearson correlation between the two arenas'
participant-level signed margins $\Vs_P$ and $\Vs_J$, after residualizing each on its
own arena's control-load margin.
\item \textbf{Estimation of $F$ and $\Delta$.} Per subject, per arena, by plug-in
from that subject's trial-level response proportions: $F$ from the three within-context
correlations against the fixed facet, $\Delta$ as the sum of the three absolute
context-to-context marginal differences. Both arms and the control are scored this way;
scoring the control differently is worse than changing nothing (table above).
\item \textbf{Facet.} Fixed in advance at the all-negative sign pattern
$\varepsilon^{\star}=(-1,-1,-1)$, never selected in-sample.
\item \textbf{Nuisance variables.} One per arena: that arena's control-load $\Vs$. No
others enter the primary model.
\item \textbf{Model.} Ordinary least squares residualization within each arena, then the
standardized Fisher-$z$ statistic $\sqrt{N-3}\,\operatorname{atanh}(r_{\mathrm{res}})$ of
the residual correlation $r_{\mathrm{res}}$, the scale on which item 7's cutoff is stated.
\item \textbf{Missingness and exclusions.} Subjects completing fewer than $80\%$ of
trials in either arena are excluded before scoring; no imputation. Exclusions are
applied blind to the cross-domain correlation.
\item \textbf{Null generator and critical value.} The $95$th percentile of the
statistic's \emph{own} simulated null at the realised design point --- realised trial
counts, realised control reliability, pure-confound world --- not a textbook cutoff.
The simulation takes the estimated reliability as known and does not carry that estimate's own
sampling error, so size conditional on the estimate is not established here. At
the planned design, with $\Delta$ scored and the control's reliability set at $0.88$, that
value is near $1.72$--$1.76$ and barely moves with $N$
(\code{scripts/p32\_calibrate\_vstar\_test.py}, \code{scripts/p32\_delta\_included\_rerun.py});
with the control simulated trial by trial at the $640$ trials/context floor and scored with its
own plug-in $\Delta$, it is $1.74$ at $N=200$ and $1.77$ at $N=500$ (Figure~\ref{fig:calibration}),
and at $320$ it rises with $N$ from $1.84$ to $1.94$,
which is why it must be computed at the realised design point rather than copied from here.
\item \textbf{Minimum effect of interest.} Not fixed by this paper; the preregistration
must set $r_{\min}$ before collection, and the null reading is an upper bound rather
than a refutation until it does (falsifier paragraph above).
\item \textbf{Multiplicity --- allocated across tiers, not pooled across all three.} The
cross-domain test is this paper's only primary hypothesis, and it is tested at a fixed
one-sided $\alpha_{\mathrm{P}}=0.05$ against the simulated null of item 7. Its critical
value is the $95$th percentile of that null at the realised design point (item 7:
realised trial counts, realised control reliability). It is therefore fixed \emph{before
unblinding} rather than before collection, and --- this is the property the
non-contingency claim actually needs --- it is computed without reference to any test's
observed $p$-value. So the event ``reject the primary'' is a function of the primary
statistic and the realised design point alone, and its verdict is not contingent on the other two. The
weaker timing is sufficient here: non-contingency requires only that the threshold not
depend on the other two hypotheses' $p$-values, which holds for any value fixed ahead of
unblinding, and the preregistration is required to record the simulated value at that
point (\S\ref{sec:tests}). The two
within-arena tests form a \emph{separate, secondary} family of two hypotheses, and Holm
correction is applied \emph{within that family only}, at a one-sided
$\alpha_{\mathrm{S}}=0.05$: the smaller of their two $p$-values is compared to $0.025$ and
the larger to $0.05$. Holm is not run across the two families, and neither family gates
the other. \textbf{The cost of that allocation is stated rather than hidden.} Type-I error
is controlled at $0.05$ for the primary hypothesis on its own under item 7's simulated
stationary pure-confound null, the only null it is calibrated against (nulls with a shared
disturbance trait or a session-level state fall outside it), and at $0.05$
family-wise within the secondary pair provided the two within-arena tests deliver valid
$p$-values, which they will not until they are calibrated (see below); it is \emph{not} controlled at $0.05$ simultaneously over all
three. Once both families deliver valid $p$-values, a simultaneous reading of all three is bounded at $0.10$ (union bound over the
two families), the simulated rate under a global null is $0.0965$
(\code{scripts/p32\_multiplicity\_scheme\_check.py}), and this paper makes no simultaneous
three-way claim.
\end{enumerate}

\noindent\textbf{Holm across all three tests, as item 9 once required, would make the
primary verdict contingent on the other two.} Earlier drafts of this item asked
for Holm correction over all three tests \emph{and} asserted that the primary's verdict
was not contingent on the other two. Holm is a step-down procedure: it ranks the whole
family's $p$-values and compares the $j$th smallest to $\alpha/(m-j+1)$, stopping at the
first failure, so \emph{every} hypothesis's rejection threshold is a function of the other
hypotheses' $p$-values. One exact counterexample at $m=3$, $\alpha=0.05$ settles it. Hold
the primary's $p$-value at $0.02$. With the other two at $(0.90,0.95)$ the smallest
$p$-value in the family is the primary's own $0.02$, which already fails
$\alpha/3=0.0167$, Holm stops, and the primary is \textbf{not} rejected. With the other
two at $(0.001,0.95)$ the smallest is now $0.001$, which passes $\alpha/3$; the procedure
steps down, and $0.02$ then passes $\alpha/2=0.025$, so the primary \textbf{is} rejected.
Identical primary evidence, opposite verdicts, decided entirely by a test the primary was
said not to depend on. The scheme in item 9 buys the non-contingency claim by
\emph{allocating} error across tiers instead of pooling it, at the stated price of a
$0.10$ rather than $0.05$ bound on any-false-rejection across all three. Both the
counterexample and the invariance of the new scheme are checked, each against a negative
control that fires, in \code{scripts/p32\_multiplicity\_scheme\_check.py}.

\noindent\textbf{Calibrated and uncalibrated parts of the procedure.} Items 1--7 describe a
procedure whose size and power have been simulated at the design point
(Figure~\ref{fig:calibration}, the reliability-$0.86$ table below, whose $\CF$ rows leave $\Delta$ out, and the
design-point power table after it), with the control fixed at the noncontextual $c=+0.50$ and the disturbance $\Delta$ estimated
by plug-in and subtracted on both arms and the control, as item~2 prescribes
(\code{scripts/p32\_delta\_included\_rerun.py}). That calibration covers only the simulated
stationary pure-confound world, whose true $\Delta$ is zero; a null in which a shared
disturbance trait loads on both arenas is not covered (see below). Item 9's allocation is a
specification rather than a simulated quantity; its one simulated ingredient, the
primary's critical value, is item 7's, and the error rates it claims are confirmed by
simulation in \code{scripts/p32\_multiplicity\_scheme\_check.py}. Item 8 has not been fixed, and
the secondary procedures have not been calibrated at all --- in particular, no type-I
error or power is reported anywhere in this paper for the disturbance-adjusted statistic
or for the SEM likelihood-ratio test. \textbf{The experimental programme is therefore
described here as proposed, with a preregistration to be finalized before data
collection, and not as preregistered.} The calibration that would be needed to promote
the adjusted statistic --- estimating $F$ and $\Delta$ jointly in a hierarchical
trial-level model, carrying measurement error through to the cross-domain association,
and recalibrating under nulls containing shared disturbance, noisy controls, drift,
fatigue and control-only method traits --- is stated here as the next piece of work on
this design, not as something already done.

\noindent\textbf{$\Vs$ must replace $\CF$ on the control as well as on the frustrated
arms.} Scoring the arms on $\Vs$ while leaving the control scored as a clamped $\CF$ is
\emph{worse than changing nothing} --- $0.143$ against $0.098$ at $N=200$, and worse at every
control reliability and sample size measured. At the control reliability this design targets ($0.86$), scoring the control on $\Vs$ as
well largely removes the bias that made the error rate grow with the sample. In this
simulation, which sets the control's reliability directly, the margin's error rate does not
grow across the planned range, although its null mean still rises, from $0.07$ to $0.09$
(\code{scripts/p32\_delta\_included\_rerun.py}, part~A, $\Delta$ scored on every arm and absorbed
into the control's set reliability; three seeds, $24{,}000$ cohorts per cell, control reliability
$0.86$, pure-confound null, one-sided $5\%$ comparator $1.645$). The trial-level simulation of
Figure~\ref{fig:calibration}, with the control at $640$ trials/context (reliability $0.87$) and
scored with its own plug-in $\Delta$, shows the same pattern: the margin's remaining bias stays
between $0.05$ and $0.09$ over $N=150$--$500$ and its rate between $0.059$ and $0.066$. At $320$
trials/context (reliability $0.77$) the same simulation is less forgiving: bias $0.17$ rising to
$0.27$, rate $0.074$ to $0.087$. The values at reliability $0.86$ ($\CF$ rows from
\code{scripts/p32\_vstar\_control\_noise\_sweep.py}, four seeds, $1000$ cohorts per cell, with
$\Delta$ left out):

\begin{center}\small
\begin{tabular}{lccc}
\hline
$N$ & $200$ & $300$ & $400$\\
mean of the null statistic, $\CF$ (should be $0$) & $0.209$ & $0.247$ & $0.280$\\
mean of the null statistic, $\Vs$ & $0.071$ & $0.076$ & $0.087$\\
false-positive rate, $\CF$ & $0.098$ & $0.094$ & $0.101$\\
false-positive rate, $\Vs$ & $0.062$ & $0.063$ & $0.065$\\
\hline
\end{tabular}
\end{center}

\noindent With the confounding variable measured \emph{exactly}, the margin's null statistic
is centred outright --- mean $0.003$, $-0.003$ and $-0.006$ at the three sample sizes, against
$0.144$, $0.175$ and $0.199$ for $\CF$ (with $\Delta$ left out). The residual bias in the table above is therefore what
imperfect control measurement leaves behind, not a property of the statistic.

\noindent\textbf{One calibration step survives, and it is a smaller one.} The margin's null
statistic is centred but about $2$--$3\%$ too wide, so the textbook one-sided cutoff $1.645$
leaves the realised rate at $0.060$--$0.063$ rather than $0.050$. The honest critical value is
the $95$th percentile of the margin's own simulated null, and with the control's reliability
held at $0.88$ \emph{it barely moves with the sample}: $1.72$, $1.74$ and $1.76$ at $N=200$, $300$
and $400$, against $2.11/2.29/2.39$ for the statistic it replaces
(\code{scripts/p32\_delta\_included\_rerun.py}, part~A, $\Delta$ scored). With the control
simulated trial by trial and scored with its own plug-in $\Delta$, the cutoff is $1.73$--$1.79$
at $640$ trials/context and $1.84$--$1.94$, rising with $N$, at $320$ (item~7), so a cutoff is
valid only at the design point it was simulated for. \textbf{The primary analysis must
therefore compute $\Vs$ on both arms and the control, and compare it against the $95$th
percentile of its own simulated null at the realised design point.} Calibration costs the
margin about one percentage point of power ($0.938$ to $0.929$ at $N=200$) against $8.1$
points for the $\CF$-scored test ($0.826$ to $0.745$, $\Delta$ scored in both), so the change
improves both error rate and power rather than trading one for the other. At $N\approx200$, with
${\ge}80$ trials/context on the frustrated loads, this injected-reliability simulation detects a modest shared component
(30\% of the target trait's variance) with power $0.929$; the planned $640$-trial control gives $0.919$ (table below). $\CF$ remains the within-arena contextuality measure, and the $\CF$--$\CF$
correlation is reported as a secondary descriptive analysis so the original proposal stays
visible.

\noindent\textbf{Power at the design point.} The table gives the primary test's power at
the design as the budget below fixes it --- a real $640$-trial noncontextual control per
arena, $80$ trials/context on the frustrated loads, $\Delta$ scored on every arm and control
score --- for sample sizes $N$ and shared components $\rho_\theta$ (the share of the target
trait's variance the two arenas share; $0.30$ is the modest component above). Each row's
cutoff is the $95$th percentile of its own simulated null, so the size column is $0.050$ by
construction and is omitted; power is the share of alternative cohorts past that cutoff
($3$ seeds $\times$ $2{,}000$ cohorts per cell; binomial standard error at most $0.007$, plus about $0.005$ from the simulated cutoff;
\code{scripts/p32\_delta\_included\_rerun.py}, part~E). A $\rho_\theta$ of $0.30$ at $N=200$
gives $0.919$ here against $0.929$ under part~A's injected-reliability control. The one-point
difference lies within Monte Carlo error (each cell's cutoff is itself a simulated percentile,
which adds about $0.005$ to the standard error) so these runs cannot separate it from a difference between the two control
models. Dropping the control arm altogether, the unresidualised
test rejects $0.64$--$1.00$ of the null cohorts across these $N$, which is the control arm's
whole job made visible.

\begin{center}\small
\begin{tabular}{lccccc}
\hline
$N$ & cutoff & $\rho_\theta=0.10$ & $0.20$ & $0.30$ & $0.45$\\
\hline
$100$ & $1.75$ & $0.164$ & $0.403$ & $0.695$ & $0.952$\\
$150$ & $1.74$ & $0.212$ & $0.538$ & $0.840$ & $0.991$\\
$200$ & $1.77$ & $0.250$ & $0.652$ & $0.919$ & $0.999$\\
$300$ & $1.74$ & $0.357$ & $0.799$ & $0.986$ & $1.000$\\
$400$ & $1.73$ & $0.433$ & $0.896$ & $0.997$ & $1.000$\\
$500$ & $1.75$ & $0.491$ & $0.948$ & $1.000$ & $1.000$\\
\hline
\end{tabular}
\end{center}

\noindent The table says what the budget buys and what it does not: $N=200$ is adequate
for a shared component of $0.30$ or more (power $0.919$) and inadequate at $0.20$ (power $0.652$),
where $N=400$ gives $0.896$ and $N=500$ only $0.948$; at $0.10$ even $N=500$
gives $0.491$. A preregistration that expects a weaker shared component than $0.30$ needs a
larger $N$ than $200$. The table varies only $N$, holding arm trials at $80$ and control trials
at $640$, so it does not say what more trials would buy; the control-trial floor below sets
the control's reliability, and each row's simulated cutoff sets the size. Each cutoff is a simulated percentile with standard error about $0.03$, so the $1.77$
at $N=200$ here and Figure~\ref{fig:calibration}'s $1.74$ are two runs of the same quantity.

\noindent\textbf{The control arm needs about eight times a frustrated arm's trials, not twice
or four times.} The control's between-subject variance comes only from the general-consistency factor,
a smaller source than real spread in frustration, so at equal trial counts it is the
\emph{less} reliable of the two. Measured split-half, the noncontextual control scored on $\Vs$ with its own plug-in $\Delta$, as
item~2 prescribes, reaches reliability $0.62$ at $160$ trials/context --- the $2\times$ this
design previously specified --- then $0.77$ at $320$ and $0.87$ at $640$
(\code{scripts/p32\_delta\_included\_rerun.py}, part~C; $0.76$, $0.87$ and $0.93$ with $\Delta$ left
out). Eight times a frustrated arm's trials ($640$ per context, reliability $0.87$) is
therefore the smallest multiple measured that reaches the $0.86$--$0.90$ the design asks of the
control; four times ($320$) reaches only $0.77$, and the budget below plans $640$.
Scoring that control on $\Vs$ is necessary, not a refinement: its clamped $\CF$ is zero for
every subject. Even on the past-threshold control the superseded comparisons use, clamping
discards variance, so its clamped score is the less reliable of the two at every trial count
($0.68$ against $0.77$ at $160$, $0.90$ against $0.93$ at $640$). The measured disturbance makes the requirement
stricter rather than looser: estimating $\Delta$ from finite trials is an absolute value of
sampling noise, so it is positively biased under a true null ($0.378$ where the truth is $0$, at
$80$ trials/context)
and it enters the control score as extra error. Carried on the \emph{frustrated arms} it
costs nothing measurable --- $0.056$ with and without it at $N=200$ with an exactly measured
control, over $24{,}000$ cohorts (\code{scripts/p32\_delta\_included\_rerun.py}) --- because it
is uncoupled from the confound ($r=-0.006$ over $40{,}000$
draws). Carried on the \emph{control} it costs error control like any other loss of control
precision, taking an $80$-trial control from $0.13$ to $0.26$ with $\Delta$ on the defined
scale ($0.21$ with $\Delta$ carried on the arms as well; $1{,}500$ cohorts, standard error
about $0.01$; \code{scripts/p32\_vstar\_robustness\_checks.py}, part \code{delta}).
The disturbance is therefore not a second
defect of the kind the clamp was; it is one more reason the control arm needs its trials.

\noindent\textbf{Total burden and the two-session split.} Summed across both arenas, three
contexts each: $2\times3\times80=480$ target-context presentations plus
$2\times3\times640 = 3{,}840$ control-context presentations, $4{,}320$
presentations in total, each eliciting two binary judgments ($8{,}640$ responses).
No single sitting carries that load: the perceptual arena (target $+$ control) and the
judgment arena (target $+$ control) are run in separate sessions, session order
counterbalanced across subjects, putting $2{,}160$ presentations in each. Report
the achieved session duration and trial index/block in the analysis to check for time drift;
if a session still runs long, split the control arm itself across two sittings before
shortening it.

\noindent\textbf{Splitting the study into two sessions does not make that load harmless,
and the design has to treat nonstationarity as part of the measurement problem.} Three
routes matter, and each attaches to a quantity this paper already relies on. Because
direct influence is estimated through absolute marginal differences, criterion drift
across a session can masquerade as $\Delta$ and so depress $\Vs$ without any change in
frustration. Because participant-level correlations pool trials within a subject,
adaptation can alter $F$ over the course of an arena. And because the two arenas sit in
different sessions, a difference in fatigue or practice between them can create or erase
cross-domain covariance that has nothing to do with a shared tolerance. The required
design work follows from those three, and none of it is done in this paper:
\textbf{(a)} pilot the actual session duration and attrition rather than inferring them
from trial counts; \textbf{(b)} carry trial index, block and session as covariates in the
analysis \emph{and} as factors in the null generator, so the calibration in
Figure~\ref{fig:calibration} is recomputed in a world that drifts; \textbf{(c)} schedule
planned breaks and counterbalance pairing order locally as well as across subjects;
\textbf{(d)} \emph{test} exchangeability and stationarity within each arena --- for
instance by comparing first-half and second-half estimates of $F$ and $\Delta$ --- rather
than assuming them; \textbf{(e)} consider several shorter nuisance indicators in place of
one extremely long control arm, which is where most of the burden sits; and \textbf{(f)}
recalibrate under dropout and time-varying psychometric parameters before the adjusted
statistic is promoted to confirmatory use. Item (b) and item (f) are the two that would
change a reported error rate, and until they are run the size figures quoted above hold
only for a stationary world.

\begin{figure}[htbp]\centering
\includegraphics[width=0.82\linewidth]{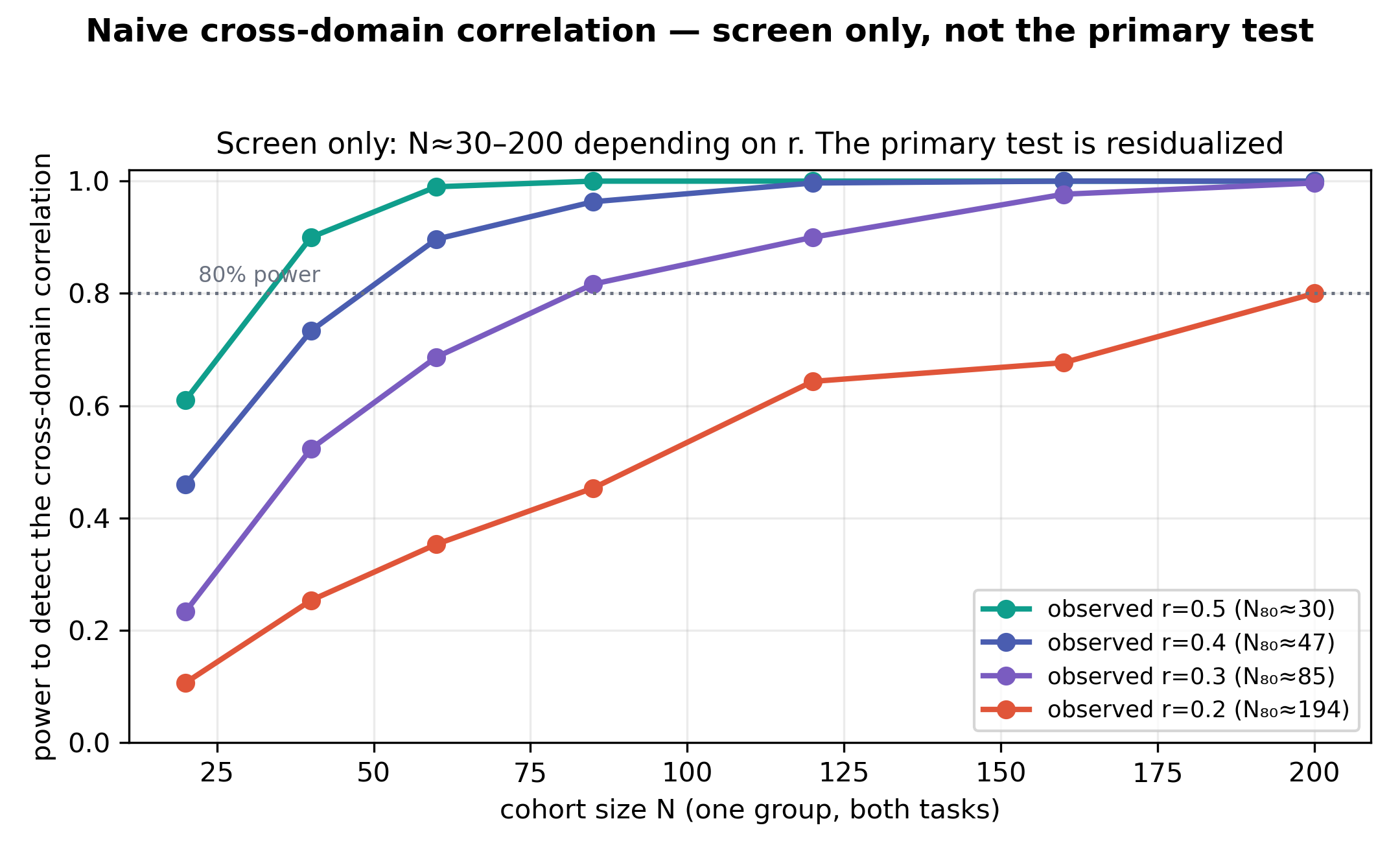}
\caption{Power for the \emph{naive} cross-domain correlation --- the screen, not the
primary test. \S\ref{sec:tests} argues this statistic is confounded and
non-diagnostic; the control-adjusted test needs $N\approx200$ with ${\ge}80$
trials/context and a control arm at about eight times that trial count. Detection
power vs $N$ (one group, both tasks) for the perceptual-loop-tolerance
$\leftrightarrow$ judgment-$\CF$ correlation: 80\% power at $N\approx85$ for an observed $r=0.3$,
$N\approx47$ for $r=0.4$ (Fisher-$z$, simulation-confirmed; the null $r=0$ holds
its 5\% false-positive rate). Measurement reliability attenuates the latent
correlation --- a latent $r=0.5$ at reliabilities $0.80$ and $0.70$ reads as
$r\approx0.37$, which needs $N\approx54$ for the same 80\% power. Code:
\code{crossdomain\_power.py}.}
\label{fig:crossdomain}
\end{figure}

The two within-arena tests, once run, would establish a cyclic contextuality violation in
each substrate; the cross-domain test, in its confound-residualized form, is the one that puts a
\textbf{shared residual association} at risk, with the
single-latent (SEM) form as the fuller extension. \textbf{No figure shows the primary test's power}:
Fig.~\ref{fig:crossdomain} is the \emph{naive} screen, which this section argues against,
and Fig.~\ref{fig:calibration} is the primary statistic's \emph{size} rather than
its power. The primary test's power at the planned design, $0.919$ against a shared component of $30\%$ of the target
trait's variance at $N=200$ with $\Delta$ scored, comes from \code{scripts/p32\_delta\_included\_rerun.py}
and is quoted above; \code{crossdomain\_estimability.py} and \code{pilot\_simulation.py} simulate the
superseded $\CF$-scored statistic under a linear confound. Neither form warrants the word
``same'' for the underlying mechanism, and the observational-equivalence and rival-trait
results above are why.

\paragraph{The honesty condition. {[}R caveat{]}}
Behavioural data \emph{signal}: a judgment's marginal depends on its context
(order effects). The plain contextual fraction assumes no signalling, so the
within-arena scoring uses the signalling-robust \textbf{Contextuality-by-Default}
statistic (Kujala et al. 2015), stated formally through maximal noncontextual
couplings and read here as the cyclic inequality corrected by the measured
disturbance. The prior nulls are \textbf{consistent with} the theory
rather than threatening it --- though a predicted null is only weak corroboration
(equally consistent with there being little behavioural contextuality), so
the theory's decisive falsifiers remain the unrun positives below, and it has
\textbf{no confirmed novel prediction yet}: Dzhafarov et al. (2016) found no true
CbD contextuality in the behavioural datasets they examined --- but \textbf{none of those was a
frustrated cyclic-relational design past $c^\ast$} (they were public-opinion polls,
visual illusions, conjoint choices, word combinations and psychophysical
matching), so the framework \emph{predicts}
$V\le0$ --- equivalently $\CF=0$ --- there, exactly as it does for the sub-threshold even
control (\S\ref{sec:ncycle}) and
the star-graph 3-cue design (for which acyclicity alone is \emph{not} sufficient --- the
prediction needs the restrictions toward the fused target to be surjective, which a factorial
cue-conflict design guarantees by construction). A dataset cannot make a cohomology
\emph{group} appear or vanish --- $\Hone(C_n;\Zt)\cong\Zt$ for every cycle
(\S\ref{sec:framework}) --- so every data-level prediction in this paper is a statement
about $V$ or $\CF$, never about $\Hone$.

\noindent\textbf{Perception is not a null literature, and an earlier draft of this
paragraph wrongly implied it was.} The two analyses of one double-detection experiment (Cervantes \&
Dzhafarov 2017a, 2017b) did return null, and a double-detection is not a frustrated
cyclic-relational percept, so $V\le0$ is what the framework expects of them. But
Cervantes \& Dzhafarov (2019) report \textbf{true contextuality in each of three individually analysed
psychophysical observers}, with repeated presentations of visual stimuli in randomly
varying contexts, established through the Contextuality-by-Default dichotomization
theorem on five-option responses. Within-person perceptual contextuality is therefore an
existing finding, not something this design would establish first. The unrun claim is narrower, and \S\ref{sec:tests} tests it: a \emph{binary cyclic} perceptual task
matched to a binary cyclic judgment task, scored on one scale, in one cohort.

\noindent Cervantes \& Dzhafarov's (2018) ``Snow Queen,'' one of the cyclic positives in the
judgment literature (Basieva et al.\ 2019 report five more, tabled in \S\ref{sec:tests}),
\emph{is} a purpose-built cyclic scenario past threshold, and an
\textbf{even 4-cycle}: its contextuality comes from the sign structure of the
correlations, not from odd parity, exactly as \S\ref{sec:ncycle}(c) requires. The
falsifying criterion
stays explicit: if a genuinely frustrated (past-$c^\ast$) relational cycle yields
no CbD excess, the instantiation fails. The theory tells you where to look --- a
sufficiently frustrated relational cycle --- and why the field, looking at
sub-threshold or non-cyclic designs, found nothing there.

\paragraph{Published-statistic reproduction and scope check. {[}R for the arithmetic{]}}
The paper's \emph{novel} claim is unrun, and this exercise does not test it. It checks two things:
that our implementation reproduces published CbD statistics, and that the within-arena
scoping rule makes nontrivial calls on the existing literature
(\code{published\_validation.py}). The rule reads: a system is contextual iff it is
\textbf{sufficiently frustrated and cyclic} ($V>0$), whatever its parity. It is \textbf{consistent with five of the six} published verdicts, and the
one disagreement traces to the source's arithmetic, not to the rule. That consistency is not a risky
postdiction: the six category-level verdicts are classified by whether they satisfy the
same cyclic criterion used to define contextuality in the first place, so the exercise is
a scope and implementation check on author-assigned category labels, not a validation. The contextual cases the rule
scores are frustrated cyclic: Cervantes \& Dzhafarov's (2018) ``Snow Queen'' and Cervantes \&
Dzhafarov's (2020) impossible-figures analysis in its cyclic representation (three disjoint
rank-2 cycles, their Option~1; their other representations are not cyclic and fall outside
the rule). Bruza et al.\ (2025) report nine of twenty face sets contextual, but in their
forced-choice design every within-context correlation is $-1$, an even sign pattern on a
4-cycle, for which $s_{\max}=n-2=2$ rather than the $3$ their \S2.2 uses; their own
criterion then gives $-\Delta\le0$, the rule's noncontextual call. The nulls are all non-cyclic or
sub-threshold: the (2017a, 2017b) double-detection, Dzhafarov et al.\ (2016) behavioural
sets, and Wang et al.'s (2014) order effects, which Dzhafarov et al.\ (2016) reanalyse as
signalling rather than true contextuality. The rule makes the \textbf{right call where a naive ``odd $=$
contextual'' heuristic fails}: Snow Queen is a contextual \emph{even} 4-cycle. The
check is also \textbf{quantitative, across eight real behavioural
systems}, with its data provenance split rather than summarised. Our code reproduces
published CbD statistics to rounding error ($\le0.002$), spanning $n=3$ and $n=4$.
\textbf{Two rows are reconstructed from raw contingency tables; six use published choice
proportions and design-specific formulas.} All eight are forced-choice systems, and in
seven of them the format drives the within-context correlations to their extremes, so those
seven rows primarily check transcription and arithmetic. This exercise does not validate the novel cross-domain
hypothesis. The table reproduces the authors' own
signalling-corrected cyclic-violation statistic on their reported scale, not $\CF$;
$\CF\in[0,1]$, so values above $1$ or below $0$ in the table are on that other scale, not
a contextual fraction. Every value below is regenerated by
\code{published\_validation.py}; the recomputation is ours, not a quotation, and is
listed so a reader can audit each cell. The input table holds the proportions we
recompute from; the reported values are printed in Cervantes \& Dzhafarov's (2018) Results
text (p.~198) and in Basieva et al.'s (2019) Table~5 (p.~1934).

\begin{center}\small
\begin{tabular}{llrrr}
\hline
System & Input table & Reported & Ours & Diff \\
\hline
Snow Queen, correct  & C\&D 2018, Table 2 & $0.452$  & $0.452$  & $0.000$ \\
Snow Queen, all      & C\&D 2018, Table 3 & $0.279$  & $0.277$  & $0.002$ \\
Meals ($n{=}3$)      & Basieva 2019, Tab.\ 3 & $1.361$  & $1.360$  & $0.001$ \\
Clothes ($n{=}3$)    & Basieva 2019, Tab.\ 3 & $1.440$  & $1.440$  & $0.000$ \\
Presents ($n{=}3$)   & Basieva 2019, Tab.\ 3 & $1.548$  & $1.548$  & $0.000$ \\
Exercises ($n{=}3$)  & Basieva 2019, Tab.\ 3 & $1.223$  & $1.224$  & $0.001$ \\
Directions ($n{=}4$) & Basieva 2019, Tab.\ 4 & $0.758$  & $0.756$  & $0.002$ \\
Coloured figures ($n{=}4$) & Basieva 2019, Tab.\ 4 & $-0.984$ & $-0.984$ & $0.000$ \\
\hline
\end{tabular}
\end{center}

\noindent Worst discrepancy $0.002$, which is the authors' own three-decimal rounding;
four reproduce exactly. Seven contextual, one genuine null.

\noindent\textbf{The reproduction mostly checks transcription and arithmetic.} Only the two Snow Queen rows are computed from raw $2\times2$ contingency
tables; the six Basieva et al.\ systems are recomputed from the authors' reported choice
proportions through their own design-specific disturbance formulas, for which
$V=2-\Delta$. Moreover in seven of the eight systems the forced-choice format drives every
within-context correlation to $\pm1$, so $s_{\max}\equiv n$ identically and $V$ reduces to
a linear function of the transcribed proportions. Agreement in those rows can fail only
through mis-transcription: it checks the arithmetic and the transcription, not the
odd-parity search or the linear program. Only ``Snow Queen, all'' has non-degenerate
correlations ($0.901$, $0.894$, $0.838$, $-0.712$) and genuinely exercises the machinery.

\noindent The measure reproduced is the authors'
signalling-corrected violation $V$ and, on the contextual side $V>0$, its
normalization $\CNT=V/4$. The
$\CF=2\,\CNT$ identity is not confined to non-signalling systems: Cervantes
(2023, Thm.~1) proves it for cyclic systems generally, taking the consistification
when a system signals, and demonstrates it on the Snow Queen data, which signal
heavily. Agreement to $0.002$ validates the reported transcription and arithmetic, and
shows the scoping rule coincides with standard CbD; only the nondegenerate ``Snow Queen,
all'' row exercises the full correlation calculation. Across randomly sampled cyclic
systems the \emph{rate} of
contextuality --- how often a system is contextual at all, not the contextual fraction $\CF$
within one that is --- falls from $0.66$ at $n=3$ to below $0.001$ by $n\ge9$. Dzhafarov,
Kujala \& Cervantes (2021) derive the bound $2^{n-1}/n!$ on that rate; it is attained, not
merely respected, for the unbiased marginals sampled here
(\code{scripts/p32\_sampling\_rate\_check.py}). (One
caveat: the impossible-figures case is an analytic epistemic-mixing construction,
not behavioural data.) None of this tests the \emph{cross-domain} prediction ---
no published dataset has the same-subject perceptual+judgment structure --- and it
shows only that the scoping rule agrees with the published verdicts where data exist.
The rarity result does not apply to this design. Both arenas are $3$-cycles, where
two-thirds of uniformly drawn systems are contextual, and in this design's own simulated
population ($c$ centred at $-0.55$ with SD $0.18$) the share past $c^\ast$ is $0.886$, so
contextuality is the expected state of a subject, not a rare one. The cross-domain
prediction concerns a correlation, not whether contextuality occurs.

\section{Relation to prior work and limitations}\label{sec:prior}
\textbf{Inherited work.} The operators are Seely's (2025); the cohomological
treatment of contextuality is Abramsky \& Brandenburger's (2011) and Abramsky et
al.'s (2015) \emph{Contextuality, Cohomology and Paradox}; the computable
contextual fraction is Abramsky, Barbosa \& Mansfield's (2017); the $n=3$ triangle
inequalities are Suppes \& Zanotti's (1981) and the general $n$-cycle inequalities
Araújo et al.'s (2013); the signalling-robust CbD criterion is
Dzhafarov and colleagues'; and the $\Hone$ formalization of the impossible figure
is Penrose's own (1992). Nonzero $\Hone$ has already been placed on \emph{visual
perception}: Ghrist \& Cooperband (2025) cast impossible figures as torsor
obstructions with an $\Hone$ invariant, and L.\ \& R.\ Ghrist (2026) build an
$H^0$--$H^2$ hierarchy for bistable percepts. Inoué (2026) models brain function
as a sheaf over neural state spaces, reading pathologies as obstructions to a global
section. We therefore do \textbf{not} claim priority for ``sheaf cohomology of
perception,'' nor for the sheaf reading of neural state spaces. Inoué's work is
conceptual and pathology-oriented rather than an empirical clinical study, and carries
no contextuality, cyclic system or experiment, so it does not reach the cross-domain
claim below; but it is independent evidence that the framing itself is in the air.

\noindent\textbf{Perceptual contextuality is an existing positive finding, and the
novelty claim below is scoped around that.} Cervantes \& Dzhafarov (2017a, 2017b) ran
psychophysical CbD and found no contextuality in their double-detection experiment under
either of their two analyses --- but Cervantes \& Dzhafarov (2019) then found \emph{true} CbD
contextuality in each of three observers' individually analysed psychophysical responses to repeated visual
stimuli in randomly varying contexts. Bruza et al.\ (2025) report CbD contextuality in
face-realness judgments, a verdict that does not survive their own criterion once its
arithmetic is corrected (\S\ref{sec:tests}). Zhan et al.\ (2024) report a violation of a contextual
reformulation of the Leggett--Garg inequality in bistable object perception, which is
adjacent quantum-like perceptual evidence rather than an instance of the same test: the
Leggett--Garg inequality is temporal in its original form and rests on different assumptions, including forms of noninvasiveness. Measuring
perceptual contextuality is therefore not new, and neither is measuring it within a
single person.

\noindent\textbf{On the relation to Sheaf-Theoretic Contextuality and to CbD.} The two
frameworks are not a theory plus an honesty correction bolted onto it. In behavioural
data inconsistent connectedness is the default observational structure, and CbD defines
the comparison couplings that separate direct influences from contextuality proper;
Dzhafarov (2023) works that comparison out directly and is the bridge this paper's
positioning rests on. The layering this paper uses is his. Abramsky and Brandenburger's sheaf formulation is
where the global-section language and the deterministic global-assignment structure come
from; Abramsky et al.'s cohomology contributes witnesses that are useful but not generally
complete; and convex geometry settles the quantitative obstruction completely for the
cyclic empirical models used here. Those three layers are kept apart throughout and are not collapsed into one.

\noindent\textbf{On the relation to quantum cognition.} This paper does not argue that
human cognition is quantum. It uses contextuality machinery as a theory-neutral test of
whether specified context-indexed variables admit a global coupling once direct
influences are accounted for, and the cross-domain hypothesis concerns covariance in the
\emph{degree} of cyclic incompatibility, not a quantum neural substrate. Quantum
cognition is in any case a broader field than contextuality tests: it also covers order
effects, interference, incompatibility, quantum probability models, and an active
model-selection debate over whether quantum formalisms outperform classical stochastic
alternatives (Pothos \& Busemeyer 2022; Busemeyer \& Bruza 2025). Nothing here should be
read as making contextuality that field's defining criterion.

\textbf{The new content} is narrower than perceptual contextuality or within-person
contextuality, both of which already exist, and we argue it is still load-bearing. Four
items, in the order of how much each is at risk.

\textbf{(i) An exact applied-mathematical bridge} from deterministic signed-cycle
frustration to a convex distance and contextual-fraction scale: parity fixes the
vertices, convexification defines the noncontextual region, and facet slack measures
violation, with the closed form $d_p=\Minc^{\star}n^{1/p-1}$ and
$\CF=\Minc^{\star}/2$ (\ref{app:prop15a}). The $\ell_p$ closed form itself is Dzhafarov,
Kujala \& Cervantes's (2020, Eq.~(56)); what is new is the signed-cycle-to-cut-polytope
route to it and a nearest point that attains it in every norm at once.
\textbf{(ii) A construct-validity result}: the one-bit forced-choice encoding, which is
the design the intransitivity literature actually runs, pins the frustration term and
inverts the intended control contrast, so two judgments per context are required rather
than preferred (\ref{app:encodingb}).
\textbf{(iii) A proposed same-subject, cross-domain individual-differences test} using
matched cyclic systems: a confound-residualized, load-specific correlation between the
two arenas' signed obstruction margins $\Vs$, the pre-clamp quantity behind $\CF$
(\S\ref{sec:tests}). We are not aware of an existing measurement of it; that statement
rests on the literature cited here and not on a systematic database search, so it is a
statement about our reading rather than a priority claim.
\textbf{(iv) An explicit account of the limits} on what such a test could show --- how
direct influence, control measurement error and exact nonidentification bound the
reading of a positive result (\S\ref{sec:tests}).

\noindent The four carry different weight. Item (i) is proved and item (ii) is
computed, so both are settled; the binding-sheaf framing that motivates (i) is
\textbf{motivational} rather than at risk, since the $\Mcap$/$\Minc$ split is Seely's
Hodge decomposition under a precision weight and none of the three designs measures
$\Minc$. Placing the two arenas on a shared cyclic footing is, once both are
\emph{defined} as cyclic tasks, close to definitional. The genuinely at-risk novelty is
\textbf{(iii)} together with the quantitative linking hypothesis --- a shared
obstruction-tolerance moving a person's signed margin $\Vs$ in both arenas, that is,
their position relative to each arena's threshold $c^\ast$ --- which a bare ``two things
share a diagram'' reading cannot satisfy.

\textbf{Limitations.} All three tests are designed but unrun [C], and the theory
has no confirmed novel prediction yet; a bare cross-domain correlation is confounded
(\S\ref{sec:tests}), so only the residualized, load-specific pattern is diagnostic. The
$\Mcap$/$\Minc$ binding-sheaf severity split is \textbf{motivational}: none of the
three designs measures $\Minc$ --- all score the contextual fraction $\CF$ or its
pre-clamp margin $\Vs$ --- so
the binding sheaf is not itself put at empirical risk here. The identification of
$\Minc$ with $\CF$ is settled in \ref{app:prop15a} (Prop 15a): a
\emph{theorem} \textbf{[R]} for the
convex correlation-polytope obstruction $\Minc^{\star}$, and a \emph{refuted} claim
for the fixed-coboundary seminorm. The perceptual arena is built from Penrose's
own $\Zt$ figure, extended by Ghrist \& Ghrist (2026); his continuous
$\R^{+}$-holonomy tribar plays no role in the cross-domain test
(\S\ref{sec:arenas}). Intransitivity can be mere noise (Regenwetter et al. 2011), which
the perceptual arena must rule out; the neural identification of stalks is [A]; and
the judgment-side $\CF$ and the CbD statistic $\CNT$ coincide up to the verified
factor $\CF=2\,\CNT$ (Cervantes 2023, for cyclic systems generally, via the
consistification when a system signals). We model the structure of
binding, not phenomenal experience.

\noindent\textbf{The limitation that bounds every positive reading, stated here and not
only in the analysis section.} Under the six-score covariance model this design
produces, a genuine shared target component and a shared control-method component
generate the \emph{identical} population covariance matrix (\S\ref{sec:tests};
\code{p32\_control\_trait\_identifiability.py}, exact in rational arithmetic and
re-derived in an independent computer-algebra engine). The covariance matrices agree exactly,
so no sample size resolves the difference for any statistic computed from that matrix.
Gaussian latent variables and errors are sufficient for the same to hold of every statistic
of the six scores. With skewed latents, third moments carry a weak signal
(\S\ref{sec:tests}). For covariance-based analysis the
practical consequence is a hierarchy the design sits at the bottom of: with \emph{two} controls, neither generic detection nor
point identification of a method component is available; with \emph{three}, generic
nonproportional contamination becomes detectable from the target--control and
control--control covariances but is still not point-identified; with \emph{four}, one
method factor is generically point-identifiable under the stated loading and error
restrictions; and at any number, exactly proportional method loadings remain
unidentified. A validated pure anchor or randomized crossed forms would identify only
under explicit exclusion and invariance assumptions. This design carries two controls.
A positive cross-domain result is therefore reportable as a shared residual association
consistent with the framework, together with the contamination share that would be required
to forge it, computed from the observed scores by the formula of \S\ref{sec:tests} ($26.5\%$
is that section's worked case only), and as nothing stronger.

\section{Conclusion}
A perceptual illusion and a framing effect look like different failures --- one of
the senses, one of reason. We have argued they are failures of the same kind: a
set of locally-coherent commitments that cannot be made globally coherent,
measured by the same cyclic global-consistency obstruction. If either matched test
comes back positive, that arena carries a signature --- a behaviour outside the
noncontextual polytope --- which no
model constrained to a single global joint distribution can produce. If both do, the
same mathematical signature is present whether the parts are cues or claims. A shared
pattern across the two arenas does not by itself show that one mechanism drives them:
only the cross-domain association puts that claim at risk, and even a positive result
there would not identify a common mechanism (\S\ref{sec:tests}).

\noindent\textbf{The contribution, and what it leaves out.} None of the proposed experiments
has been run. The paper contributes a theoretical claim about two literatures --- that a binding-motivated
perceptual task and judgment contextuality are scored against the same cyclic
noncontextuality inequalities and share one convex global-consistency geometry --- together
with the machinery that would let a study decide it: a named primary pre-specified
statistic, the signed margin $\Vs$; the simulated size and power of that statistic at the
design point, and the demonstration that the statistic it replaces does not hold its
nominal rate; the five pass/fail pilot gates the perceptual arena must clear before it is
fielded at all; and the proof that a positive result would license a shared residual
association and not a common mechanism, at any sample size. Three things a confirmatory
report would have to supply are absent, and are named rather than implied: the minimum
effect of interest $r_{\min}$ that makes the null reading a refutation rather than an upper
bound, the discriminant indicators that would narrow a positive reading past ``some trait
loads on both arenas,'' and a feasibility pilot for a response burden this paper computes
but has not imposed on anyone.

\noindent A fourth absence is a stronger design together with the assumptions that would make it informative. One proposed route varies the structural property expected to create the obstruction against a sham matched in task load and response demand, with both arenas measured in each randomly assigned arm. Randomization makes the two arm samples comparable; it does not prevent a stable participant trait from responding differently to the two forms. Sufficient conditions for identification are that the sham switches off the target component, that the total non-target cross-arena covariance is invariant across arms, that the target response is uncorrelated with the remaining nuisance score components, and that cross-arena measurement errors are excluded. Under those conditions the structural-minus-sham cross-arena covariance block is exactly the target component's covariance, so one control level and one active level per arena suffice for a control-to-active response contrast. A third level offers checks of the loading profile but is neither necessary for that contrast nor sufficient for identification on its own. The cross-arena block identifies loading products; separate standardized loadings are identified given a valid scale anchor, or given invariance of the entire nuisance and measurement-error covariance across arms, and in the latter case the full covariance contrast identifies a single additive loading vector up to sign. A within-subject crossover can instead remove unchanged nuisance scores before taking covariance and thereby relax the target--baseline orthogonality assumption, provided stability, period handling, absence of carryover, and differenced measurement errors that do not cross arenas are justified. A randomized $2\times2$ design assigns structural or sham form per arena, observing both scores in every cell. With $B_{ab}$ the centered cross-arena covariance in cell $(a,b)$, the double contrast $F=B_{11}-B_{10}-B_{01}+B_{00}$ equals the covariance of individual own-arena treatment responses under \emph{arena separability} (no spillover): each arena's score depends only on that arena's assignment. Separability is sufficient for this equality, not necessary, and neither randomization nor invariance checks establish it. Attributing $F$ to a shared target further requires invariant nuisance responses, response errors uncorrelated with the target and across arenas, and no rival trait responding to the manipulation. The estimand is a response covariance, not the shared residual association that \S\ref{sec:tests} tests. At equal cell variability, the balanced design needs four times a two-arm comparison's participants for equal precision, more as mixed cells shrink. Each route identifies a component attributed to the structural manipulation only under its stated exclusions; calling it a common obstruction mechanism still requires excluding rival traits whose response changes with that manipulation. Those exclusions and the sham's feasibility remain work for the proposed experiment, not consequences of randomization. The proposal is refutable in the specific sense that matters:
it says which number, from which statistic, against which calibrated cutoff, would count
against it --- once a preregistration fixes the one quantity \S\ref{sec:tests} leaves open.

\section*{Data and code availability}
No new empirical data are reported; all quantitative claims are reproducible from
code. \code{code/} in the project repository holds the contextual-fraction linear
program and the general odd-$n$ closed form
(\code{check\_contextuality.py}, \code{check\_gm003\_verify.py}), the CbD/$\CNT$
and its non-monotonicity counterexample (\code{check\_cbd\_monotonicity.py}), the
survey design/power (\code{survey\_pipeline.py}), the cross-domain power analysis
(\code{crossdomain\_power.py}), the per-subject-$\CF$ estimability and
confound-discrimination simulation (\code{crossdomain\_estimability.py}), the
linking reality-check (\code{crossdomain\_linking.py}: why the single-latent claim
needs a latent-variable model, not disattenuation), and the end-to-end simulation pilot
(\code{pilot\_simulation.py}). The same directory holds the latent-variable
leg-(iii) estimator (\code{crossdomain\_sem.py}, the method-of-moments screen, a
\emph{secondary} analysis alongside the primary residualized-$\Vs$ correlation specified
in \S\ref{sec:tests};
\code{crossdomain\_sem\_ml.py}, the full
maximum-likelihood or generalised-least-squares model with its likelihood-ratio
test, planned and not yet calibrated), the control-trait exposure of that screen
(\code{p32\_control\_trait\_forgery.py}) and the proof that no check on the covariances of the design's own
scores can rule that trait out (\code{p32\_control\_trait\_identifiability.py}),
the vanishing-equivalence verification
(\code{check\_vanishing\_equivalence.py}), the derivation of
$\Minc^{\star}=V=2\,\CF$ for a general correlation vector
(\code{p32\_minc\_cf\_general\_gamma.py}), the consistified contextual-fraction
check (\code{p32\_verify\_consistified\_cf.py}), the monodromy multiplicity
witnesses (\code{p32\_monodromy\_multiplicity.py}), the published-data
reproduction (Snow Queen $+$ Basieva 2019; \code{published\_validation.py}), and
the figure generators (\code{make\_unification\_fig.py},
\code{make\_survey\_power\_fig.py}, \code{make\_crossdomain\_fig.py},
\code{make\_local\_global\_gluing\_fig.py}, \code{make\_vstar\_cf\_hinge\_fig.py},
\code{make\_encoding\_ab\_curves\_fig.py}, \code{make\_monodromy\_intuition\_fig.py},
\code{make\_normball\_intuition\_fig.py}, \code{make\_polytope\_distance\_fig.py}, and the
independent geometric check it is distinct from, \code{p32\_cut\_c3\_independent\_check.py},
whose result \ref{app:prop15a} reports).
\code{scripts/} holds the calibration studies \S\ref{sec:tests} relies on
(\code{p32\_verify\_residualization.py}, together with
\code{p32\_mcverify\_asymptote\_0058.py}, the independent re-derivation of its asymptote
through a separate random-number path that \S\ref{sec:tests} reports disagreeing with it;
\code{p32\_calibrate\_residualized\_test.py}
for the superseded plug-in statistic; \code{p32\_vstar\_control\_noise\_sweep.py} and
\code{p32\_calibrate\_vstar\_test.py} for the signed margin the primary pre-specified test now uses),
the robustness checks behind \S\ref{sec:tests}'s misspecification results
(\code{p32\_vstar\_robustness\_checks.py}), the rerun of the primary test's size and power
with the disturbance term $\Delta$ scored and the power table that follows from it
(\code{p32\_delta\_included\_rerun.py}), the signalling-trait forgery check
\S\ref{sec:tests} cites (\code{p32\_confound\_and\_sensitivity\_check.py}), the
attainment check behind the same section's $2^{n-1}/n!$ rate claim
(\code{p32\_sampling\_rate\_check.py}, which reuses the exact-rational simplex routine of
\code{p32\_verify\_contextual\_fraction.py}), the exact-rational linear program over the
$2^n$ global assignments that the appendix checks its cyclic-system specialisation of
$\CF$ against (\code{p32\_exact\_lp.py}), the generator for
Fig.~\ref{fig:calibration} (\code{p32\_fig\_calibration.py}), the multiplicity-scheme
check behind item 9 of \S\ref{sec:tests}'s estimand specification
(\code{p32\_multiplicity\_scheme\_check.py}, which exhibits the three-way-Holm
counterexample and confirms the adopted scheme's invariance and error rates, each with a
negative control that fires), the presubmission gate
(\code{p32\_presubmission\_check.py}), the computer-algebra check separating the vertex
and edge projectors of \S\ref{sec:framework} and establishing the $y=0$ restriction on
the cue-fusion identity (\code{p32\_projector\_direction\_check.wl}, eleven checks with
six negative controls), and the two encodings of the perceptual task:
\code{p32\_forced\_choice\_encoding\_check.py} establishes the degeneracy that
\S\ref{sec:arenas} cites, and \code{p32\_encoding\_b\_full\_analysis.py} with its
computer-algebra companion \code{p32\_encoding\_b\_wolfram.wl} reproduce every claim in
\ref{app:encodingb} --- the general closed form, the geometry, Table~\ref{tab:Bpower},
and the cross-domain consequence --- each paired with a negative control that is reported
together with whether it fired, except the five figures of B.5(c), which come from
\code{p32\_appendixB\_P\_check.py}. Two further checks were added for this revision:
\code{p32\_v5\_encodingB\_linear\_ordering\_check.py} and its computer-algebra companion
\code{p32\_v5\_lop\_equivalence.wl} establish B.3(iv), the identification of Encoding B's
contextual region with the interior of the linear-ordering polytope, by two routes that
share no algebra; and \code{p32\_v4\_pilot\_gate\_scale\_check.py} fixes which scale each
standard deviation in pilot gate (4) is quoted on. A third establishes
\ref{app:encodingb}'s B.8, the mixture-of-transitive-orders bound for the
\emph{adopted} encoding: \code{p32\_mixture\_of\_orders\_encodingA\_check.py}, in exact
rational arithmetic, with three negative controls that fire and one reported non-firing
robustness case. Four more were added for this revision, for \S\ref{sec:tests}'s
treatment of a same-day state in $g$: \code{p32\_bverify\_plim.wl} derives in closed form the
limits of the method-of-moments screen and of the primary residualized test;
\code{p32\_bverify\_a\_f3\_mc.py} checks the screen's values by simulation;
\code{p32\_bverify\_f1\_mc.py} gives the primary test's rejection rates under that state, with
its two controls; and \code{p32\_bverify\_tau\_mc.py} gives the size and power of the
cross-session tetrad difference and its forgery by a control-only trait. The two
experimental protocols are in \code{protocols/}:
\code{penrose-experiment.md} (perceptual $+$ cross-domain) and
\code{contextuality-experiment.md} (survey).

All of the above, together with the mathematical appendix behind Proposition~15a, is
additionally deposited at OSF as a standalone, citable archive:
\url{https://doi.org/10.17605/OSF.IO/5NPT4}.

\section*{Declarations}
\textbf{Ethics.} This paper reports no new human-subjects data; all results are
theoretical or reproducible from code. The proposed experiments will require
prospective ethics approval before data collection. \textbf{Competing interests.}
The author declares none. \textbf{Funding.} None to declare.

\section*{Use of AI tools}
AI tools assisted under the author's direction. Claude (Anthropic) wrote simulation and
verification code to the author's specification, checked derivations, gave adversarial review and
gave drafting and editorial support. ChatGPT (OpenAI) was used for adversarial review and
independent mathematical checking, Gemini (Google) for adversarial review and independent
mathematical checking, Perplexity for literature search and citation verification,
Wolfram Mathematica for checking derivations, and the Z3 and cvc5 solvers for counterexample
searches.

\appendix

\section{The co-vanishing of $\Minc$ and $\CF$ (Proposition 15a)}
\label{app:prop15a}
\setcounter{figure}{0}\setcounter{table}{0}\setcounter{equation}{0}

Section~\ref{sec:ncycle} claims that the perceptual obstruction and the
judgmental one are scored against the same convex cyclic geometry, and immediately warns
that the co-vanishing must be read carefully. This appendix gives that careful reading. The short
version is that the identification is true, but not of the functional one would
first reach for: the fixed-coboundary seminorm $\Minc$ of \S\ref{sec:framework} does
\emph{not} co-vanish with $\CF$, and no functional of its kind can, whereas a
convex obstruction built from the same data does --- and does so exactly, with a
constant of proportionality rather than merely a shared zero set.

Throughout, fix the dichotomous $n$-cycle scenario of Section~\ref{sec:ncycle}:
contexts are adjacent pairs, marginals are unbiased, and the empirical model is
non-signalling. Such a model is summarised by its adjacent-correlation vector
$\gamma=(\gamma_0,\dots,\gamma_{n-1})\in[-1,1]^{n}$, and $\CF(e)$ is the
contextual fraction of Abramsky, Barbosa \& Mansfield (2017). Write
$\mathrm{CUT}(C_n)$ for the cut polytope of the $n$-cycle.

\medskip\noindent\textbf{Proposition 15a ($\Minc$/$\CF$ vanishing-equivalence).}
\emph{The literal linear form is false; the convex form is a theorem.}

\emph{(i) The literal biconditional fails for the geometric residual.} With scalar
vertex stalks, the frustrated odd-$n$ binding coboundary $\dz$ --- every edge
demanding anti-alignment --- is invertible, so $\Hone_{\R}=\mathrm{coker}\,\dz=0$ and
hence $\Minc=\|\Pi_{\mathrm{coker}\,\dz}\,d\|\equiv0$ for \emph{every} conflict $d$
\emph{in the hard-binding limit $\lambda\to\infty$}; at finite $\lambda$ the
$\lambda$-dependent form is strictly positive for $d\neq0$ (the two forms coincide
only in that limit), while $\CF$ sweeps $(0,1]$ as $c\to-1$. The statement
$\Minc=0\Leftrightarrow\CF=0$ is therefore false as written even in the limit where
the residual vanishes.

\emph{(ii) No fixed linear sheaf can do better, so (i) is not a bad choice of
stalks.} Any fixed linear residual $b\mapsto\|\Pi_{\mathrm{coker}\,\dz}\,b\|$ is a
seminorm, and a seminorm vanishes on a linear \emph{subspace}. The zero set of
$\CF$ is $\mathrm{CUT}(C_n)$, which is full-dimensional. No choice of stalks or
restriction maps can make the two sets agree.

\begin{figure}[htbp]\centering
\includegraphics[width=0.75\linewidth]{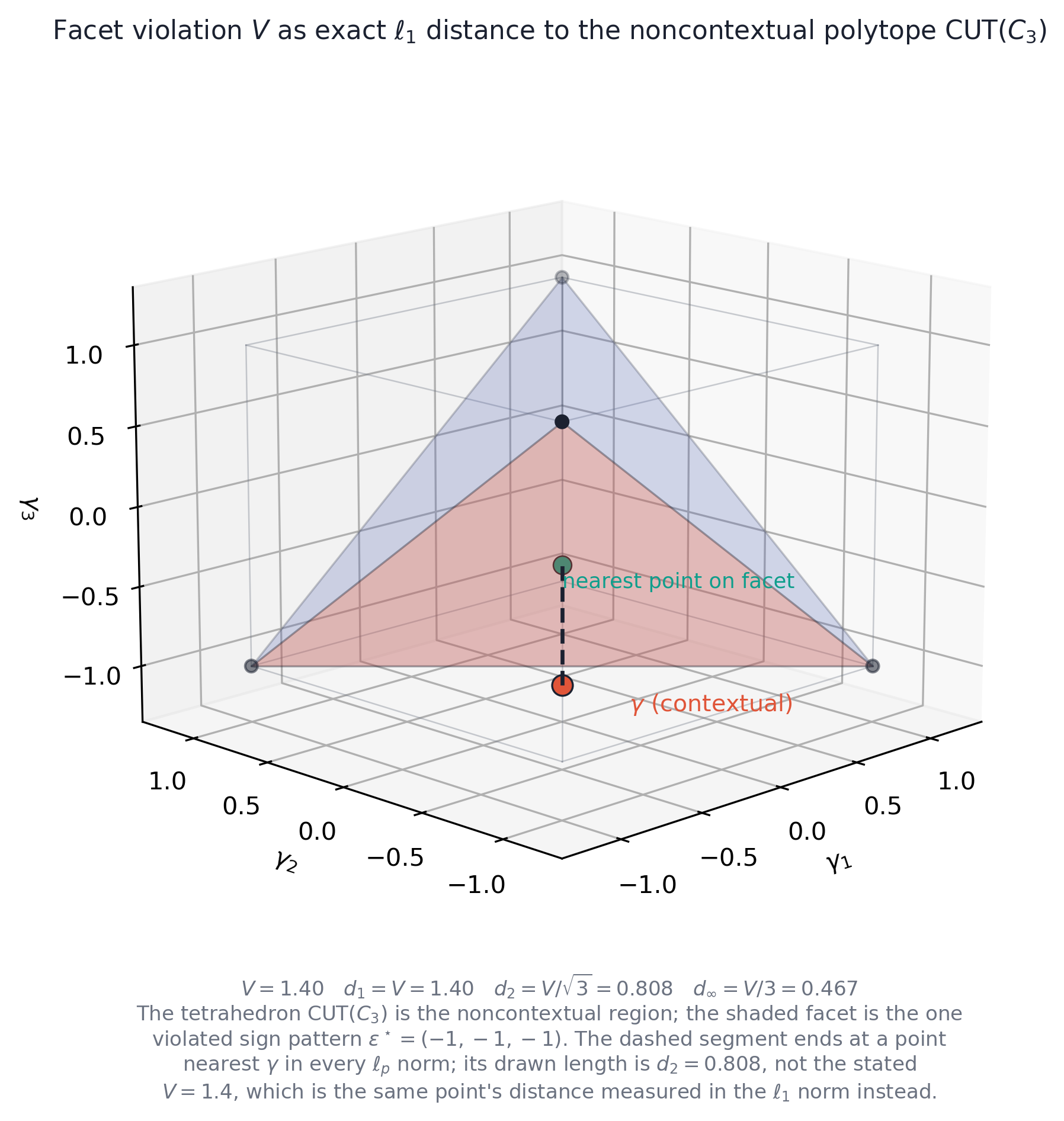}
\caption{The geometry behind Equation~\eqref{eq:lpdistance}, drawn for $n=3$.
$\mathrm{CUT}(C_3)$ is the tetrahedron with vertices at the four even sign patterns
$(\varepsilon_1,\varepsilon_2,\varepsilon_3)$ with $\prod_i\varepsilon_i=+1$ --- the
noncontextual region. A correlation vector $\gamma$ outside one odd-parity facet (shaded) is
exactly $V$ away from it in the $\ell_1$ norm, with the other standard $\ell_p$ distances scaled
by the fixed factor $n^{1/p-1}$; they are not independent quantities.
The dashed segment ends at $x=\gamma-(V/n)\varepsilon^\star$, a point nearest $\gamma$ in
\emph{every} $\ell_p$ norm (the H\"older argument proving Equation~\eqref{eq:lpdistance} in
this appendix); its drawn Euclidean length is
$d_2=0.808$, not the stated $V=1.4$, which is the same point's distance measured in the
$\ell_1$ norm instead. \code{code/p32\_cut\_c3\_independent\_check.py} rebuilds
$\mathrm{CUT}(C_3)$ from the cut definition (not from Equation~\eqref{eq:lpdistance} or the
figure script) and computes $d_1,d_2,d_\infty$ by brute-force optimisation over the polytope
itself, confirming all three match Equation~\eqref{eq:lpdistance} and that the nearest point
found lies inside $\mathrm{CUT}(C_3)$, not merely on the violated facet's plane; three
negative controls (an even-parity sign pattern, the wrong exponent, an out-of-box point)
each correctly fail.}
\label{fig:polytope-distance}
\end{figure}

\emph{(iii) The convex obstruction is equivalent.} Define
$\Minc^{\star}(e):=\max\{0,\,s_{\max}(\gamma)-(n-2)\}$ --- the amount by which the
correlation vector overshoots the cycle inequality that defines $\mathrm{CUT}(C_n)$,
i.e.\ the amount of the correlation demand that no global $\pm1$ assignment can meet.
This is the \emph{unnormalised facet slack}, and it is in fact exact $\ell_1$ distance
to $\mathrm{CUT}(C_n)$ --- not merely proportional to it under one choice of norm, but
equal to it under every standard $\ell_p$ norm up to that norm's own scale factor. For
$\gamma\in[-1,1]^n$ (a hypothesis that is load-bearing here, not decorative: outside the
box more than one facet can be violated at once and the argument below fails), write
$\epsilon$ for the unique violated odd sign pattern and $V=\Minc^{\star}(e)$. Then for
every $1\le p\le\infty$,
\begin{equation}
d_p\bigl(\gamma,\mathrm{CUT}(C_n)\bigr)=V\,n^{1/p-1},
\label{eq:lpdistance}
\end{equation}
In words: for the $\ell_1$ norm the raw facet violation $V$ \emph{is} the distance itself;
every other $\ell_p$ norm measures the same one-facet departure, scaled by the fixed factor
$n^{1/p-1}$. In particular $d_1=V=\Minc^{\star}$ and, using~(v) below, $\CF=\tfrac12d_1$.
\textbf{Equation~\eqref{eq:lpdistance} restates a published result, and this appendix
claims its derivation rather than its content.} Dzhafarov, Kujala \& Cervantes (2020)
already formulate cyclic contextuality measures as an $\ell_1$ distance to the
noncontextual polytope; their Theorem~15 identifies that $\ell_1$ distance with the excess
of the cyclic inequality's left side over its right, which is $V$ in the coordinates used
here; and their Eq.~(56) gives the all-$p$ scaling directly,
$\mathrm{CNT}_2^{(p)}=n^{(1-p)/p}\,\mathrm{CNT}_2$, whose exponent $(1-p)/p$ equals the
$1/p-1$ of Equation~\eqref{eq:lpdistance}. Cervantes (2023, Thm.~1, cited above)
proves $\CF=2\CNT$ for cyclic systems generally via the same consistification. The new content here is narrower, and is stated as such: the derivation runs through the cut polytope
$\mathrm{CUT}(C_n)$ and its facet characterisation rather than through the
Contextuality-by-Default box-and-polytope construction, and the proof below exhibits an
explicit nearest point $x=\gamma-(V/n)\epsilon$ attaining the bound simultaneously in
every $\ell_p$ norm; Figure~\ref{fig:polytope-distance} draws that polytope and that
nearest point for $n=3$. Then $\Minc^{\star}(e)=0$ if and only if
$\CF(e)=0$, unconditionally under non-signalling.

\emph{(iv) Hypotheses: both are load-bearing, and neither is necessary.} The
equivalence is proved under two conditions, labelled (C1)--(C2) to avoid colliding
with the cohomology $\Hone$ or with this appendix's own Equation~(A.1): (C1) non-signalling
and (C2) a correlation-determined scenario, in which pairwise correlations fix
contextuality. Both are \textbf{sufficient} hypotheses for the proof as given, and it is
worth saying plainly that neither is claimed \textbf{necessary}.

\textbf{(C1) is explicitly not necessary.} Under signalling, replace membership in
$\mathrm{CUT}(C_n)$ by the Contextuality-by-Default noncontextuality criterion of
Kujala, Dzhafarov \& Larsson (2015, Theorem~2), under which the biconditional again holds. So
(C1) buys the simpler polytope, not the result.

\textbf{(C2) is sufficient, and \emph{some} condition of its kind is needed --- which is
a weaker statement than (C2) itself being necessary.} Without any such condition the
equivalence can fail: for higher-moment or more-than-two-outcome scenarios a model can be
contextual while its pairwise correlations are jointly realisable, and then even the
convex form fails. That argument shows the hypothesis cannot simply be dropped; it does
not show that (C2) is the weakest condition that would do, and this paper does not claim
it is. The $n$-cycle used throughout sits inside the safe regime by construction, so
nothing here turns on finding the weakest one.

\emph{(v) The link is quantitative, not merely a shared zero set.} In the
contextual regime the raw cycle-inequality violation $V$ satisfies
\begin{equation}
\Minc^{\star} \;=\; V \;=\; 2\,\CF .
\label{eq:severity}
\end{equation}
Equation~\eqref{eq:severity} says the two obstructions are not just co-vanishing
but strictly proportional: the same severity, up to a fixed factor of two. It
holds for every correlation vector $\gamma\in[-1,1]^n$, not only on the uniform-$c$
family, and does not require the marginals to be unbiased (proved symbolically for
general, not necessarily unbiased, marginals, and checked numerically for biased
marginals on both sides of the facet: \code{p32\_minc\_cf\_general\_gamma.py}); what it
does require is the non-signalling and correlation-determined regime of~(iv).

\medskip\noindent\emph{Proof.} (i) The all-anti coboundary is $\dz=I+S$ with $S$
the cyclic shift, $(Ss)_i=s_{i+1}$ (written $S$, not $P$: in \S\ref{sec:framework}
the matrix $P$ is the precision metric, and this shift is neither that nor an arena label). Its determinant is
$\det\dz=\prod_k(1+\omega_k)=1-(-1)^n$ over the $n$-th roots of unity $\omega_k$,
which equals $2$ for odd $n$ and $0$ for even $n$. So for odd $n$ the coboundary is
an isomorphism, its cokernel is trivial, the projection $\Pi_{\mathrm{coker}}$ is the
zero map, and $\Minc\equiv0$. For balanced cycles $\dz$ instead has the
two-colouring kernel, rank $n-1$, and cokernel $\cong\R$.

(ii) The seminorm $\|\Pi_{\mathrm{coker}}\cdot\|$ vanishes exactly on
$\ker \Pi_{\mathrm{coker}}=\mathrm{im}\,\dz$, a linear subspace. $\mathrm{CUT}(C_n)$
contains a neighbourhood of the origin, since sub-threshold correlations are
noncontextual with slack, and is therefore full-dimensional. A subspace coincides
with a full-dimensional convex set only if it is the whole space. But
$\mathrm{CUT}(C_n)\subseteq[-1,1]^n$ is bounded and so is not $\R^n$; the two zero sets
therefore still disagree even in the terminal case $\Pi_{\mathrm{coker}}=0$, where the
seminorm vanishes identically while $\CF$ does not. (The argument assumes, as
``fixed linear residual'' is meant to convey, that the conflict $d$ is a fixed linear
function of $\gamma$, so that the two zero sets live in one space.)

(iii) $\CF(e)=0$ holds exactly when a global section of the empirical presheaf
exists (Abramsky, Barbosa \& Mansfield 2017), which holds exactly when $\gamma$
lies in the noncontextual polytope. For the $n$-cycle that polytope's facets are
the cycle inequalities (Araújo et al.\ 2013), and those are precisely the
\emph{nontrivial} facets of $\mathrm{CUT}(C_n)$ (Barahona \& Mahjoub 1986); the trivial
box facets $\lvert\gamma_i\rvert\le1$ hold automatically on the domain fixed above, so on
that domain membership in $\mathrm{CUT}(C_n)$ is equivalent to the cycle inequalities. The slack
$\max\{0,s_{\max}-(n-2)\}$ vanishes exactly when every cycle inequality holds, i.e.\
exactly on $\mathrm{CUT}(C_n)$, which gives the biconditional.

For the distance claim~\eqref{eq:lpdistance}: set $x=\gamma-(V/n)\epsilon$. Then
$\epsilon^Tx=\epsilon^T\gamma-V=n-2$, so $x$ sits on the violated facet. Every other constraint also holds at $x$, because the
move toward that facet uses up only part of each coordinate's slack. Write
$t_i=1-\epsilon_i\gamma_i$, non-negative by the box hypothesis; then $\sum_i t_i=2-V$ and
$\epsilon_ix_i=1-t_i-V/n$. Since $0\le t_i\le2-V$, each $\epsilon_ix_i$ lies in $[-1,1]$,
so every box constraint holds. Any other odd sign pattern $\epsilon'$ differs from
$\epsilon$ on an even set $S$ of $k\ge2$ coordinates, and
${\epsilon'}^{T}x=(n-2)-2\sum_{i\in S}\epsilon_ix_i$, where
$\sum_{i\in S}\epsilon_ix_i=k-\sum_{i\in S}t_i-kV/n\ge k-2+V(1-k/n)\ge0$. So
${\epsilon'}^{T}x\le n-2$, every cycle inequality holds, and $x\in\mathrm{CUT}(C_n)$. For any $y\in\mathrm{CUT}(C_n)$,
H\"older's inequality applied to $V\le\epsilon^T(\gamma-y)$ gives
$\|\gamma-y\|_p\ge Vn^{1/p-1}$ for every $p$, and $x$ attains this bound exactly, since
$\|\gamma-x\|_p=\|(V/n)\epsilon\|_p=Vn^{1/p-1}$. Hence $x$ is a nearest point in every
$\ell_p$ norm and~\eqref{eq:lpdistance} is exact.

(v) Fix the maximising pattern $\epsilon$ and write $q_i=(1-\epsilon_i\gamma_i)/2$ for the
probability, in context $i$, of the outcome pair that breaks it. Then
$\sum_i q_i=(n-\sum_i\epsilon_i\gamma_i)/2$, so the cycle inequality reads $\sum_i q_i\ge1$
and $s_{\max}-(n-2)=2(1-\sum_i q_i)$; the claim is $\CF=1-\sum_i q_i$. Since
$\prod_i\epsilon_i=-1$ while $\prod_i g_ig_{i+1}=+1$ for every global $\pm1$ assignment $g$,
each $g$ breaks an odd number of the constraints, so every noncontextual model has
$\sum_i q_i\ge1$; applied to a decomposition $e=\lambda e^{\mathrm{NC}}+(1-\lambda)e'$ this
gives $\lambda\le\sum_i q_i$ and hence $\CF\ge1-\sum_i q_i$.

\emph{The converse direction is a sketch, not a reader-checkable certificate, and the
formal basis is the cited theorem.} Take the $2n$ global assignments that break exactly
one constraint --- for each $i$ there are exactly two, since on a cycle the identity of the
single broken edge fixes the labelling up to a global flip. Give the assignment breaking
constraint $i$ and realizing outcome pair $s$ in context $i$ the weight $b_g=e_i(s)$. The
two assignments for a given $i$ therefore carry total weight $q_i$, and the whole family
carries mass $\sum_i q_i$; on every \emph{breaking} outcome the domination constraint
holds with equality by construction. One family of domination inequalities is left
unwritten here: those on the \emph{holding} outcomes of each context. A full primal
certificate would have to enumerate the context--outcome incidence map and check every
one of them. We therefore do not claim the
argument above as a proof. The identity $\CF=\tfrac12(s_{\max}-(n-2))$ for cyclic systems
is Cervantes's (2023, Thm.~1) --- stated there as $\mathrm{CNTF}=2\,\mathrm{CNT}_2$,
including the consistified case --- together with the facet characterisation of
Araújo et al.\ (2013), and what is given here is the specialisation of that result to the
correlation coordinates used throughout this paper. The specialisation is confirmed
numerically against an independent exact-rational linear program over the $2^n$ global
assignments (\code{p32\_exact\_lp.py}, \code{check\_vanishing\_equivalence.py}). The
bounds meet, so
$\CF=\tfrac12(s_{\max}-(n-2))$ in the contextual regime for every $\gamma$ and every
non-signalling choice of marginals, and $\Minc^{\star}=V=2\,\CF$. Two facets cannot be
violated at once: if $\epsilon\neq\epsilon'$ both have odd parity they differ in $k\ge2$
places, whence $\sum_i\epsilon_i\gamma_i+\sum_i\epsilon'_i\gamma_i\le2(n-k)\le2(n-2)$.

(iv) The substitution under signalling is the definition of the
Contextuality-by-Default polytope. The higher-moment failure is the standard gap
between correlation consistency and full noncontextuality. $\square$

\medskip\noindent\textbf{Remark 15a.1 (why the odd cycle was invisible: a monodromy
law).} For a cycle whose restriction maps are isomorphisms, $\dim\Hone_{\R}$ equals the
\emph{geometric} multiplicity of eigenvalue $1$ --- the dimension of the fixed
subspace $\ker(M-I)$, which is strictly less than the algebraic multiplicity exactly
when $M$ is non-diagonalisable \emph{at eigenvalue $1$} --- in the monodromy $M$, the composite of the restriction
maps once around the loop. This one law explains part~(i) without computing any
determinant. The frustrated odd cycle demands anti-alignment on every edge, so its
monodromy is $(-1)^n=-1$ for odd $n$; there is no eigenvalue $1$, so
$\Hone_{\R}=0$, so $\Minc\equiv0$. The invertibility of $\dz$ in the proof is not
an artefact of the sign convention --- it is this law. The frustrated odd cycle is,
in other words, \emph{already twisted, by sign rather than by angle}.

\begin{figure}[htbp]\centering
\includegraphics[width=\linewidth]{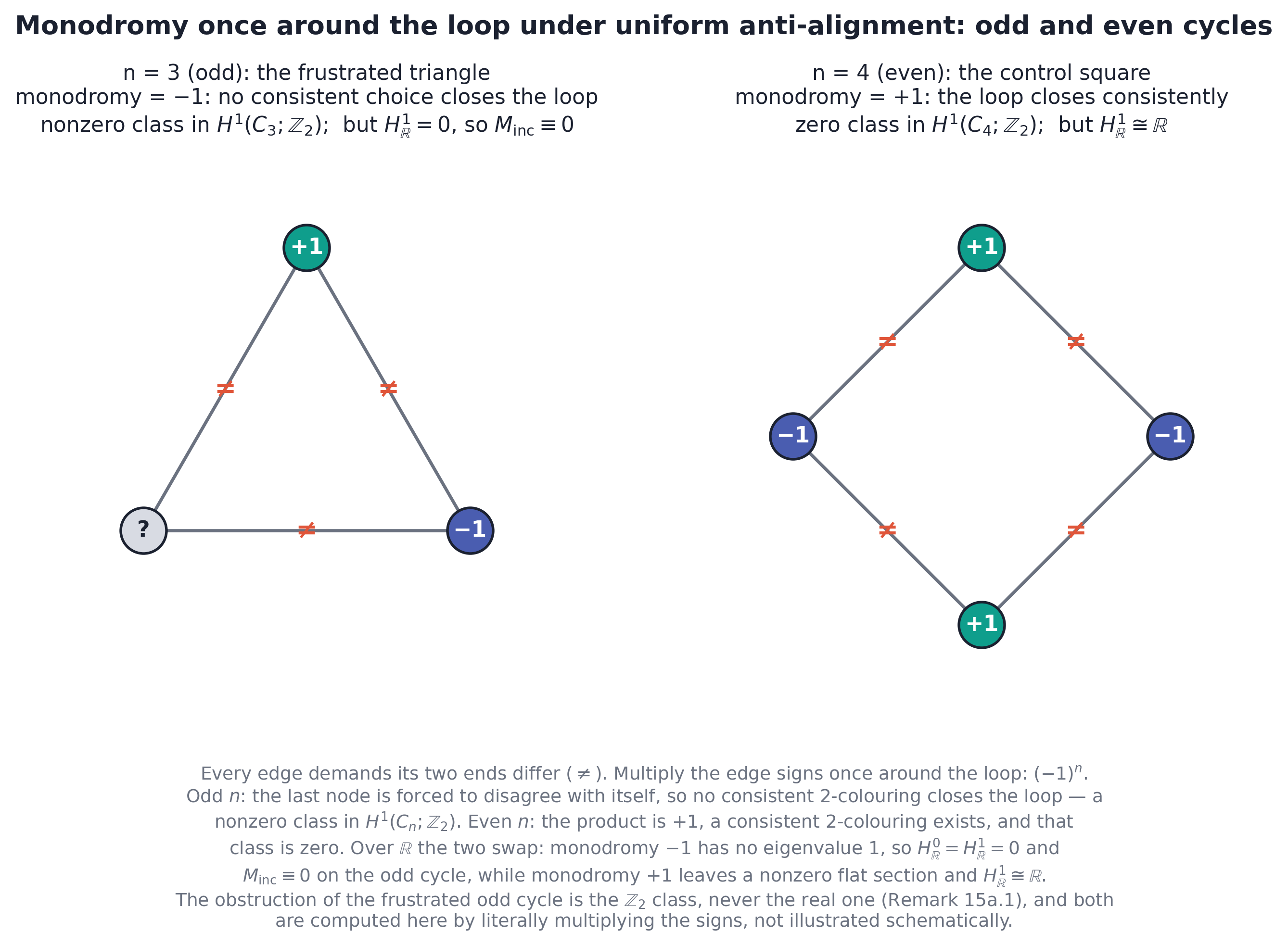}
\caption{The monodromy law of Remark~15a.1, computed by literally multiplying the
edge signs once around the loop. For a local system on a cycle, $H^0_{\R}\cong\ker(M-I)$
and $\Hone_{\R}\cong\operatorname{coker}(M-I)$ for the monodromy $M$, so the two always
have the same dimension. \textbf{Left:} the frustrated $3$-cycle, monodromy
$(-1)^3=-1$: there is no nonzero flat real section --- equivalently, no consistent
$\pm1$ anti-alignment colouring closes the loop --- so both $H^0_{\R}$ and
$\Hone_{\R}$ vanish and $\Minc\equiv0$. \textbf{Right:} the control $4$-cycle,
monodromy $(-1)^4=+1$: a nonzero flat section exists and the loop closes
consistently. This is the structural reason behind the deterministic, uniform-$c$
skeleton of Figure~\ref{fig:unification}'s specific $n=3$ versus $n=4$ instance.}
\label{fig:monodromy}
\end{figure}

The scope of the remark sharpens Proposition~15a rather than
softening it. It is a real-coefficient statement only. The contextuality of that same
odd cycle is not a fact about $\R$-cohomology at all; it is the $\Zt$ and
cut-polytope fact of parts~(ii) and~(iii). So the monodromy law is the structural
reason a real-cokernel seminorm cannot see contextuality --- an independent route
to the refutation, arriving from holonomy rather than from the
seminorm-versus-cone argument.

\medskip\noindent\textbf{Contribution and inherited results.} The
polytope machinery is prior art: that the noncontextual correlation
polytope is the cut polytope is established in polyhedral combinatorics (Barahona \& Mahjoub
1986); the contextual fraction is Abramsky, Barbosa \& Mansfield's (2017); its reading, up to a
constant, as an $\ell_1$ distance to the noncontextuality polytope follows from Dzhafarov,
Kujala \& Cervantes's (2020) Theorem~15 with Cervantes's (2023) $\CF=2\,\CNT$; and the all-$p$
scaling of that distance is Dzhafarov, Kujala \& Cervantes's (2020, Eq.~(56)). This paper adopts all of it. Its own contribution
is narrower: the \emph{refutation} that the binding
sheaf's geometric $\Minc$ co-vanishes with $\CF$, together with the
seminorm-versus-cone argument showing that no fixed linear readout can; the exact
severity link of Equation~\eqref{eq:severity} in the perceptual binding setting;
the cut-polytope route to Equation~\eqref{eq:lpdistance} with an explicit
simultaneously-optimal nearest point $x=\gamma-(V/n)\epsilon$; and the observation that
the correct co-vanishing partner is the convex obstruction.

The consequence for the title is a scoping condition rather than a retraction.
``One obstruction'' is supported provided $\Minc$ denotes the convex
correlation-polytope obstruction $\Minc^{\star}$. The seminorm of
\S\ref{sec:framework} is a different functional and must not be conflated with this one.

All numerical claims in this appendix are reproduced by
\code{check\_vanishing\_equivalence.py}: zero co-vanishing mismatches over $500$
random correlation vectors at each of $n=3,4,5$; and, separately, at each of
$n=3,5,7$, the odd-$n$ thresholds $c^{*}=-(n-2)/n$ coinciding for $\CF$ and
$\Minc^{\star}$, with the ratio in Equation~\eqref{eq:severity} returning $2.000$
throughout.

\section{The one-bit forced-choice encoding, worked in full}
\label{app:encodingb}
\setcounter{figure}{0}\setcounter{table}{0}\setcounter{equation}{0}

Section~\ref{sec:arenas} rejects the obvious version of the perceptual task --- one
forced choice per pairing, ``is A nearer than B?'' --- and adopts two separately
calibrated binary judgments instead. That rejection deserves more than an assertion,
because the one-bit task is the natural design, it is what the intransitivity
literature actually runs, and it is what an earlier draft of this paper proposed. This
appendix therefore works the one-bit encoding out completely: its exact obstruction
margin for arbitrary, non-uniform response probabilities, the geometry of the region
where it is contextual, what its within-arena test can and cannot detect at realistic
trial counts, and what becomes of the cross-domain correlation under it. The
one-bit encoding is tractable, and its mathematics is cleaner than the two-judgment
version's. Its measured quantity, however, is not the one the design is trying to
measure, and three separate consequences of that mismatch are severe enough to decide
the question.

Throughout, write the two encodings as \textbf{A} (two binary judgments per pairing,
the design adopted in \S\ref{sec:arenas}) and \textbf{B} (one forced choice per
pairing, expanded into complementary winner/loser outcomes). Every numerical statement
in this appendix is reproduced by \code{p32\_encoding\_b\_full\_analysis.py}, with two
exceptions named where they arise: the five figures in B.5(c) come from
\code{scripts/p32\_appendixB\_P\_check.py}, and B.3(iv) from
\code{scripts/p32\_v5\_encodingB\_linear\_ordering\_check.py}. Every
algebraic statement is independently re-derived in a computer-algebra engine by
\code{p32\_encoding\_b\_wolfram.wl}, and B.3(iv) additionally by
\code{scripts/p32\_v5\_lop\_equivalence.wl}.

\subsection*{B.1 The system, and why only one parameterisation survives}

In pairing $\{A,B\}$ the subject picks one stimulus. Encoding B reads that single
response as two complementary $\pm1$ outcomes: the chosen stimulus scores $+1$, the
unchosen one $-1$. The design then has exactly three free parameters, the win
probabilities
\[
  p_1=\mathbb{P}(A\!\succ\!B),\qquad p_2=\mathbb{P}(B\!\succ\!C),\qquad
  p_3=\mathbb{P}(C\!\succ\!A),
\]
which we carry in the \emph{signed dominance strengths} $q_i=2p_i-1\in[-1,1]$. A
strength of $q_i=0$ means the subject is at chance on that pairing; $q_i=1$ means the
first-named stimulus always wins. In plain terms: the whole perceptual arena is
described by how reliably each of the three comparisons goes one way.

The first consequence is immediate and it is the source of everything that follows.
Because the two outcomes in a pairing are complementary by construction, the
within-context correlation is $-1$ identically, whatever $p_i$ is. The
frustration term of \S\ref{sec:ncycle} --- the maximum over odd sign patterns, minus
$n-2$ --- is therefore \emph{pinned}:
\begin{equation}
  F \;=\; s_{\max} - (n-2) \;=\; 3-1 \;=\; 2
  \qquad\text{for every }(p_1,p_2,p_3).
  \label{eq:Bfrust}
\end{equation}
Encoding A leaves the within-context correlation $c$ free, and it is precisely the
spread in $c$ that the power analysis of \S\ref{sec:tests} is indexed by. Encoding B
removes that parameter. \textbf{No individual differences in frustration survive the
one-bit encoding}, because frustration has no room left to vary. Whatever varies
between subjects must therefore enter through the other term.

\subsection*{B.2 The exact margin, for arbitrary non-uniform strengths}

The direct-influence term $\Delta$ is what remains: it measures how far a content's mean
shifts between the two pairings it appears in. Under Encoding B each content is the
winner-indicator in one pairing and the loser-indicator in the next, so its two
context-conditional means are $2p_i-1$ and $1-2p_j$. Summing the three absolute
differences and using \eqref{eq:Bfrust} gives the margin in closed form. Writing
$\Vs = F-\Delta$ as in \S\ref{sec:ncycle},
\begin{equation}
  \boxed{\;\Vs \;=\; 2 \;-\; \bigl(\,|q_1+q_2| \;+\; |q_2+q_3| \;+\; |q_3+q_1|\,\bigr).\;}
  \label{eq:Bmargin}
\end{equation}
In words: the obstruction margin starts at its ceiling of $2$ and is reduced by how
strongly the three comparisons are decided, counted pairwise. The system is contextual
exactly when that reduction is less than the ceiling. This holds for arbitrary,
unequal $p_i$; it is not a uniform-strength special case. It is proved over the whole
unit cube by quantifier elimination
(\code{Resolve[ForAll[\ldots]]} returns \texttt{True}), and confirmed on an exact
rational grid of $1331$ triples with zero mismatches, and again --- on eight
deliberately non-uniform points --- by an exact-rational linear program over the
$2^{6}$ global assignments to the Cervantes consistification, a route that uses
neither $s_{\max}$, nor $\Delta$, nor any closed form. This encoding signals at
every $q\neq0$, because $\Delta=\lVert Mq\rVert_1$ is a norm (B.3) and vanishes only when
every comparison is a fair coin, so computing
$\CF$ on it here uses the same license already established in \S\ref{sec:tests}:
Cervantes (2023, Thm.~1) proves $\CF=2\,\CNT$ for cyclic systems generally, via the
consistification, whether or not the system signals.

Setting all three strengths equal recovers the special case quoted in
\S\ref{sec:arenas}: $\Vs = 2-6|2p-1|$, contextual exactly when
$\tfrac13<p<\tfrac23$. That is the statement the reader has already met; equation
\eqref{eq:Bmargin} is the general version it is a slice of.

\subsection*{B.3 The geometry: a norm, and a two-thirds ball}

\begin{figure}[htbp]\centering
\includegraphics[width=0.62\linewidth]{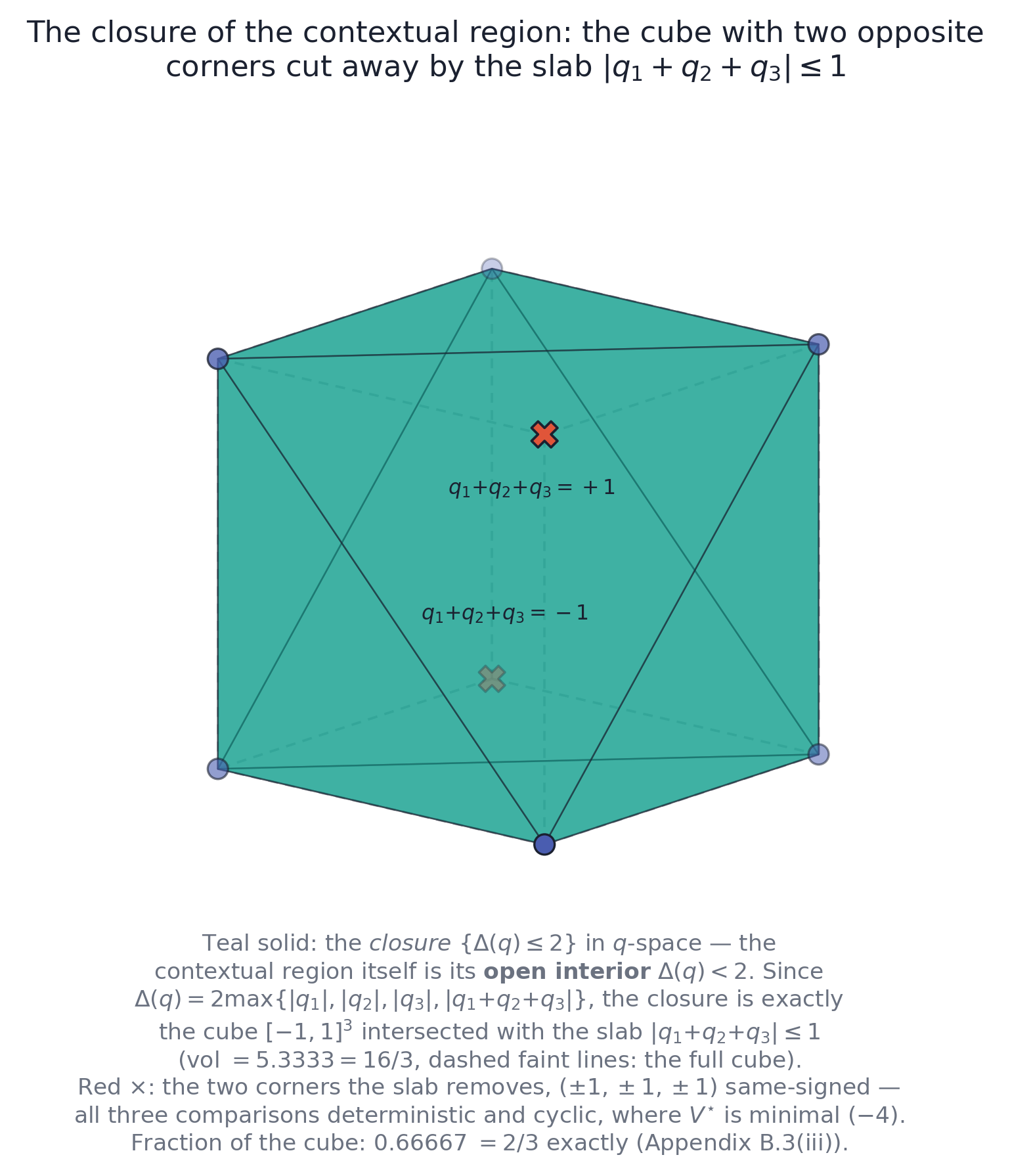}
\caption{The two-thirds ball of B.3(iii), made concrete. Since
$\Delta(q)=2\max\{|q_1|,|q_2|,|q_3|,|q_1{+}q_2{+}q_3|\}$ (dual $\ell_1$
representation of $M$), its closure $\{\Delta\le2\}$ is exactly the cube
$[-1,1]^3$ intersected with the slab $|q_1{+}q_2{+}q_3|\le1$ --- drawn here; the
\emph{contextual region itself is the open interior} $\Delta(q)<2$. Inverting $M$
sends the $u$-space octahedron's six vertices $(\pm2,0,0)$, $(0,\pm2,0)$,
$(0,0,\pm2)$ to exactly six of the cube's eight corners in $q$-space. The two the
slab removes are $(1,1,1)$ and $(-1,-1,-1)$ (red $\times$): the respondents whose
three comparisons are all deterministic \emph{and} all cyclic --- the perfect
intransitive cycle --- and they are exactly where $\Vs$ attains its minimum $-4$.
The structure the task is built to elicit is the structure Encoding B scores as
maximally \emph{non}contextual, and that is the construct-validity failure B.3(iv)
then states as a theorem: this solid is the linear-ordering polytope in disguise, carried
here by the affine change of variables $q_i=2p_i-1$. Each removed corner is a tetrahedron of volume $\tfrac{2\cdot2\cdot2}{6}=\tfrac43$,
so the retained closed solid has volume $8-2(\tfrac43)=\tfrac{16}{3}$ against the
cube's $8$ --- the $2/3$ of part~(iii) as an exact volume argument rather than
only an integral.}
\label{fig:normball}
\end{figure}

Equation~\eqref{eq:Bmargin} has more structure than it first shows. The three pairwise
sums are a linear image $u=Mq$ of the strength vector, with
$M=\bigl(\begin{smallmatrix}1&1&0\\0&1&1\\1&0&1\end{smallmatrix}\bigr)$ and $\det M=2$.
Since $M$ is invertible, $\Delta(q)=\lVert Mq\rVert_1$ is a genuine \textbf{norm} on
$\mathbb{R}^3$ --- not merely a seminorm --- and the contextual region
$\{\Vs>0\}$ is exactly its open ball of radius $2$: an open, convex, origin-symmetric
polytope. The dual $\ell_1$ representation makes this norm's shape explicit: as
$\sigma$ ranges over $\{\pm1\}^3$, the eight vectors $\tfrac12M^T\sigma$ are exactly
$\pm e_1,\pm e_2,\pm e_3,\pm(1,1,1)$, so
\begin{equation}
\Delta(q)=\lVert Mq\rVert_1=2\max\{|q_1|,|q_2|,|q_3|,|q_1+q_2+q_3|\},
\label{eq:maxform}
\end{equation}
and the closed ball $\{\Delta\le2\}$ is exactly the cube $[-1,1]^3$ intersected with
the slab $|q_1+q_2+q_3|\le1$ --- the object Figure~\ref{fig:normball} draws.

\textbf{This $M$ is Appendix~A's coboundary.} $M=I+S$ for the same cyclic shift $S$,
$(Ss)_i=s_{i+1}$, used in Proposition~15a's proof: $\dz=I+S$ there too. The same
odd-$n$ invertibility, $\det(I+S)=1-(-1)^n=2$, has opposite-looking consequences in
the two appendices. In Appendix~A it kills the real cokernel, so $\Pi_{\mathrm{coker}\,\dz}$
is the zero map and $\Minc\equiv0$ on the frustrated odd cycle. Here the same
invertibility is exactly what makes $\Delta=\lVert(I+S)q\rVert_1$ a genuine norm
rather than a seminorm, so the contextual region is a full-dimensional ball instead
of collapsing to a subspace. One operator, two measurement spaces: the reason the
real-linear residual cannot see contextuality on the odd cycle (Appendix~A) is the
same invertibility that makes Encoding B's contextual region full-dimensional here.

The geometry has four consequences.

\textbf{(i) The contextual region lies strictly inside the cube.} Equation
\eqref{eq:maxform} gives $2|q_i|\le\Delta$ for every $i$, hence $|q_i|\le\Delta/2$
directly, rather than only via the inverted-$M$ formula $q_1=(u_1-u_2+u_3)/2$. If
$\Delta<2$ then $|q_i|<1$ throughout. The physical constraint $|q_i|\le1$ is
therefore never binding, and the contextual set is an unclipped polytope.

\textbf{(ii) A single reliable comparison ends contextuality.} Read the same bound
backwards: if even one pairing is judged deterministically, $|q_i|=1$, then
$\Delta\ge2$ and $\Vs\le0$. One comparison the subject always gets the same way is
enough to make the whole system noncontextual, no matter how the other two behave.

\textbf{(iii) Two-thirds of the parameter space is contextual.} The ball has volume
$\operatorname{vol}\{\lVert u\rVert_1<2\}/|\det M| = (32/3)/2 = 16/3$, against the
cube's $8$. So under a uniform prior on the three win probabilities, a randomly drawn
respondent is contextual with probability exactly $2/3$ --- confirmed to four decimal
places by a four-million-point Monte-Carlo run ($0.66659$, standard error $0.00024$,
against $2/3=0.66667$). A criterion that two-thirds of all possible respondents
satisfy is not a demanding one.

\textbf{(iv) The contextual region is exactly the interior of the linear-ordering
polytope, so Encoding B measures representability by a total order --- from the inside.
{[}R, \code{p32\_v5\_encodingB\_linear\_ordering\_check.py}{]}} The one-variable-per-context
reading of the same forced choices, discussed in \S\ref{sec:arenas}, has its own
classical obstruction: the choice probabilities $p_i$ fail to be a mixture of strict
orders on $\{A,B,C\}$, that is, they fall outside the \emph{linear-ordering polytope}
$\mathrm{LOP}_3$ --- the mixture-of-orders model that Regenwetter, Dana \&
Davis-Stober (2011) argue is the right null for apparent intransitivity. That polytope
is by definition the convex hull of the six strict orders' deterministic choice vectors.
For three objects its facet description needs no citation, because the six vectors can
simply be listed: they are exactly the six vertices of the unit cube whose coordinates
sum to $1$ or to $2$, the omitted pair being $(0,0,0)$ and $(1,1,1)$. The planes
$\sum_i p_i=1$ and $\sum_i p_i=2$ each meet the cube in a triangle whose corners are
themselves cube vertices, so cutting the cube with $1\le\sum_i p_i\le2$ introduces no new
vertex and removes exactly those two corners. Hence
$\mathrm{LOP}_3=[0,1]^3\cap\{1\le\sum_i p_i\le2\}$. Since $q_i=2p_i-1$ gives
$\sum_i q_i = 2\sum_i p_i - 3$, those two inequalities are precisely
$\lvert\sum_i q_i\rvert\le1$; and by \eqref{eq:maxform}, $\Delta(q)\le2$ is precisely
the conjunction of $\lvert q_i\rvert\le1$ for each $i$ with $\lvert\sum_i q_i\rvert\le1$
--- the cube and the slab. Each step is an equivalence, so, writing $\phi(p)=2p-\mathbf{1}$
for the affine bijection that carries choice probabilities to strengths,
\begin{gather*}
  \{q:\Delta(q)\le 2\}\;=\;\phi(\mathrm{LOP}_3),\\
  \Vs>0 \iff p \text{ lies in the \emph{interior} of }\mathrm{LOP}_3 .
\end{gather*}
In words: under Encoding B a respondent is scored contextual if and only if their
pairwise choices \emph{can} be written as a mixture of strict orders, strictly. This is
a biconditional, not a one-way implication --- the two boundaries coincide, and
B.3(ii)'s deterministic points $\lvert q_i\rvert=1$ sit on both of them at once. It also
explains (iii): the $2/3$ there is the known volume fraction of $\mathrm{LOP}_3$, the
same object reached a second way.

\textbf{Scope of the equivalence.} \emph{Sufficient and necessary:} Encoding B
contextuality and violation of $\mathrm{LOP}_3$ --- stochastic intransitivity in the
linear-ordering sense --- are mutually exclusive, and exhaustively so up to the shared
boundary. A respondent whose pairwise choices admit no representation as a mixture of
total orders is scored \emph{non}contextual. \emph{Not licensed:} the result does not
extend to \emph{weak} stochastic transitivity, and the difference is not academic.
$q=(\tfrac15,\tfrac15,\tfrac15)$, i.e.\ $p=(\tfrac35,\tfrac35,\tfrac35)$, is a strict
majority cycle $A\!\succ\!B\!\succ\!C\!\succ\!A$ with $\Vs=\tfrac45>0$: contextual under
Encoding B and weakly intransitive at once, because $\mathrm{LOP}_3$ admits majority
cycles as far as the Condorcet mixture $\sum_i p_i=2$. Any reading of this result as
``Encoding B cannot fire on an intransitive respondent'' is therefore wrong; the
statement is about the polytope.

\textbf{The sharpest of the four findings.} B.1, B.4 and~(iii) say the
one-bit encoding has no between-subject variance in frustration, inverts its own control,
and fires on two-thirds of all possible respondents. This one says what it measures
instead: after an affine change of variables, $\Vs$ is the margin by which a respondent
\emph{succeeds} at being a mixture of total orders. The statistic is not a weak measure
of cyclic frustration --- it is a measure of the opposite quantity, and the maximum
$\Vs=2$ at $q=0$ (every comparison a fair coin) is where order-representability is
least strained rather than most.

\textbf{Independent checks and negative controls.} Two routes that share
no algebra. \code{p32\_v5\_encodingB\_linear\_ordering\_check.py} decides
membership in the interior of $\mathrm{LOP}_3$ by exact-rational linear programming over
the six order vertices, with six exact probes off each point, never using the facet
description: zero mismatches against $\{\Vs>0\}$ on two
incommensurate exact grids ($9261$ and $3375$ points). \code{p32\_v5\_lop\_equivalence.wl}
never uses the linear program, proving the set identity by containment plus equal volume
($2/3$ both ways). Six negative controls, all of which fired: a slab bound of $1/2$
instead of $1$ produces $2200$ grid mismatches; the slab without the cube fails; the
whole cube has volume $1$, not $2/3$; adding $(1,1,1)$ to the vertex set breaks
containment; the weak-transitivity misreading above is returned as a mismatch rather
than passed over; and the gate's counterexample to the retracted claim
($q=0$, $c=-0.70$, contextual under both encodings) still comes back false, confirming
this result does not smuggle that claim back in.

The maximum of $\Vs$ over the cube is $2$, attained at the single point $q=0$: every
comparison a fair coin. The minimum is $-4$, attained when all three comparisons are
deterministic and cyclic. \textbf{Maximal contextuality under Encoding B is exactly
maximal indecision}, and this is now a statement about the unique maximiser, not an
observation about one slice.

\subsection*{B.4 The control inversion --- the finding that decides the question}

Section~\ref{sec:arenas} already notes that Encoding B inverts the psychological
reading along the uniform-strength line. The general form is worse than that, and it
falls on the part of the design that carries the discriminant evidence.

\begin{figure}[htbp]\centering
\includegraphics[width=\linewidth]{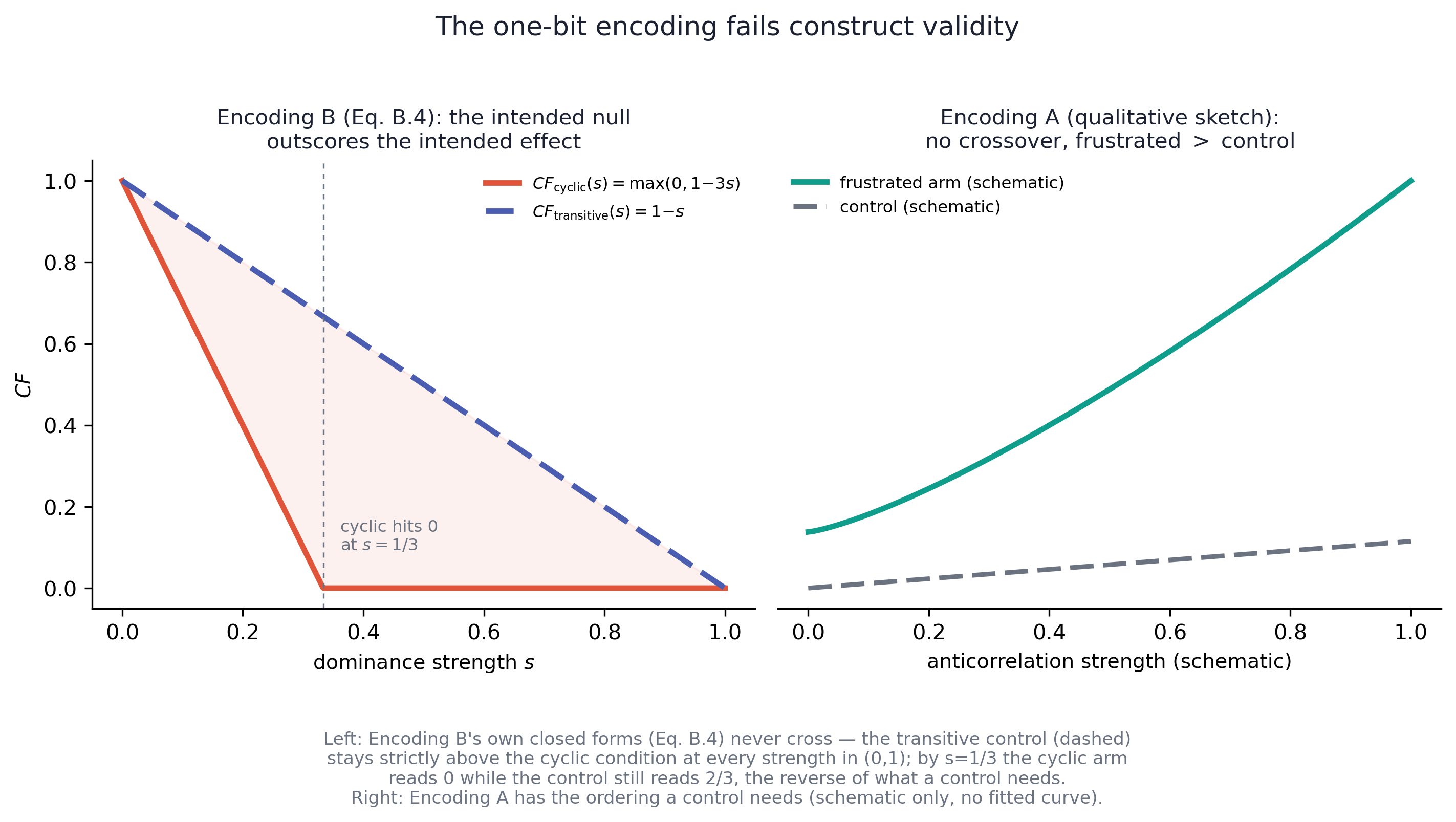}
\caption{\textbf{Left:} Equation~\eqref{eq:Binversion}'s two closed forms plotted against
dominance strength, making the inversion stated above visible: the two curves never cross,
and the intended null (dashed) sits strictly above the intended cyclic effect at every
strength between $0$ and $1$ --- by $s=1/3$ the cyclic arm has already fallen to
$\CF=0$ while the null still reads $2/3$. The control outscores the effect it exists to
control for. \textbf{Right:} the ordering Encoding A is designed to give, the frustrated
arm above the control throughout --- shown schematically (no closed form is fitted or
claimed here; contrast with the exact curves on the left). This is why the one-bit encoding is rejected on
construct-validity grounds even though every quantity in it is well defined and
correctly computed.}
\label{fig:encodingAB}
\end{figure}

Compare the two arms the within-arena test needs, at a matched dominance strength
$s\in(0,1)$. The \emph{frustrated} cyclic triple is $q=(s,s,s)$: A beats B, B beats C,
C beats A, each with probability $(1+s)/2$. The \emph{transitive control} is
$q=(s,s,-s)$: A beats B, B beats C, and A beats C, the coherent total order that is
supposed to score zero. Equation~\eqref{eq:Bmargin} gives both in closed form:
\begin{equation}
  \CF_{\text{cyclic}}(s)=\max\bigl(0,\,1-3s\bigr),
  \qquad
  \CF_{\text{transitive}}(s)=1-s .
  \label{eq:Binversion}
\end{equation}
The difference is $\min(2s,\,1-s)$, strictly positive for every $s$ in $(0,1)$.
\textbf{The intended null control scores strictly higher than the intended
experimental arm at every dominance strength there is.} At $s=0.6$ the frustrated
triple reads $\CF=0$ and the transitive control reads $\CF=0.4$. Both closed forms are
confirmed point by point against the exact-rational linear program, and
Figure~\ref{fig:encodingAB} plots them against the ordering Encoding A is designed to
give: the two curves never cross, and their order is the wrong way round.

The inversion comes from the encoding, not from an artifact of the statistic. Under
Encoding B, $\Delta$ measures whether each stimulus wins at the same rate against both
of its opponents. In a transitive triple only the middle element is inconsistent ---
B loses to A and beats C --- so one content contributes and $\Delta=2s$. In a cyclic
triple every content wins one and loses one, so all three contribute and
$\Delta=6s$. The one-bit encoding turns the contextual fraction into a measure of
\emph{transitivity-induced marginal shift}, and cyclic structure is the configuration
that maximises it. The quantity is well defined and correctly computed; it simply
points the other way.

A negative control confirms this is a property of the encoding and not of $\CF$: run
the identical comparison under Encoding A, where the marginals are balanced and
$\Delta=0$, and the frustrated arm out-scores the transitive control at every
anticorrelation strength tested, which is the ordering the design requires.

\subsection*{B.5 The within-arena test and what it can detect}

Encoding B still supports a pre-specified within-arena test, and it is worth being precise about
what that test would be, because it is not the one \S\ref{sec:tests} describes. The
null is composite --- $H_0\!:\Delta\ge2$ against $H_1\!:\Delta<2$ --- so size must be
evaluated at the least-favourable boundary $\Delta=2$, not at a convenient interior
point. Estimating each $p_i$ from $T$ trials per pairing and forming $\Vs$ with a
delta-method standard error gives the operating characteristics in
Table~\ref{tab:Bpower}.

\begin{table}[htbp]
\centering
\footnotesize
\setlength{\tabcolsep}{4pt}
\begin{tabular}{lrrrrrr}
\hline
 & $\Delta$ & $T{=}40$ & $T{=}80$ & $T{=}160$ & $T{=}320$ & $T{=}640$\\
\hline
\multicolumn{7}{l}{\emph{Size} (nominal $0.05$, one-sided)}\\
\quad uniform bdry $(\tfrac23,\tfrac23,\tfrac23)$ & $2.00$ & $0.051$ & $0.044$ & $0.047$ & $0.046$ & $0.050$\\
\quad smooth bdry $(.75,.75,.50)$ & $2.00$ & $0.050$ & $0.048$ & $0.049$ & $0.049$ & $0.048$\\
\quad \textbf{near-kink} bdry $(.90,.60,.50)$ & $2.00$ & $0.131$ & $0.097$ & $0.055$ & $0.049$ & $0.048$\\
\quad deep null $(.80,.80,.80)$ & $3.60$ & $0.000$ & $0.000$ & $0.000$ & $0.000$ & $0.000$\\
\hline
\multicolumn{7}{l}{\emph{Power}}\\
\quad fair coins $(.50,.50,.50)$ & $0.00$ & $0.967$ & $1.000$ & $1.000$ & $1.000$ & $1.000$\\
\quad weak cyclic $(.55,.55,.55)$ & $0.60$ & $0.846$ & $0.978$ & $1.000$ & $1.000$ & $1.000$\\
\quad moderate cyclic $(.60,.60,.60)$ & $1.20$ & $0.460$ & $0.677$ & $0.912$ & $0.996$ & $1.000$\\
\quad near bdry $(.64,.64,.64)$ & $1.68$ & $0.154$ & $0.207$ & $0.325$ & $0.527$ & $0.787$\\
\quad \textbf{transitive control} $(.80,.80,.20)$ & $1.20$ & $0.648$ & $0.841$ & $0.980$ & $1.000$ & $1.000$\\
\hline
\end{tabular}
\caption{Encoding B, within-arena test of $\Vs>0$: rejection rates at $40{,}000$
replicates per cell, one-sided $\alpha=0.05$. The test is well behaved in the ordinary
sense --- it holds its size at boundary points well clear of a kink and it has power. Its power runs
against the difficulty: the most readily detected configuration is
fair-coin responding, the transitive control that is supposed to be null is detected
about as easily as a genuine weak cyclic effect, and the strong cyclic dominance the
perceptual arm is built to elicit sits in the deep null and is never rejected. Size at
the uniform boundary is stable across three independent seeds
($0.0454$, $0.0453$, $0.0456$).}
\label{tab:Bpower}
\end{table}

Two entries in that table need comment, and one of them is a genuine statistical
obstacle rather than a design complaint.

\textbf{The estimator is biased against the effect.} $\Delta$ is a sum of absolute
values, so by Jensen's inequality its plug-in estimate is biased upward and $\Vs$ is
biased downward. The bias is largest exactly where the design would sit --- near
$q=0$, where the margin is greatest --- and is approximately $3\,\mathbb{E}|N(0,2/T)| = 6/\sqrt{\pi T}$ there, the large-$T$ approximation to the exact binomial value $6\binom{2T}{T}4^{-T}$: $0.378$ at $T=80$ against a maximum
margin of $2$, falling as $1/\sqrt{T}$ to $0.134$ at $T=640$ (simulated $0.3787$ and
$0.1339$, and the observed $T{=}80$ to $T{=}320$ ratio is $1.994$ against the
$\sqrt{4}=2$ the rate predicts). A negative control confirms the bias is a kink
effect and not a coding error: far from the kinks, at $q=(0.6,0.6,0.6)$ with $T=640$,
it is $+0.0004$.

\textbf{The test over-rejects near --- not at --- the non-differentiable part of the
null boundary.} $\Delta(q)=\lVert Mq\rVert_1$ has kinks wherever a pairwise sum $u_i$
vanishes, so $\Delta$ is differentiable at $q$ exactly when every $u_i\ne0$. The point
$p=(0.90,0.60,0.50)$ is \emph{not} such a kink: there $q=(0.8,0.2,0)$ and
$Mq=(1.0,0.2,0.8)$, so no absolute-value argument vanishes and $\Delta$ is
differentiable. One argument is only $0.2$ from zero, and it is that nearby kink ---
not non-differentiability at the population point --- that the small-$T$ behaviour
reflects. All three boundary rows of
Table~\ref{tab:Bpower} sit on $\Delta=2$ and differ only in how close they come to a
kink, $\min_i\lvert u_i\rvert = 2/3$, $1/2$ and $1/5$; the over-rejection is ordered the
same way. At $\min_i\lvert u_i\rvert=1/5$ the delta method understates the variance and
the true size reaches $0.131$ at $T=40$ --- $2.6$ times nominal --- decaying to nominal
by $T=320$, which is the signature of a finite-sample effect rather than an asymptotic
one. A parametric bootstrap does not repair it either: at the same point it rejects at
$0.198$, $0.184$ and $0.148$ for $T=40,80,160$, while holding $0.047$ and $0.052$ at
the \emph{uniform} boundary point $p=(\tfrac23,\tfrac23,\tfrac23)$, where
$\min_i\lvert u_i\rvert=2/3$, for $T=80,160$. At a \emph{genuine} kink the standard
bootstrap need not be consistent, because $\Delta$ is there only directionally
differentiable; at this point the distortion is instead sampling variation crossing the
nearby kink, and it shrinks with $T$. (At $T=40$ the bootstrap is mildly liberal at that
same uniform point too, $0.082$; that is reported here rather than folded into the
comparison, because at $T=40$ the two cases do not separate.)

The location of the genuine kinks matters for the design, and the kinks are more confined than they
first appear: on the boundary $\Delta=2$, a kink forces $\lvert q_i\rvert=1$ for some
$i$, i.e.\ $p_i\in\{0,1\}$. If $u_j=0$ then $q_{j+1}=-q_j$, and the two surviving terms
give $\lvert q_a+q_b\rvert+\lvert q_a-q_b\rvert=2\max(\lvert q_a\rvert,\lvert
q_b\rvert)=2$. So no respondent with strictly interior response probabilities sits on a
kink of the null boundary, and excluding the kink set by design buys almost nothing. The
exposure a pre-specified test under Encoding B actually has to price is the near-kink
band: calibrating at the least-favourable point of the whole boundary, and paying power
for it, is the remedy that survives. Verified in exact rational arithmetic with negative
controls by \texttt{scripts/p32\_b5\_kink\_verify.py}.

Finally, the score has usable between-subject spread only near the boundary. On a
population straddling it ($p\sim N(2/3,0.10)$) the true contextual fraction has
standard deviation $0.198$ and split-half reliability climbs from $0.33$ at $T=20$ to
$0.83$ at $T=160$ and $0.95$ at $T=640$ --- workable, at roughly twice the trial
budget Encoding A needs for the same reliability. But on the population the
perceptual task is actually built to produce --- strong cyclic dominance,
$p\sim N(0.80,0.06)$ --- the score has almost no between-subject variance. Population
standard deviation is $0.0005$ under independently drawn pairings
($\Pr(\CF>0)=5.930\times10^{-5}$ by closed-form calculation; a $40$M-draw Monte Carlo gives
$5.8\times10^{-5}$, consistent with the closed form to within the Monte Carlo's own
$\sim\!2\%$ relative sampling error at that event rate --- not four significant figures,
which would require upward of $2\times10^{12}$ draws). It is $0.019$ instead if a
subject-level shared latent couples the three pairings ($\Pr(\CF>0)=1.3\%$) --- the case a
cross-domain trait study actually posits. (The $0.0000$ figure
reported from a single $N=400$ sample is a finite-sampling artifact, not the population
value.) Against the boundary population's $0.198$ that is a range restriction of roughly
$396$-fold under independently drawn pairings and $10$-fold under the shared latent ---
severe enough on either reading to matter, and it is the paper's own range-restriction objection
(\S\ref{sec:tests}, leg (i)) turned against its own perceptual arm.

\subsection*{B.6 Consequences for the cross-domain test}

The cross-domain correlation is the test of the thesis, and Encoding B changes it in
two ways, one exact and one simulated.

\textbf{The perceptual score becomes pure signalling.} By \eqref{eq:Bfrust} the
perceptual frustration term $F_P\equiv2$ is a constant, so every covariance involving
it vanishes and the decomposition of \S\ref{sec:tests} collapses:
\begin{equation}
  \operatorname{Cov}(\Vs_P,\Vs_J)
  \;=\;\operatorname{Cov}(\Delta_P,\Delta_J)-\operatorname{Cov}(\Delta_P,F_J)
  \;=\;-\operatorname{Cov}(\Delta_P,\Vs_J),
  \label{eq:Bcross}
\end{equation}
and because $\Vs_P=2-\Delta_P$ is an exact decreasing affine function of $\Delta_P$,
the same holds for correlations, whenever both are defined, at every sample size rather
than asymptotically:
$\operatorname{Corr}(\Vs_P,\Vs_J)=-\operatorname{Corr}(\Delta_P,\Vs_J)$. The
shared-mechanism claim, stated above as a shared \emph{obstruction}
tolerance, would under Encoding B be a claim that judgment contextuality tracks
perceptual \emph{response-bias inconsistency across pairings} --- a different
hypothesis, and one with a much more ordinary confound. Equation \eqref{eq:Bcross} is
confirmed numerically to $7\times10^{-18}$, and a negative control in which $F_P$ is
allowed to vary and to couple to the judgment arena breaks the identity by exactly the
predicted amount $\operatorname{Cov}(F_P,F_J)-\operatorname{Cov}(F_P,\Delta_J)$.

\textbf{Residualisation stops controlling the confound.} Residualization gives the
design its discriminance (\S\ref{sec:tests}, leg (ii)) by regressing each arena's frustrated-load
score on a control-load score that estimates the general response-consistency factor
$g$. That works only to the extent that a linear adjustment on the control can track how
$g$ moves the frustrated load; the sign of that dependence does not matter, only how
well the control measures it. Under Encoding B
careless responding drives every $p_i$ toward $\tfrac12$, which by B.3 raises the
score on \emph{both} arms, monotonically, toward the same maximum. The transitive control
carries that shift through one pairwise sum only, so at $160$ trials per pairing its score has
reliability about $0.21$ and the regression under-adjusts. Simulating a cohort
in which $g$ acts as a lapse rate, with no shared obstruction at all, the residualised
correlation stays positive and the false-positive rate of the cross-domain test climbs
with sample size instead of holding at its nominal level: $0.16$ at $N=40$, $0.30$ at
$N=85$, $0.64$ at $N=200$, $0.91$ at $N=400$ ($20{,}000$ cohorts per sample size, standard
error at most $0.004$), against the $0.023$ that the one-sided $r>2/\sqrt N$ rule allows.
\textbf{The failure is control under-adjustment}, the defect \S\ref{sec:tests} prices for
Encoding A, made severe by how little $g$ the one-bit control carries. Measuring the control
better repairs it, slowly: at $N=200$ the rate falls to about $0.04$ with $2{,}560$ control
trials (reliability $0.81$) and comes within $0.005$ of the nominal rate only at $10{,}240$
(reliability $0.95$; $4{,}000$ cohorts per control length), about sixty-four times the
simulated control length. Encoding A shows the same
dependence on a smaller scale: its own $\CF$-scored test with an $80$-trial control runs at
$0.121$ against a nominal $0.025$ (\S\ref{sec:tests}). A rate of $0.02$ for Encoding A in the
same independent world comes from a linear score-level model whose control has reliability
about $0.97$, so it compares two control reliabilities, not two encodings. At the simulated
control length the failure is still the sharpest form of the problem: under Encoding B a
larger study is more likely to produce a false confirmation of the paper's central claim, not
less.

\subsection*{B.7 The verdict on the one-bit encoding}

The one-bit forced-choice design is not intractable and it is not ill-posed. It has a
clean closed form for arbitrary response probabilities, \eqref{eq:Bmargin}; an
elegant geometry, a norm ball occupying exactly two-thirds of the parameter space; and
a within-arena test with honest size at boundary points well clear of a kink and good
power. Four
findings nonetheless decide against it for this study, three of them on their own, and they
are ordered here by how much of the design each one removes.

\begin{enumerate}
\item \textbf{The frustration term is constant} (\eqref{eq:Bfrust}), so the quantity
the paper is about has no between-subject variance left to measure. Everything the
design would observe enters through direct influence instead.
\item \textbf{The control arm out-scores the experimental arm} at every dominance
strength (\eqref{eq:Binversion}). The transitive triple that is supposed to be null
reads $\CF=1-s$ while the frustrated triple reads $\max(0,1-3s)$, so the within-arena
contrast runs backwards.
\item \textbf{The cross-domain test loses false-positive control} and loses it
faster in larger samples, rising to $0.91$ at $N=400$ in a world with no shared
mechanism at all. This one is a cost rather than a verdict: about sixty-four times the
simulated control trials bring the size close to nominal (B.6).
\item \textbf{The statistic measures the wrong quantity}, and this is the one finding
that names what it measures instead. The contextual region is exactly the interior of
the linear-ordering polytope (B.3(iv)), so $\Vs$ is the margin by which a respondent
\emph{succeeds} at being a mixture of total orders --- maximal at indifference on all
three comparisons, negative on the perfect deterministic cycle. A respondent who cannot
be written as such a mixture is scored noncontextual, so this reading and the
one-variable-per-context reading of the same data never fire together. The result is
about the polytope, not about weak stochastic transitivity, which Encoding B can violate
while calling the respondent contextual.
\end{enumerate}

Each of these is a statement about Encoding B specifically; none is a criticism of the
contextual fraction, of the $n$-cycle mathematics of \S\ref{sec:ncycle}, or of the
published forced-choice datasets that \S\ref{sec:prior} recomputes --- those are
analysed as they were collected, and the same degeneracy is what
\S\ref{sec:prior} reports finding in seven of eight of them. The finding is narrower
and more useful than ``the one-bit task does not work'': it is that a single forced
choice per pairing cannot separate frustration from direct influence, because it fixes
one and leaves only the other, and the second binary judgment is precisely what buys
that separation back.

\subsection*{B.8 The same noise model under the \emph{adopted} encoding}

B.3(iv) works the mixture-of-transitive-orders null out for Encoding B, the one-bit
encoding this paper rejects. That leaves a question the rejection does not answer, and
the one \S\ref{sec:arenas} actually needs: whether a mixture of transitive orders, which
is the standard null for apparent intransitivity (Regenwetter, Dana \& Davis-Stober 2011),
can produce a positive margin under \textbf{Encoding A}, the two-judgment encoding the
design adopts. It cannot, and a short proof shows it without appeal to the Encoding B
result. It belongs here because the machinery is set up here, not because it is a statement
about Encoding B.

\medskip
\noindent\textbf{Proposition B.8 \textbf{[R]}.} \emph{Present three contents $A,B,C$ in the
three cyclic contexts $c_1=\{A,B\}$, $c_2=\{B,C\}$, $c_3=\{C,A\}$, and suppose each trial's
pair of binary judgments is generated by}
\begin{enumerate}
\item[(i)] \emph{a per-trial draw of a strict total order $\pi$ on $\{A,B,C\}$ from an
arbitrary subject-level mixture $\lambda$ over the six orders --- the mixture-of-transitive-orders
null;}
\item[(ii)] \emph{a readout that turns $\pi$ into a truth value for each content
\textbf{separately and identically in every context}: the judgment ``this content is at or
above the criterion'' is $\mathbf{1}[\mathrm{rank}_\pi(x)\le k]$ for a cut $k$ fixed across
contexts;}
\item[(iii)] \emph{independent per-judgment lapses, flipping each response with a
content-specific, context-independent rate $\epsilon_x$.}
\end{enumerate}
\emph{Then the observed empirical model admits a global joint distribution over the three
contents. Hence $\Delta=0$, $s_{\max}\le n-2$, and}
\[
  \Vs \;\le\; V \;\le\; 0, \qquad \CF \;=\; 0 ,
\]
\emph{at every $\lambda$, every cut and every lapse profile. No model of this form is
scored contextual under Encoding A.}

\medskip
\noindent\emph{Proof.} Draw $\pi\sim\lambda$ once and set
$X_x=\mathbf{1}[\mathrm{rank}_\pi(x)\le k]\oplus e_x$, with $e_A,e_B,e_C$ independent
lapses at rates $\epsilon_A,\epsilon_B,\epsilon_C$. Fix a context $\{x,y\}$. The observed
joint there is the law of the pair $\bigl(\mathbf{1}[\mathrm{rank}_\pi(x)\le k]\oplus e_x,\;
\mathbf{1}[\mathrm{rank}_\pi(y)\le k]\oplus e_y\bigr)$, because within a context the two
contents' lapses are independent of each other and of $\pi$ --- and that is exactly the law
of $(X_x,X_y)$, since the three global lapses are independent too. So $(X_A,X_B,X_C)$ is a
global section whose marginal on every context matches what is observed. By
\S\ref{sec:ncycle} (Fine 1982; Abramsky \& Brandenburger 2011) a cyclic system with a global
joint satisfies every cyclic noncontextuality inequality, i.e.\ $s_{\max}\le n-2$.
Non-signalling is immediate: neither $\lambda$, nor $k$, nor $\epsilon_x$ depends on the
context, so each content's marginal is the same in both contexts it appears in and
$\Delta=0$. The fixed facet's sum is at most $s_{\max}$, so $\Vs\le V=s_{\max}-(n-2)-\Delta\le0$. $\square$

\medskip
\noindent\textbf{Scope of Proposition B.8.} \emph{Stronger than the statement
needs:} transitivity of the drawn orders is never used. Any latent state whose readout is
context-independent produces a global section, so the whole class of latent-state models
with context-independent readout --- of which the mixture of transitive orders is one
member --- is excluded, and independent response noise of any severity cannot rescue it.
\emph{The load-bearing assumption is (ii), and it is a design feature rather than a
convenience.} Each judgment in this arena is anchored to an external criterion, not to the
co-presented stimulus (\S\ref{sec:arenas}). Drop that and the bound fails at once: if the
respondent instead answers ``does this content beat its partner?'', the two responses in a
context are mirror images, $\gamma_i=-1$ for every $i$, and $s_{\max}=3$ lets $\Vs$ reach $2$,
attained whenever $\Delta=0$, as under a uniform mixture of the six orders --- the same one-bit degeneracy \S\ref{sec:arenas} rejects on construct-validity
grounds, reached here from the noise model's side. \emph{Not licensed:} this closes the
mixture-of-orders account, not every noise account. A criterion that drifts \emph{within} a
session breaks (ii)'s fixed cut. Drift that shifts a content's marginal between contexts
enters as measured direct influence $\Delta$, which $\Vs$ subtracts; drift that tracks the
co-presented stimulus can change the joint response with $\Delta=0$ (the comparative readout
above is the extreme case, $\Vs=2$), so the subtraction does not exclude it. That exposure
is part of $F$, not of $\Delta$, so the shared-disturbance adjustment of \S\ref{sec:tests} does
not reach it, and no score in the design separates it from the pairing-dependent appearance
the arena is built to elicit. Across arenas it is one of the rival traits of \emph{A positive
result corroborates; it does not confirm} (\S\ref{sec:tests}); within an arena nothing in
this paper prices it. Lapse rates that differ between the two arenas' sessions leave (iii)
intact within each arena; they bear on the cross-domain comparison instead.

\medskip
\noindent\textbf{Exact checks and negative controls}
(\code{scripts/p32\_mixture\_of\_orders\_encodingA\_check.py}). The proposition was
evaluated in exact rational arithmetic over $4{,}000$ random order mixtures crossed with
both cuts and random lapse profiles: $\max\Delta=0$ and $\max\Vs=-0.0112$ across all draws,
with $s_{\max}$ reaching $0.9888$ against the noncontextual bound of $1$ --- close enough to
the bound that a false claim would have been visible. Three negative controls fire.
\textbf{(1)} The comparative readout of \emph{the same} order mixture gives $\Vs=0.9796$.
\textbf{(2)} A general context-dependent deterministic readout, again on the same mixtures,
reaches $V=1.1579$ on its maximising facet, with $\Delta=0.42$ (on the fixed facet
$\Vs=-2.84$ there), and gives $V>0$ in $58$ of $4{,}000$ random rules; $\Vs$ exceeds zero in
$13$ of them, at most $2/3$, and no positive-$V$ draw has $\Delta=0$. \textbf{(3)} A
genuinely contextual model, $\gamma=(-0.6,-0.6,-0.6)$ with $\Delta=0$, gives $\Vs=0.8$. One
further case is reported and is \textbf{not} a firing control: letting the criterion cut
vary by context while keeping the readout a rank threshold leaves $\Vs\le0$ in all $4{,}000$
draws. That is recorded as non-discriminating rather than as support --- what it suggests is
narrower: in these draws a merely shifting threshold was not enough, and a positive margin
appeared only once the readout became context-dependent in the co-presented-item sense.

\bigskip
\begin{center}\itshape
The stalks align as moments meet the eye,\\
each fiber holds what only parts can see.\\
When local truths refuse to coincide,\\
the seam still itches with lost geometry.
\end{center}

\end{document}